%% file: hdp.tex
\documentclass[12pt]{article}

\usepackage[default]{jasa_harvard}    % for formatting citations in text
\usepackage{JASA_manu}

\newcommand {\ctn}{\citeasnoun} % change to \citet if using natbib
\usepackage{graphicx,subfigure,amsmath,latexsym,amssymb}
\usepackage{float,epsfig,multirow,rotating,times}
\usepackage{upgreek,wrapfig}
\usepackage{comment}

\newcommand{\boeta}{\boldsymbol{\eta}}

\newcommand{\bbeta}{\boldsymbol{\beta}}

\newcommand{\bphi}{\boldsymbol{\phi}}

\newcommand{\bXi}{\boldsymbol{\Xi}}
\newcommand{\bxi}{\boldsymbol{\xi}}

\newcommand{\bzeta}{\boldsymbol{\zeta}}

\newcommand{\bG}{\boldsymbol{G}}

\newcommand{\bH}{\boldsymbol{H}}

\newcommand{\bA}{\boldsymbol{A}}

\newcommand{\bE}{\boldsymbol{E}}
\newcommand{\bp}{\boldsymbol{p}}
\newcommand{\bP}{\boldsymbol{P}}

\newcommand{\bv}{\boldsymbol{v}}

\newcommand{\bx}{\bm{x}}
\newcommand{\bX}{\boldsymbol{X}}
\newcommand{\by}{\boldsymbol{y}}
\newcommand{\bY}{\boldsymbol{Y}}

\newcommand{\bm}{\mathbf}

\numberwithin{equation}{section}
\numberwithin{algo}{section}
\numberwithin{table}{section}
\numberwithin{figure}{section}

\begin{document}

\normalsize

\title{\vspace{-0.8in}
%\LARGE\bf ORIGINAL RESEARCH PAPER
%\\[10mm]
{\bf A Non-Gaussian, Nonparametric Structure for
Gene-Gene and Gene-Environment Interactions in Case-Control Studies Based on Hierarchies of
Dirichlet Processes}}
\author{Durba Bhattacharya and Sourabh Bhattacharya\thanks{
Durba Bhattacharya is an Assistant Professor in St. Xavier's College, Kolkata, pursuing PhD in 
Interdisciplinary Statistical Research Unit, Indian Statistical
Institute, 203, B. T. Road, Kolkata 700108. Sourabh Bhattacharya is an Associate Professor in
Interdisciplinary Statistical Research Unit, Indian Statistical Institute, 203, B. T. Road, Kolkata 700108.
Corresponding e-mail: sourabh@isical.ac.in.
}}
\date{\vspace{-0.5in}}
\maketitle

\begin{abstract}

It is becoming increasingly clear that complex interactions among genes and environmental
factors play crucial roles in triggering complex diseases. Thus, understanding such interactions
is vital, which is possible only through statistical models that adequately account for such
intricate, albeit unknown, dependence structures. 

%\ctn{Bhattacharya16} attempt such modeling, relating finite mixtures composed of Dirichlet processes
%that represent unknown number of genetic sub-populations through a hierarchical matrix-normal structure
%that incorporates gene-gene interactions, and possible mutations, induced by environmental variables.
%However, the product dependence structure implied by their matrix-normal model seems to be too simple
%to be appropriate for general complex, realistic situations. 

In this article, we propose and develop a novel nonparametric Bayesian model for case-control genotype data 
using hierarchies of Dirichlet processes that offers a more realistic and nonparametric dependence 
structure between the genes, induced by the environmental variables. In this regard, we propose a novel
and highly parallelisable MCMC algorithm that is rendered quite efficient by the combination
of modern parallel computing technology, effective Gibbs sampling steps, retrospective sampling and 
Transformation based Markov Chain Monte Carlo (TMCMC). We use appropriate Bayesian hypothesis testing
procedures to detect the roles of genes and environment in case-control studies.

Applying our ideas to 5 biologically realistic case-control genotype datasets simulated under 
distinct set-ups, we obtain encouraging results in each case.
We finally apply our ideas to a real, myocardial infarction dataset, and obtain interesting results 
on gene-gene and gene-environment interaction, that broadly agree with the results reported in the literature.
\\[2mm]
{\bf Keywords:} {\it Case-control study; Hierarchical Dirichlet process; Gene-gene and gene-environment interaction;  
Myocardial Infarction; Parallel processing; Transformation based MCMC.} 
\end{abstract}

%\tableofcontents

%\pagebreak

\section{{\bf Introduction}}
\label{sec:intro}

Thanks to intense research on gene-gene interaction, including genome-wide association studies (GWAS), 
it has become increasingly clear that gene-gene interaction alone is insufficient for explaining
most complex diseases. Investigating environmental factors independently of the genetic factors 
is not sufficient either -- biomedical research points towards the importance of interactions between genes and the environment
in explaining complex diseases. Indeed, according to \ctn{Hunter05} (see aso \ctn{Mather76}), 
considering only the separate contributions of genes and environment to a disease, ignoring their interactions, 
will lead to incorrect estimation of the disease proportion (the ``population attributable fraction'') that is 
explained by genes, the environment, and their joint effect. In particular, environmental exposures are expected
to influence gene-gene interactions of the individuals. 
A comprehensive overview of gene-environment interaction with various examples is provided in \ctn{Bhattacharya16}.  

Since no simple relationship exists between the genes and environment, it is clear that linear or additive models, as are
mostly used so far, are inadequate for modeling gene-environment interactions. Also, the logistic model based approaches, (see for example  \ctn{Ahn13}, \ctn{Wen14} and \ctn{Liu15}) resting on Fisher's definition of interaction result in the inclusion of a large number of interaction terms even with a moderate number of genetic and environmental factors. 
The existing methods construct various measures of main effects and interaction
effects from the genotype data. For instance, \ctn{Larson13}, use kernel-based methods, \ctn{Wan10}, interpret interactions in terms of Kullback-Leibler divergence,
\ctn{Li15}, use entropy-based information
gain to interpret interactions, \ctn{Yi11},
use common homozygote, heterozygote and rare homozygote to model interactions, to
name only a few. Thus, there is no unique notion of such measures of main and interaction effects, and since the final results can be very much dependent on the measures of interactions employed, it does not seem to facilitate comparison with our method
of genuinely modeling interactions in clear, well-established, statistical terms based on
Hierarchical Dirichlet Process based nonparametric methodologies.
The existing methods, including those cited above, try
to model interactions through pairwise SNP-SNP interaction kernels and logistic regression, ignoring genes as functional units and are hence not scalable for higher order
interactions. Two stage approaches like the BOOST software are also SNP based, which
test interactions only on the SNPs found marginally significant in the first stage, leading to higher rate of false positives. While almost all the classical methods are based
on Fisher’s idea of GxG, existing Bayesian techniques like BEAM, EpiBN study interaction by identifying the SNPs that influence the disease risk given particular allele
combinations, ignoring the genes as functional units. In a nutshell, unlike our Bayesian approach, none of the existing methods, classical
or Bayesian, attempts simultaneous modelling of the uncertainties associated with the
genes as the functional units along with the interactions, both at SNP and gene level
through unified statistical models.
The fact that the genetic data may arise from a stratified population with an unknown number of subpopulations makes the problem all the more demanding.
\ctn{Wen14}, in their attempt to study the genetic association with respect to genetic data arising from multiple potentially-heterogeneous subgroups, assume the number of subgroups to be known in advance.
Also, the problem of quantifying the strength of heterogeneity, as acknowledged by \ctn{Wen14}, remains unanswered due to the above considerations and the need of an appropriate prior.
The Bayesian semiparametric model proposed by \ctn{Bhattacharya16} takes care of the above mentioned problems by proposing a model based on Dirichlet Processes (DP) and 
a hierarchical matrix-normal distribution, which encapsulates the mechanism of dependence among genes under environmental effects with respect to genotype data arising out of a possibly stratified population. 

We now elaborate on a possible drawback of the dependence structure induced by the modeling strategy 
of \ctn{Bhattacharya16}, which motivated us to develop our present work based on Hierarchical Dirichlet Processes. 

In their model, the relevant gene-gene covariance matrix for individual $i$ is $\tilde\sigma_{ii}\bA$, where
$\bA$ is the gene-gene interaction matrix common to all the individuals in the absence of environmental variables,
and $\tilde\sigma_{ii}=\sigma_{ii}+\phi$, with $\sigma_{ii}$ being the $i$-th diagonal element of a symmetric,
positive definite matrix not associated with the environmental variable, and $\phi$ is a non-negative parameter,
to be interpreted as the effect of the environmental variable $\bE$ on gene-gene interaction. Note that \ctn{Bhattacharya16} assumed that the covariance matrices for 
all the individuals are affected in the same way by the environmental variable, which
seems to be a limitation of the covariance structure. The enviromental variables may affect the gene-gene interactions of individuals differently depending on the extent and type of their exposure to the environmental factors.% Note that it is more
%appropriate to replace $\phi$ with $\phi_i$ to reflect the effect of the $i$-th environmental factor $\bE_i$ on gene-gene %interaction,
%but this would make the marginal distribution of the genotype dependent on $\bE_i$, which is not desirable biologically,
%except in the rare mutational case.
%Moreover, from the biological perspective, $\bE_i$ can affect only gene-gene interactions, not the marginal
%genetic effect: with $\phi_i$ replacing $\phi$, $\tilde\sigma_{ii}=\sigma_{ii}+\phi_i$, so that even the marginal variances
%are influenced. 
%Observe that, since $\sigma_{ii}$ may differ subject-wise, the marginal distribution of the genotype of the
%$i$-th individual does depend upon $i$, but as noted in \ctn{Bhattacharya16}, this realistically considers
%the issue that the subjects represent further sub-divisions of a given sub-population, and these sub-divisions
%may correspond to slightly different genotypic distributions. In fact, $\sigma_{ii}$ 
%may encapsulate a negligible effect of $\bE_i$.

In this article, we introduce a novel Bayesian nonparametric model for gene-gene and gene-environment interactions
for case-control genotype data that solves the issues detailed above. Our model represents the individual genotype data
as finite mixtures based on Dirichlet processes as before, but instead of the hierarchical matrix normal 
distribution, we introduce a hierarchy of Dirichlet processes that create appropriate nonparametric dependence
among the genes induced by the environment, case-control dependence, and dependence among the individuals.
As we show, our modeling strategy satisfies all the desirable properties, bypassing the drawbacks of the 
matrix-normal based model of \ctn{Bhattacharya16}. Although our hierarchical Dirichlet process (HDP) model has parallels
with the HDP introduced by \ctn{Teh06}, our HDP has one more level of hierarchy compared to \ctn{Teh06}. Moreover,
we develop a novel and highly parallelisable Markov Chain Monte Carlo (MCMC) methodology that combines the 
efficiencies of modern parallel computing infrastructure, Gibbs steps, retrospective sampling methods, 
and Transformation based Markov Chain Monte Carlo (TMCMC).
For the hypothesis testing procedures, we essentially adopt and extend the ideas provided in \ctn{Bhattacharya16}. 
Application of our model and methods to five simulation experiments for the validation purpose 
yielded quite encouraging results, and application to a real myocardial infarction (MI) case-control type dataset yielded
results that are broadly in agreement with the results reported in the literature, but provided new and
interesting insights into the mechanisms of gene-gene and gene-environment interactions.

The rest of our paper is structured as follows. 
We introduce our HDP-based Bayesian nonparametric gene-gene and gene-environment interaction model
in Section \ref{sec:proposal}, and in Section \ref{sec:hdp_dependence} discuss the relevant dependence
structures induced by our model. 
%In Section \ref{sec:inferential_procedure} we propose an MCMC methodology
%for fitting our model, and in Section \ref{sec:computation} we propose a parallel algorithm for implementing 
%the proposed MCMC method.
%The methodology combines efficient Gibbs sampling strategies with an efficient TMCMC scheme, while
%exploiting the conditional independence structure of our model to build a highly efficient parallel MCMC
%algorithm.
In Section \ref{sec:detection} we extend the Bayesian hypothesis testing procedures proposed in \ctn{Bhattacharya16} 
to learn about the roles of genes, environmental variables and their interactions in case-control studies, with respect to our current HDP model.
In Section \ref{sec:simulation_study} we briefly discuss the results of application of our model and methodologies to $5$ biologically realistic simulated
data sets, the details of which are provided in the supplement, described below.
%we apply the developments
%to five biologically realistic simulated data sets associated with five different set-ups, and obtain encouraging
%results, thereby validating our model and methods. 
In Section \ref{sec:realdata} we analyze the real MI dataset using
our ideas, demonstrating quite interesting and insightful outcome.
Finally, we summarize our work with concluding remarks in Section \ref{sec:conclusion}.

Additional details are provided in the supplement, whose sections and figures have the prefix
``S-" when referred to in this paper.

\section{{\bf A new Bayesian nonparametric model for gene-gene and gene-environment interactions}}
\label{sec:proposal}

\subsection{{\bf Case-control genotype data}}
\label{subsec:data}

For $s=1,2$ denoting the two chromosomes, let $y^s_{ijkr}=1$ and $y^s_{ijkr}=0$ indicate 
the presence and absence of the minor allele of 
the $i$-th individual belonging to the $k$-th group (either control or case), for $k=0,1$, with $k=1$ denoting case; at the $r$-th locus of $j$-th gene, where $i=1,\ldots,N_k$; $r=1,\ldots,L_j$ and $j=1,\ldots,J$; let $N=N_0+N_1$.
Let $\bE_i$ denote a set of environmental variables associated with the $i$-th individual.
In what follows, we model this case-control genotype and the environmental data using our
Bayesian semiparametric model, described in the next few sections.

%In this paper, we shall concern ourselves with data sets of the aforementioned type. However,
%for our model, which we introduce below, it is obvious that data sets consisting of only minor allele counts at each locus
%contains exactly the same information as the above described data type.

\subsection{{\bf Mixture models based on Dirichlet processes}}
\label{subsec:mixtures}
Let $\by_{ijkr}=(y^1_{ijkr},y^2_{ijkr})$, and if $L=\max\{L_1,\ldots,L_J\}$, let
$\bY_{ijk}=(\by_{ijk1},\by_{ijk2},\ldots,\by_{ijkL_j})$
and $\tilde{\bY}_{ijk}=(\tilde\by_{ijk,{L_j+1}},\ldots,\tilde\by_{ijkL})$,
where $\tilde{\bY}_{ijk}$ %$(\tilde\by_{ijk,{L_j+1}},\ldots,\tilde\by_{ijkL})$ 
are unobserved and assumed to be missing. We introduce these unobserved variables 
to match the number of loci for all the genes, which is required so that the vectors of minor allele
frequencies come from the distribution having the same dimension. This ``dimension-matching" is required
for the theoretical development of our modeling ideas; see (\ref{eq:dp1}) and (\ref{eq:dp2}).

We assume that for every triplet $(i,j,k)$, 
$\bX_{ijk}=(\bx_{ijk1},\ldots,\bx_{ijkL)}=(\bY_{ijk},\tilde\bY_{ijk})$ 
%are independently distributed with
%mixture probability mass function with a {\it maximum} of $M$ components, given by 
have the mixture distribution
\begin{equation}
[\bX_{ijk}]=\sum_{m=1}^M\pi_{m ijk}\prod_{r=1}^{L}f\left(\bx_{ijkr}\vert p_{m ijkr}\right),
\label{eq:mixture1}
\end{equation}
where $f\left(\cdot\vert p_{m ijkr}\right)$ 
%is the probability mass function of independent Bernoulli
%distributions, given by
is the Bernoulli mass function given by
\begin{equation}
f\left(\bx_{ijkr}\vert p_{m ijkr}\right)=
\left\{p_{m ijkr}\right\}^{x^1_{ijkr}+x^2_{ijkr}}
\left\{1-p_{m ijkr}\right\}^{2-(x^1_{ijkr}+x^2_{ijkr})}.
\label{eq:pmf1}
\end{equation}
In the above, $M$ denotes the {\it maximum} number of mixture components and $p_{mijkr}$ stands for the minor allele frequency at the $r$-th locus of the $j$-th
gene for the $i$-th individual of the $k$-th case/control group. Note that minor allele frequency is the
frequency at which the second most common allele occurs in a given population.

Allocation variables $z_{ijk}$, with probability distribution
\begin{equation}
[z_{ijk}=m]=\pi_{m ijk},
\label{eq:alloc_z}
\end{equation}
for $i=1,\ldots,N_k$ and $m=1,\ldots,M$, allow representation of (\ref{eq:mixture1}) as
\begin{equation}
[\bX_{ijk}|z_{ijk}]=\prod_{r=1}^{L}f\left(\bx_{ijkr}\vert p_{z_{ijk} ijkr}\right).
\label{eq:x_given_z}
\end{equation}
Following \ctn{Majumdar13}, \ctn{Bhattacharya15}, we set $\pi_{m ijk}=1/M$, for $m=1,\ldots,M$,
and for all $(j,k)$.
%We may assume appropriate Dirichlet distribution priors on $\left(\pi_{1ijk},\ldots,\pi_{Mijk}\right)$
%for $j=1,\ldots,J$; $i=1,\ldots,N_k$; $k=0,1$. 
%However, as investigated in \ctn{Majumdar13}, the Dirichlet distribution often yields very small values
%of the probabilities $\bpi_{m ijk}$, thereby tending to underestimate the true number of mixture components.
%On the other hand, setting $\bpi_{m ijk}=1/M$ exhibited much better performance.
%Therefore, in this work, we set $\pi_{m ijk}=1/M$, for $m=1,\ldots,M$,
%and for all $(j,k)$. 

Letting $\bp_{m ijk}=\left(p_{m ijk1},p_{m ijk2},\ldots,p_{m ijkL}\right)$, we next assume that 
\begin{align}
\bp_{1ijk},\bp_{2ijk},\ldots,\bp_{Mijk}&\stackrel{iid}{\sim} \bG_{ijk};\label{eq:dp1}\\
\bG_{ijk}&\sim \mbox{DP}\left(\alpha_{G,ik}\bG_{0,jk}\right),\label{eq:dp2}
\end{align}
where $\mbox{DP}\left(\alpha_{G,ik}\bG_{0,jk}\right)$ stands for Dirichlet process
with expected probability measure $\bG_{0,jk}$ having precision parameter $\alpha_{G,ik}$, with
\begin{equation}
\log(\alpha_{G,ik})=\mu_{G}+\bbeta^T_{G}\bE_{ik},
\label{eq:env_regression1}
\end{equation}
where $\bE_{ik}$ is a $d$-dimensional vector of continuous environmental variable for the $i$-th individual
in the $k$-th group,
$\bbeta_G$ is a $d$-dimensional vector of regression coefficients, and $\mu_G$ is the intercept term.
The model can be easily extended to include categorical
environmental variables along with the continuous ones.

We clarify a further property of our $M$-component mixture model (\ref{eq:mixture1}). It follows
directly from \ctn{Sabya13} that conditional on $G_{ijk}$, (\ref{eq:mixture1})
has the same form as the traditional infinite-dimensional Dirichlet process mixture model
independent of $M$, but unlike the latter case, where the data are $iid$, in our case the data
have a joint dependence structure involving $M$.

\subsection{{\bf Hierarchical Dirichlet processes to introduce dependence between the genes
and case-control status}} 
\label{subsec:hdp}

We further assume that for $k=0,1$, %for some positive integer $\tilde M$,
\begin{align}
\bG_{0,jk}\stackrel{iid}{\sim}& DP\left(\alpha_{G_0,k} \bH_k\right);~j=1,\ldots,J,\label{eq:hdp0}
%\bG_{0,k\ell}\stackrel{iid}{\sim}&DP\left(\alpha_{G_0,k} \bH_k\right);~\ell=1,\ldots,\tilde M;\label{eq:hdp1}
\end{align}
where
\begin{equation}
\log(\alpha_{G_{0,k}})=\mu_{G_0}+\bbeta^T_{G_0}{\bar\bE}_k,
\label{eq:env_regression2}
\end{equation}
with
\begin{equation}
{\bar\bE}_k=\frac{1}{N_k}\sum_{i=1}^{N_k}\bE_{ik}.
\label{eq:env_average1}
\end{equation}
We postulate the last level of hierarchy as
\begin{equation}
\bH_k\stackrel{iid}{\sim}DP\left(\alpha_H \tilde \bH\right);~k=0,1,\label{eq:hdp2}
\end{equation}
where
\begin{equation}
\log(\alpha_H)=\mu_H+\bbeta^T_{H}\bar{\bar\bE},
\label{eq:env_regression3}
\end{equation}
with
\begin{equation}
\bar{\bar\bE}=\frac{\bar\bE_0+\bar\bE_1}{2}.
\label{eq:env_average2}
\end{equation}

We specify the base probability measure $\tilde\bH$ as follows: for $m=1,\ldots,M$, $i=1,\ldots,N_k$,
$k=0,1$, and $r=1,\ldots,L$, 
\begin{equation}
%\bG_{0,jk}=\prod_{r=1}^{L_j}\mbox{Beta}\left(\nu_{1jkr},\nu_{2jkr}\right).
p_{mijkr}\stackrel{iid}{\sim} \mbox{Beta}\left(\nu_1,\nu_2\right),
\label{eq:dp3}
\end{equation}
under $\tilde\bH$, where $\nu_1,\nu_2>0$.

This completes the specification of a hierarchy of Dirichlet processes to build dependence
between the genes and the distributions of genotypes of cases-controls given data. Note that our model consists of one more level of hierarchy
of Dirichlet processes than considered in the applications of \ctn{Teh06}, who introduce hierarchical Dirichlet processes (HDP).
Moreover, our likelihood based on Dirichlet processes ensuring at most $M$ mixture components, 
is significantly different from those
considered in the applications of \ctn{Teh06}, which are based on the traditional DP mixture;
see \ctn{Sabya11}, \ctn{Sabya12}, \ctn{Sabya13} for details on the conceptual, computational and asymptotic 
advantages of our modeling style over the traditional DP mixture. 
%Coincidences among $\bP_{Mijk}=\left\{ \bp_{1ijk},\bp_{2ijk},\ldots,\bp_{Mijk}\right\}$, 
%which occur with positive probability, is the issue that we exploit to learn about the actual
%number of mixture components. 
%in (\ref{eq:mixture1}) falls below $M$, the maximum number of components, the
%mixing probabilities taking the form $M^*/M$, where $1\leq M^*\leq M$. See \ctn{Majumdar13},
%\ctn{Sabya11}, \ctn{Sabya12}, \ctn{Bhattacharya08}, for the details. 

\subsection{{\bf The Chinese restaurant analogy}}
\label{subsec:chinese_restaurant}
An extended version of the Chinese restaurant metaphor used by \ctn{Teh06} may be considered to illustrate our model.
For $k=0,1$, the set of random probability measures $\left\{\bG_{0,jk};j=1,\ldots,J\right\}$ can be associated with 
$J$ restaurants. Letting $\tau_{ijk}$ denote the number of tables at the
$j$-th restaurant associated with the $i$-th individual, we denote by $\bphi_{lijk}$ 
the dish being served at table $l$ of the $j$-th restaurant for the $i$-th individual.
Note that $\left\{\bphi_{lijk};l=1,\ldots,\tau_{ijk};i=1,\ldots,N_k\right\}$ is a set of $iid$ realizations from $\bG_{0,jk}$.
Thus, we have different sets of realizations from $\bG_{0,jk}$ for different individuals $i$. 

For $k=0,1$, we also let $\bXi_{R_kk}=\left\{\bxi_{1k},\ldots,\bxi_{R_kk}\right\}$ denote a set of $R_k$ $iid$ realizations
from $\bH_k$. Then it follows that for $l=1,\ldots,\tau_{ijk}$, $i=1,\ldots,N_k$, and for $j=1,\ldots,J$, 
$\bphi_{lijk}\in\bXi_{R_kk}$.
In other words, $\bXi_{R_kk}$ is the set of distinct elements in the set 
$\{\bphi_{lijk};l=1,\ldots,\tau_{ijk}; i=1,\ldots,N_k; j=1,\ldots,J\}$,
and, from the Chinese restaurant perspective, is the set of global dishes among all the restaurants, given $k$.

Finally, let $\bzeta_S=\left\{\boeta_1,\ldots,\boeta_S\right\}$ denote a set of $S$ $iid$ realizations from $\tilde \bH$.
Then it follows that $\bzeta_S$ is the set of distinct elements in $\left\{\bXi_{R_kk}:k=0,1\right\}$. In other words,
$\bzeta_S$ is the set of global dishes served in all the restaurants, irrespective of $k=0$ or $k=1$.

\section{{\bf Discussion of the dependence structure induced by our HDP-based model}}
\label{sec:hdp_dependence}

\subsection{{\bf Dependence among individuals}}
\label{subsec:dependence_individuals_mutation}

It follows from the discussion in Section \ref{subsec:chinese_restaurant} that 
$\left\{\bphi_{lijk};l=1,\ldots,T_{mijk};i=1,\ldots,N_k\right\}\in\left\{\bxi_{1k},\ldots,\bxi_{R_{mk}k}\right\}$,
where $\bxi_{1k},\ldots,\bxi_{R_{mk}k}\stackrel{iid}{\sim}\bH_k$.
This shows that $\left\{\bphi_{lijk};l=1,\ldots,T_{mijk};i=1,\ldots,N_k\right\}$ in (\ref{eq:polya_urn}) 
are shared among the individuals,
thus creating dependence among the subjects.

For more precise insights regarding the dependence structure, let us first marginalize over $\bG_{ijk}$ to obtain the joint
distribution of 
$\bP_{Mijk}=\left\{\bp_{1ijk},\ldots,\bp_{Mijk}\right\}$ using the following Polya urn distributions:
given $\bG_{0,jk}$, $\bp_{1ijk}\sim \bG_{0,jk}$, and 
for $m=2,\ldots,M$,
\begin{equation}
%[\bp_{mijk}|\bp_{lijk};l<m]=\frac{\alpha_{G,ik}}{\alpha_{G,ik}+m-1}\bG_{0,jk}\left(\bp_{mijk}\right)
%+\frac{1}{\alpha_{G,ik}+m-1}\sum_{l=1}^{m-1}\delta_{\bp_{lijk}}\left(\bp_{mijk}\right).
%
[\bp_{mijk}|\bp_{lijk};l<m]=\frac{\alpha_{G,ik}}{\alpha_{G,ik}+m-1}\bG_{0,jk}\left(\bp_{mijk}\right)
+\frac{1}{\alpha_{G,ik}+m-1}\sum_{t=1}^{T_{mijk}}\tilde n_{tmijk}\delta_{\bphi_{tijk}}\left(\bp_{mijk}\right),
\label{eq:polya_urn}
\end{equation}
where $\sum_{t=1}^{T_{mijk}}\tilde n_{tmijk}=m-1$. 
Here $\tilde n_{tmijk}=\#\left\{l<m:\bp_{lijk}=\bphi_{tijk}\right\}$.
%Now, $\bphi_{lijk}\stackrel{iid}{\sim}\bG_{0,jk}$, for fixed
%$(j,k)$, so that the disribution of $\bphi_{lijk}$ is independent of the individuals, given $(j,k)$. 

%The subscript $i$ in $\bphi_{lijk}$ indicates that it is a simulation associated with the $i$-th individual
%which can take value different from that associated with any other individual, although both arise from the same
%distribution. Moreover, the subscript is also useful to distinguish between the number of such simulations 
%$\tau_{ijk}$, for the $i$-th individual, and the number for other individuals. 

%It follows from the Polya urn scheme (\ref{eq:polya_urn}) that 
Since conditionally on $\bG_{0,jk}$, the marginal 
distribution of $\bp_{mijk}$, for $m=1,\ldots,M$
and $i=1,\ldots,N_k$, is $\bG_{0,jk}$, the marginal is unaffected by the environmental variable,
but the joint distribution of $\bP_{Mijk}$ implied by the Polya urn distributions (\ref{eq:polya_urn}) shows that the 
dependence structure of $\bP_{Mijk}$ is influenced by the regression on $\bE_{ik}$ through $\alpha_{G,ik}$. 
%(see (\ref{eq:env_regression1})). 
This is a very desirable property of our modeling 
approach, since, in reality,
the population minor allele frequencies for the case-control group are not expected to be affected by 
environmental variables, although environmental exposure is expected
to influence dependence among individuals and gene-gene interactions in individuals. 
Note that marginal distributions depending upon environmental variables may be envisaged only under mutation,
but since it is an extremely rare phenomenon and the type of case control type genotype data we are dealing with is not appropriate for such studies, we do not include mutational effects in our model. 

%Before shedding light on the dependence structure among the genes, it is important to provide  
%an appropriate measure to quantify gene-gene interaction influenced by environmental variables
%associated with our HDP based model.

\subsection{{\bf Dependence among the genes}}
\label{subsec:dependence_genes}

We now show that the gene-gene interactions of the $i$-th individual are affected by $\bE_{ik}$, 
but not the marginal effects of the genes.

Dependence among the genes for the $i$-th individual is induced by 
$\left\{\bphi_{tijk};t=1,\ldots,\tau_{ijk};j=1,\ldots,J\right\}$, where, for $t=1,\ldots,\tau_{ijk}$,
$\bphi_{tijk}\stackrel{iid}{\sim}\bG_{0,jk}$, with $\bG_{0,jk}\sim DP\left(\alpha_{G_0,k}\bH_k\right)$. 
In fact, marginalizing over $\bG_{0,jk}$ yields the following 
Polya urn scheme for $\left\{\bphi_{tijk};t=1,\ldots,\tau_{ijk}\right\}$:
\begin{equation}
[\bphi_{tijk}|\bphi_{lijk};l<t]=\frac{\alpha_{G_0,k}}{\alpha_{G_0,k}+t-1}\bH_{k}\left(\bphi_{tijk}\right)
+\frac{1}{\alpha_{G_0,k}+t-1}\sum_{l=1}^{R_{tk}}\bar n_{ltik}\delta_{\bxi_{lk}}\left(\bphi_{tijk}\right),
\label{eq:polya_urn_phi}
\end{equation}
where $\bar n_{ltik}=\#\left\{l'<t:\bphi_{l'ijk}=\bxi_{lk}\right\}$.
Note that $\sum_{l=1}^{R_{tk}}\bar n_{ltik}=t-1$.

It is clear from (\ref{eq:polya_urn_phi}) that $\left\{\bphi_{tijk};j=1,\ldots,J\right\}$ share 
$\left\{\bxi_{lk};l=1,\ldots,R_k\right\}$, so that the latter set creates dependence among the genes.
Moreover, it is also clear from (\ref{eq:polya_urn_phi}) that the dependence structure does not depend directly upon
$\bE_{ik}$, but upon $\bar\bE_k$. In other words, the gene-gene dependence structure of any individual is not 
directly influenced by the corresponding environmental variable. However, the dependence structure is also
influenced by $\bar n_{ltik}$, which depends upon the $i$-th individual in the $k$-th case-control group through
$\tau_{ijk}$, which is directly influenced by $\bE_{ik}$ through $\alpha_{G,ik}$. 
Thus, as is desirable, our modeling style induces gene-gene interactions that are specific to the individuals
and are influenced by the corresponding environmental variables and the averages of the environmental variables
within the case-control groups that the individuals belong to.

It is also interesting to observe that in spite of the individual-specific gene-gene interactions, the marginal
distributions of $\bphi_{tijk}$ remains $\bG_{0,jk}$ for the non-marginalized version and $\bH_k$ 
for the marginalized version characterized by (\ref{eq:polya_urn_phi}), signifying that the individual genes are not 
affected by $\bE_{ik}$.

\subsection{{\bf Case-control dependence}}
\label{subsec:case_control_dependence}

Finally, we note that
\begin{equation}
[\bxi_{sk}|\bxi_{lk};l<s]=\frac{\alpha_{H}}{\alpha_{H}+s-1}\tilde\bH\left(\bxi_{sk}\right)
+\frac{1}{\alpha_{H}+s-1}\sum_{l=1}^{S_{sk}}\breve n_{lsk}\delta_{\bzeta_{l}}\left(\bxi_{sk}\right),
\label{eq:polya_urn_xi}
\end{equation}
where $\breve n_{lsk}=\#\left\{l'<s:\bxi_{l'k}=\bzeta_{l}\right\}$ and $\sum_{l=1}^{S_{sk}}\breve n_{lsk}=s-1$. 
So, $\left\{\bxi_{sk};s=1,\ldots,R_k;k=0,1\right\}$ share
$\left\{\bzeta_l;l=1,\ldots,S\right\}$, creating dependence between case and control status. 
Dependence between case and control status are likely to be caused by various implicit factors and
environmental variables that are not accounted for in the study. These factors and environmental variables
may be insignificant individually, but together may exert non-negligible influence on cases and controls.

\begin{figure}%[htp]
\centering
\includegraphics[width=15cm,height=15cm]{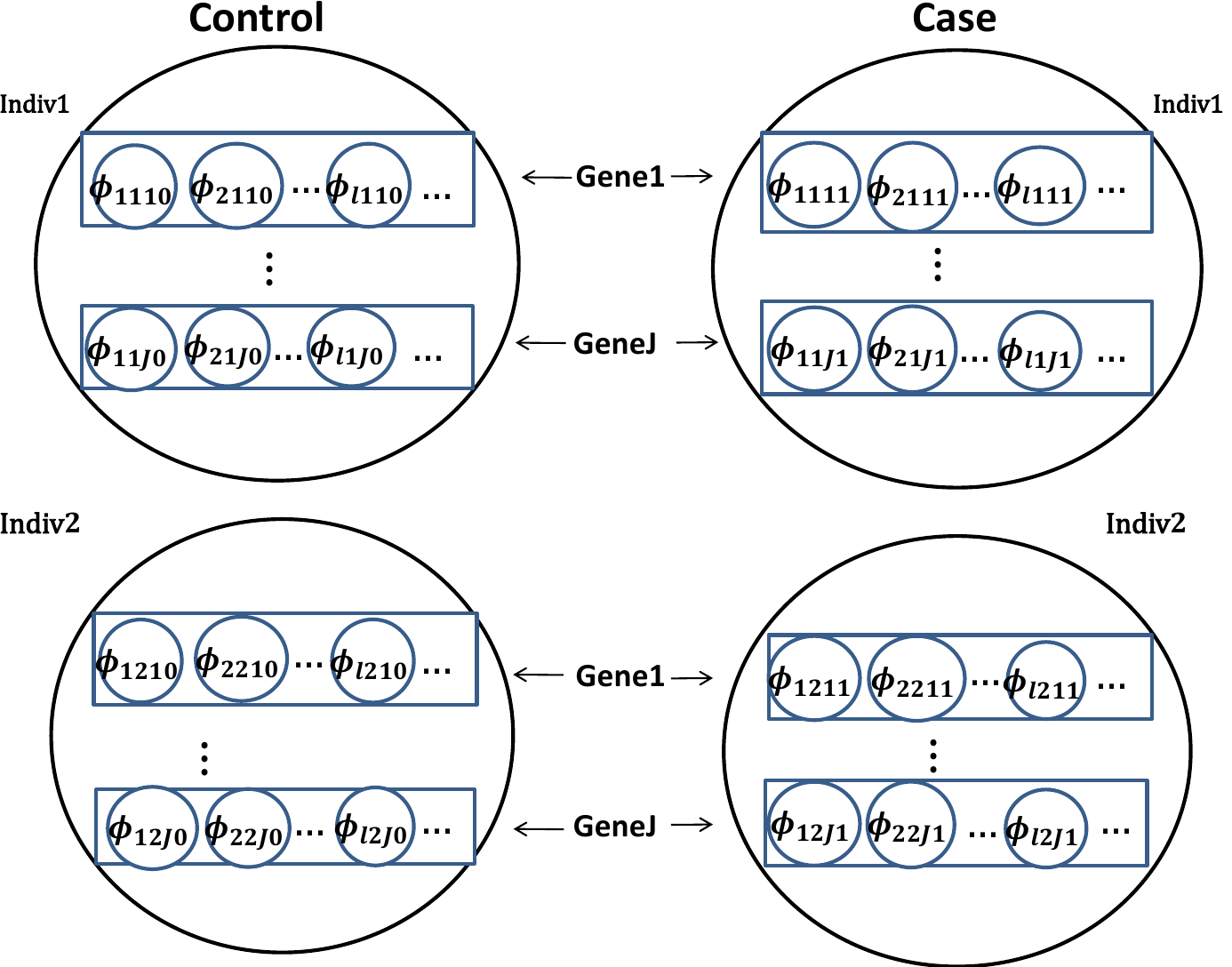}
\caption{{\bf Schematic diagram of our HDP-based Bayesian model.}}
\label{fig:HDP_schematic}
\end{figure}

A schematic diagram of our HDP-based model and the dependence structure is depicted in Figure \ref{fig:HDP_schematic}.
We remark that in a much simpler set-up, the original HDP proposed in \ctn{Teh06} has also been used by \ctn{Iorio15} for inferring population admixture,
allowing for correlations between loci due to linkage disequilibrium.

%In Section \ref{sec:inferential_procedure} we propose an MCMC methodology
%for fitting our model, and in Section \ref{sec:computation} we propose a parallel algorithm for implementing 
%the proposed MCMC method.
In Section \ref{sec:inferential_procedure} we propose an MCMC procedure for the inferential purpose, and in Section \ref{sec:computation}
we provide a parallel algorithm for implementing the MCMC method.

\section{{\bf Detection of the roles of environment, genes and their interactions with respect to our HDP based model}}
\label{sec:detection}

\subsection{{\bf Formulation of the tests and interpretation of their results}}
\label{subsec:formulation_interpretation}

\subsubsection{{\bf Bayesian test for the impact of the genes on case-control}}
\label{subsubsec:test_genes}

%Recall the definition of $\bar{w}_{ijk}$ from (\ref{eq:w_bar1}). 
%%Let $\bar{W}_{ijk}$ denote the random variable associated with the observation $\bar{w}_{ijk}$.
%Let $h_{0j}$ and $h_{1j}$ denote the distributions of $\bar{w}_{ijk=0}$ and $\bar{w}_{ijk=1}$, respectively.
%Clearly, $h_{0j}$ and $h_{1j}$ are also $M$-component mixtures, where the $m$-th mixture component of the respective
%distributions is characterized by $\bp_{mjk=0}$ and $\bp_{mjk=1}$, haing the same mixing probability $1/M$.
%%If, for case ($k=1$) and control ($k=0$), the distributions of the form (\ref{eq:mixture1}) are

%If $h_{0j}$ and $h_{1j}$ are not significantly different, then it is plausible to conclude that the role of genes is not
%significant in the case-control study. As such,
To test if genes have any effect on case-control, we formulate as in \ctn{Bhattacharya15} and \ctn{Bhattacharya16},
the following hypotheses:
\begin{equation}
H_{01}: h_{0j}=h_{1j}; ~j=1,\ldots,J,
%H_0:\sum_{m=1}^M\pi_{m jk=0}\prod_{r=1}^{L_j}f\left(\cdot\vert p^r_{m jk=0}\right)
%=\sum_{m=1}^M\pi_{m jk=1}\prod_{r=1}^{L_j}f\left(\cdot\vert p^r_{m jk=1}\right);~j=1,\ldots,J,
\label{eq:H_0}
\end{equation}
versus
\begin{equation}
H_{11}:\mbox{not}~H_{0},
\label{eq:H_1}
\end{equation}
where
%\begin{align}
%h_{0j}(\cdot)&=\prod_{i=1}^{N_0}\left\{\sum_{m=1}^M\pi_{m ijk=0}
%\prod_{r=1}^{L_j}f\left(\cdot\vert p^r_{m ijk=0}\right)\right\};\label{eq:h_0}\\
%h_{1j}(\cdot)&=\prod_{i=1}^{N_1}\left\{\sum_{m=1}^M\pi_{m ijk=1}
%\prod_{r=1}^{L_j}f\left(\cdot\vert p^r_{m ijk=1}\right)\right\}.\label{eq:h_1}
%\end{align}
%If $\underset{1\leq j\leq J}{\max}~d(h_{0j},h_{1j})$ is significantly small with
%high posterior probability, $H_{01}$ is to be accepted. 
%If $h_{0j}$ and $h_{1j}$ are not significantly different, then it is plausible 
%to conclude that the $j$-th gene is not marginally significant in the case-control study. 
\begin{align}
h_{0j}(\cdot)&=%\prod_{i=1}^{N_0}\left\{
\frac{1}{M}\sum_{m=1}^M%\pi_{m ijk=0}
\prod_{r=1}^{L_j}f\left(\cdot\vert p^r_{m i_0jk=0}\right);\label{eq:h_0}\\
h_{1j}(\cdot)&=%\prod_{i=1}^{N_1}\left\{
\frac{1}{M}\sum_{m=1}^M%\pi_{m ijk=1}
\prod_{r=1}^{L_j}f\left(\cdot\vert p^r_{m i_1jk=1}\right).\label{eq:h_1}
\end{align}
In the above, for $k=0,1$, $i_k$ is the index such that $\bP_{Mi_kjk}=\left\{\bp_{1i_kjk},\bp_{2i_kjk},\ldots,\bp_{Mi_kjk}\right\}$ is some measure of central
tendency of $\left\{\bP_{Mijk}=\left\{\bp_{1ijk},\bp_{2ijk},\ldots,\bp_{Mijk}\right\}; i=1,\ldots,N_k\right\}$. Appropriate measures of central tendency, based on clusterings,
is discussed in Section \ref{subsubsec:clustering}.

\subsubsection{{\bf Bayesian test for the significance of the environmental variables}}
\label{subsubsec:test_environment}
To check if the environmental variables are significant, we shall test the following:
for $\ell=1,\ldots,d$,
\begin{equation}
H_{02\ell}:\beta_{G,\ell}=0~\mbox{versus}~H_{12\ell}:\beta_{G,\ell}\neq 0,
\label{eq:beta_test1}
\end{equation}
\begin{equation}
H_{03\ell}:\beta_{G_0,\ell}=0~\mbox{versus}~H_{13\ell}:\beta_{G_0,\ell}\neq 0,
\label{eq:beta_test2}
\end{equation}
and
\begin{equation}
H_{04\ell}:\beta_{H,\ell}=0~\mbox{versus}~H_{14\ell}:\beta_{H,\ell}\neq 0.
\label{eq:beta_test3}
\end{equation}

\subsubsection{{\bf Bayesian test for significance of gene-gene interaction}}
\label{subsubsec:test_ggi}

In our HDP based nonparametric model there is no readily available quantification of gene-gene interaction
unlike the models of \ctn{Bhattacharya15} and \ctn{Bhattacharya16}. Thus, in order to test for gene-gene 
interaction, it is necessary to first reasonably define
such a measurement.

\subsubsection*{{\bf A measure of gene-gene interaction influenced by environmental variables}}
\label{subsubsec:ge_quantification}
For our purpose, we first define
\begin{equation}
\bar p_{mijk}=\frac{\sum_{r=1}^{L_j}p_{mijkr}}{L_j}.
\label{eq:p_bar}
\end{equation}
%the covariance between $\mbox{logit}(p_{z_{ij_1k}ij_1k})$ and $\mbox{logit}(p_{z_{ij_2k}ij_2k})$ as quantification of 
%dependence between genes $j_1$ and $j_2$. However, since membership of the $i$-th individual to the $t$-th population
%is random, we consider the posterior expectation of the covariance with respect to membership as the proper
%quantification of subject-wise gene-gene dependence. 
%In other words, 
With the above definition, for subject $i$ belonging
to case-control group $k$, 
we consider the following covariance
\begin{equation}
C(i,j_1,j_2,k)=cov\left(\mbox{logit}(\bar p_{z_{ij_1k}ij_1k}),\mbox{logit}(\bar p_{z_{ij_2k}ij_2k})\right),
\label{eq:ge_quantification}
\end{equation}
as quantification of subject-wise gene-gene dependence that accounts for population memberships of subject $i$
with respect to genes $j_1$ and $j_2$, through $z_{ij_1k}$ and $z_{ij_2k}$, where for any $p\in(0,1)$, $\mbox{logit}(p)=\log\left(\frac{p}{1-p}\right)$.
Thus, gene-gene interaction associated with our model is subject-specific. 

While implementing our model using our parallelised MCMC methodology, we simulate $C(i,j_1,j_2,k)$ at each iteration
by generating $\left\{p_{mijkr}:r=1,\ldots,L_j\right\}$ as many times as required from the respective full conditionals
holding the remaining parameters fixed, and then compute the empirical covariance corresponding to
(\ref{eq:ge_quantification}) using the generated $iid$ samples conditionally on the remaining parameters
to approximate (\ref{eq:ge_quantification}).

\subsubsection*{{\bf Formulation of the Bayesian tests for gene-gene interactions}}
\label{subsubsec:test_ggi_formulation}
To test for subject-wise gene-gene interaction, we consider the following tests:
for $i=1,\ldots,N_k$, $k=0,1$, and for $j_1,j_2\in\{1,\ldots,J\}$,
\begin{equation}
H_{05ij_1j_2k}:C(i,j_1,j_2,k)=0~\mbox{versus}~H_{15ij_1j_2k}:C(i,j_1,j_2,k)\neq 0.
\label{eq:ggi_test}
\end{equation}

\subsubsection{{\bf Interpretations of the results of the above tests}}
\label{subsubsec:interpret_tests}
The cases that can possibly arise and the respective conclusions are the following:
\begin{itemize}
\item 
For some appropriate divergence measure $d$ between two distributions, if $\underset{1\leq j\leq J}{\max}~d(h_{0j},h_{1j})$, is significantly small with
high posterior probability, then $H_{01}$ is to be accepted. 
If $h_{0j}$ and $h_{1j}$ are not significantly different, then it is plausible 
to conclude that the $j$-th gene is not marginally significant in the case-control study.

\item Suppose that $H_{01}$ is accepted (so that genes have no significant role) and 
that at least one of $\beta_{G,\ell}$ or $\beta_{G_0,\ell}$ or $\beta_{H,\ell}$ is significant, 
at least for some $\ell$.
This may be interpreted as the environmental variable $\bE$ having some altering effect on all the genes in a way 
that doesn't affect the disease status. If $C(i,j_1,j_2,k)$ turns out to be significant, then this would additionally imply that
the environmental variable $\bE$ influences interaction between genes $j_1$ and $j_2$ for the $i$-th individual, 
but not in a way that is responsible for the case/control status.
%Significance of $\phi$ and at least one $\beta_{\ell jk}$ would indicate that 
%both mutation and gene-gene interaction have been influenced by $\bE$, but not in a way that triggers the disease.

\item If $H_{01}$ is rejected, indicating that the genes are significant,
but none of the $\beta_{G,\ell}$, $\beta_{G_0,\ell}$, $\beta_{H,\ell}$ or $C(i,j_1,j_2,k)$ are significant, 
then only the genes, not $\bE$, are responsible
for the disease. In that case, one may conclude that the disease is of purely genetic nature.

\item Suppose that $H_{01}$ is rejected, none of $\beta_{G,\ell}$, $\beta_{G_0,\ell}$, $\beta_{H,\ell}$ is significant,
but $C(i,j_1,j_2,k)$ is significant for at least some $i,j_1,j_2,k$. Then the environmental variable is not significant,
and the case/control status of the individuals associated with significant gene-gene interactions can be attributed
to purely genetic causes triggered by gene-gene interactions of the individuals.

\item Now suppose that $H_{01}$ is rejected, and at least one of $\beta_{G,\ell}$, $\beta_{G_0,\ell}$, $\beta_{H,\ell}$
is significant, but none of the subject-wise gene-gene interactions is significant. Then the environmental variable $\bE$
does not significantly affect the interactions to determine the case/control status, and marginal effects of 
the individual genes are responsible for the case/control status of an individual.

\item If, on the other hand, $H_{01}$ is rejected, at least one of $\beta_{G,\ell}$, $\beta_{G_0,\ell}$, $\beta_{H,\ell}$ 
is significant, and $C(i,j_1,j_2,k)$ is significant for at least some $i,j_1,j_2,k$, 
then the environmental variable is significant and is responsible for influencing gene-gene interactions within
the individuals with significant $C(i,j_1,j_2,k)$, which, in turn, affects the case/control status of the individuals.

\end{itemize}

\subsection{{\bf Methodologies for implementing the Bayesian tests}}
\label{subsec:testing_methods}

\subsubsection{{\bf Hypothesis testing based on clustering modes}}
\label{subsubsec:clustering}

%Ideas on clusterings of the mixture distributions $h_{0j}$ and $h_{1j}$ provides us with a novel
%and computationally efficient procedure of testing $H_0$. 
%Briefly, we assess discrepancies between the two mixture distributions
%%implied by $k=0$ and $k=1$ 
%$h_{0j}$ and $h_{1j}$

As in \ctn{Bhattacharya15} and \ctn{Bhattacharya16}, here we exploit the concept of ``central" clustering
introduced by \ctn{Sabya11}. Briefly, central clustering may be interpreted as a suitable measure of central tendency
of a set of clusterings. \ctn{Sabya11} particularly consider the mode(s) of the set of clusterings, and provide
methods for appropriately obtaining the mode(s) using a suitable metric that they propose to quantify distances
between any two clusterings. Their proposed metric is also computationally inexpensive, which makes the concept
based on central clusterings extremely useful in practice.

For $k=0,1$, let $i_k$ denote the index of the central clusterings of
$\bP_{Mijk}=\left\{ \bp_{1ijk},\bp_{2ijk},\ldots,\bp_{Mijk}\right\}$, $i=1,\ldots,N_k$.
We then study the divergence between the two clusterings 
of $$\bP_{Mi_0jk=0}=\left\{ \bp_{1i_0jk=0},\bp_{2i_0jk=0},\ldots,\bp_{Mi_0jk=0}\right\}$$
and $$\bP_{Mi_1jk=1}=\left\{ \bp_{1i_1jk=1},\bp_{2i_1jk=1},\ldots,\bp_{Mi_1jk=1}\right\},$$ for $j=1,\ldots,J$.
A schematic diagram illustrating the idea can be found in \ctn{Bhattacharya16}.

Significantly large divergence between the two clusterings clearly indicates
that the $j$-th gene is marginally significant.

\subsubsection{{\bf Enhancement of clustering metric based inference using Euclidean distance}}
\label{subsubsec:clustering_shortcoming}

As argued in \ctn{Bhattacharya15}, significantly large clustering distance between 
$\bP_{Mjk=0}$ and $\bP_{Mjk=1}$ indicates rejection of $H_0$, but insignificant clustering distance 
does not necessarily provide strong evidence in favour of the null. In this regard, 
\ctn{Bhattacharya15} (see also \ctn{Bhattacharya16}) argue that the Euclidean
distance is an appropriate candidate to be tested for significance before arriving at the final conclusion.
Briefly, we first compute the averages 
$\bar{p}_{mijk}=\sum_{r=1}^{L_j}p_{m,ijkr}/L_j$, then consider their logit transformations
$\mbox{logit}\left(\bar{p}_{mijk}\right)=\log\left\{\bar{p}_{mijk}/(1-\bar{p}_{mijk})\right\}$. 
Then, we compute the Euclidean distance between the vectors
$$\mbox{logit}\left(\bar{\bP}_{Mi_0jk=0}\right)=\left\{\mbox{logit}\left(\bar{p}_{1i_0jk=0}\right),
\mbox{logit}\left(\bar{p}_{2i_0jk=0}\right),
\ldots, \mbox{logit}\left(\bar{p}_{Mi_0jk=0}\right)\right\}$$
and 
$$\mbox{logit}\left(\bar{\bP}_{Mi_1jk=1}\right)=\left\{\mbox{logit}\left(\bar{p}_{1i_1jk=1}\right),
\mbox{logit}\left(\bar{p}_{2i_1jk=1}\right),
\ldots, \mbox{logit}\left(\bar{p}_{Mi_1jk=1}\right)\right\}.$$
We denote the Euclidean distance associated with the $j$-th gene by 
$$d_{E,j}=d_{E,j}\left(\mbox{logit}\left(\bar{\bP}_{Mi_0jk=0}\right),
\mbox{logit}\left(\bar{\bP}_{Mi_1jk=1}\right)\right),$$ and denote 
$\underset{1\leq j\leq J}{\max}~d_{E,j}$ by $d^*_E$.

\subsubsection{{\bf Formal Bayesian hypothesis testing procedure integrating the above developments}}
\label{subsubsec:testing}

In our problem, we need to test the following for reasonably small choices of 
$\varepsilon$'s:
\begin{equation}
H_{0,d^*}:~d^*< \varepsilon_{d^*}\hspace{2mm}\mbox{versus}\hspace{2mm}H_{1,d^*}:~d^*\geq\varepsilon_{d^*};
\label{eq:hypothesis_d_star}
\end{equation}
%where
%\begin{equation}
%d^*=\max_{1\leq j\leq J}\hat d\left(\bP_{Mi_0jk=0},\bP_{Mi_1jk=1}\right);
%\label{eq:d_star}
%\end{equation}
\begin{equation}
H_{0,d^*_E}:~d^*_E< \varepsilon_{d^*_E}\hspace{2mm}\mbox{versus}\hspace{2mm}H_{1,d^*_E}:~d^*_E\geq\varepsilon_{d^*_E};
\label{eq:hypothesis_d_star_E}
\end{equation}
for $\ell=1,\ldots,d$,
\begin{equation}
H_{0,\beta_{G,\ell}}:~\left|\beta_{G,\ell}\right|< \varepsilon_{G,\ell}\hspace{2mm}
\mbox{versus}\hspace{2mm}H_{1,\beta_{G,\ell}}:~\left|\beta_{G,\ell}\right|\geq\varepsilon_{G,\ell},
\label{eq:hypothesis_beta1}
\end{equation}
\begin{equation}
H_{0,\beta_{G_0,\ell}}:~\left|\beta_{G_0,\ell}\right|< \varepsilon_{G_0,\ell}\hspace{2mm}
\mbox{versus}\hspace{2mm}H_{1,\beta_{G_0,\ell}}:~\left|\beta_{G_0,\ell}\right|\geq\varepsilon_{G_0,\ell},
\label{eq:hypothesis_beta2}
\end{equation}
\begin{equation}
H_{0,\beta_{H,\ell}}:~\left|\beta_{H,\ell}\right|< \varepsilon_{H,\ell}\hspace{2mm}
\mbox{versus}\hspace{2mm}H_{1,\beta_{H,\ell}}:~\left|\beta_{H,\ell}\right|\geq\varepsilon_{H,\ell},
\label{eq:hypothesis_beta3}
\end{equation}
and, for $i=1,\ldots,N_k$, $k=0,1$, $j_1,j_2\in\{1,\ldots,J\}$,
\begin{equation}
H_{0,C_{i,j_1,j_2,k}}:~\left|C_{i,j_1,j_2,k}\right|< \varepsilon_{C,ij_1j_2k}\hspace{2mm}
\mbox{versus}\hspace{2mm}H_{1,C_{i,j_1,j_2,k}}:~\left|C_{i,j_1,j_2,k}\right|\geq\varepsilon_{C,ij_1j_2k},
\label{eq:hypothesis_C}
\end{equation}

If $H_0$ is rejected in (\ref{eq:hypothesis_d_star}) or in (\ref{eq:hypothesis_d_star_E}), we could also
test if the $j$-th gene is influential by testing, for $j=1,\ldots,J$, 
$H_{0,\hat d_j}:~\hat d_j< \varepsilon_{\hat d_j}\hspace{2mm}\mbox{versus}\hspace{2mm}
H_{1,\hat d_j}:~\hat d_j\geq\varepsilon_{\hat d_j}$,
where $\hat d_j=\hat d\left(\bP_{Mi_0jk=0},\bP_{Mi_1jk=0}\right)$; we could also test
$H_{0,d_{E,j}}:~d_{E,j}< \varepsilon_{d_{E,j}}\hspace{2mm}\mbox{versus}
\hspace{2mm}H_{1,d_{E,j}}:~d_{E,j}\geq\varepsilon_{d_{E,j}}$.

\subsubsection{{\bf Null model and choice of $\varepsilon$}}
\label{subsubsec:null_model_e_choice}

To obtain the null posterior distribution, we fit our HDP-based Bayesian model to the dataset generated
from the HDP-based model where the genes are independent and not influenced by the environmental variable,
and where there is no difference between the probabilities associated with case and control.
For the null data we chose the same number of genes, the
same number of loci for each gene, and the same number of cases and controls as the non-null data.
We also choose the same value $M$ as in the non-null model, but set
$\beta_G=\beta_{G_0}=\beta_H=0$. 
To generate the data from the null model, we first simulate, independently for $j=1,\ldots,J$, 
the set $\{\bp_{m1j0}:m=1,\ldots,M\}$, using the Polya urn scheme involving
$\tilde\bH$ and $\alpha_H$, and set $\{\bp_{m1j1}:m=1,\ldots,M\}=\{\bp_{m1j0}:m=1,\ldots,M\}$,
so that there is no difference between the probabilities associated with case and control, and that the genes are independent. 
Since the simulation method is independent of the environmental variable, it is clear that the genes are
not influenced by the environment.
Given the probabilities $\{\bp_{m1j1}:m=1,\ldots,M\}$ and $\{\bp_{m1j0}:m=1,\ldots,M\}$, we then simulate the 
data using our Bernoulli model. To the data thus generated, we fit our full HDP-based Bayesian model, to obtain
the null posterior.

As in \ctn{Bhattacharya15} here also we specify $\varepsilon$'s as $F^{-1}\left(0.55\right)$, where
$F$ is the distribution function of the relevant benchmark null posterior distribution.
Recall that the choice $F^{-1}\left(0.55\right)$, rather than the median, ensures that the
correct null hypothesis is accepted under the ``$0-1$" loss. Note that, for the median, the posterior
probability of the true null is $0.5$, while under the ``$0-1$" loss, the true null will be accepted
if its posterior probability is greater than $1/2$.

%If, on the other hand, $H_{0j}$ is accepted, then it is possible that the $j$-th gene is not
%individually influential, but some interaction effect involving the $j$-th gene may be significant. To check which interactions
%are significant (we may check this even if $H_{0j}$ is rejected, since the $j$-th gene may be marginally
%significant as well as interactive with the other genes), one may conduct the tests 
%$H_{0,j,j^*}:~\left|\bA_{jj^*}\right|<\varepsilon$ 
%versus $H_{1,j,j^*}:~\left|\bA_{jj^*}\right|\geq\varepsilon$,
%for $j^*\neq j$,  
%\begin{equation}
%\rho_{jj^*} = \frac{\bA_{jj^*}}{\sqrt{\bA_{jj}\bA_{j^*j^*}}};
%\label{eq:rho}
%\end{equation}
%$\bA_{jj^*}$ being the $(j,j^*)$-th
%element of $\bA$.
%Acceptance of $H_{1,j,j^*}$ for some (or many) $j^*\neq j$, indicates which of the genes
%interact with the $j$-th gene to contribute significantly to the underlying case-control study.

\section{{\bf Simulation studies}}
\label{sec:simulation_study}

For simulation studies, we first generate realistic biological data for stratified population with known
gene-environment interaction from the GENS2 software of \ctn{Pinelli12}.  
To this data, we then apply our model and methodologies in an effort to detect gene-environment
interaction effects that are present in the data. We consider simulation studies 
in $5$ different true model set-ups: (a) presence of gene-gene and gene-environment interaction,
(b) absence of genetic or gene-environmental interaction effect,
(c) absence of genetic and gene-gene interaction effects but presence of environmental effect,
(d) presence of genetic and gene-gene interaction effects but absence of environmental effect,
and 
(e) independent and additive genetic and environmental effects.
%
%As we demonstrate, our model and methodologies successfully identify the effects of the
%individual genes, gene-gene and gene-environment interactions, and the number of sub-populations. 
%In all our applications, we set $M=30$, $\nu_1=\nu_2=1$, so that $\tilde\bH$ is the uniform distribution on $[0,1]$.
%We set $\alpha_{G,ik}=0.1\times\exp\left(100+\mu_G+\beta_G E_{ik}\right)$, 
%$\alpha_{G_0,k}=0.1\times\exp\left(100+\mu_{G_0}+\beta_{G_0} \bar E_{k}\right)$ and
%$\alpha_H=0.1\times\exp\left(100+\mu_H+\beta_H \bar{\bar E}\right)$, where we assumed
%$\mu_G,\mu_{G_0},\mu_H\stackrel{iid}{\sim}U(0,1)$ and $\beta_G,\beta_{G_0},\beta_H\stackrel{iid}{\sim}U(-1,1)$.
%This structure ensured adequate number of sub-populations and satisfactory mixing of MCMC.
The details of our simulation experiments are provided in Section \ref{sec:simstudy} of the supplement.
Here we briefly summarize the results of our experiments.

In case (a), we correctly obtained clear significance of the influence of genetic effects. Also, $\beta_H$ turned out to be very significant,
demonstrating significant overall impact of the environmental variable on tsignificant overall impact of the environmental variable on the genes.
Interestingly, as one may expect, there are more instances of significant gene-gene interactions in the case group compared
to the control group. The posteriors of the number of sub-populations gave high probabilities to the correct number of sub-populations in all the 5 simulation experiments. 
Quite importantly, we demonstrate in cases (a), (d) and (e) where the genes are relevant, that our HDP model can detect disease predisposing loci (DPL) with more precision 
compared to the matrix-normal-inverse-Wishart model 
for gene-environment interactions proposed in \ctn{Bhattacharya16}. In case (b) using our ideas in conjunction with significance testing in a simple logistic regression framework, 
we are correctly able to conclude that the genetic or gene-environmental effects are insignificant. As in \ctn{Bhattacharya16}, the right conclusion is arrived at in case (c)
by utilizing our ideas in conjunction with the Akaike Information Criterion (AIC) in the context of simple logistic regression. Using our Bayesian testing procedure along with the aid
of logistic regression, we have been able to correctly obtain insignificance of the environmental variable and significance of the genes.
In this experiment, we found no gene-gene interaction in the control group and found two (marginal) instances of gene-gene interaction among
the cases. As regards case (e), we note as in \ctn{Bhattacharya16} that additivity of genetic and environmental effects is not supported
even by our current HDP-based Bayesian model. In spite of this, we correctly obtained significance of the environmental variable and precisely obtained the DPLs.
But the lack of the additivity criterion in our model seems to have forced gene-environment interactions. \ctn{Bhattacharya16} report similar results,
who obtained, after eventually resorting to logistic regression, the AIC-based best model consisting of the additive marginal effects of the first gene and the environmental variable, along
with an additive intercept, which is broadly consistent with the data-generating mechanism.

\section{{\bf Application of our HDP based ideas to a real, case-control dataset on Myocardial Infarction}}
\label{sec:realdata}
We now consider application of our model and methods to a case-control dataset on early-onset of myocardial 
infarction (MI) from MI Gen study, obtained from the dbGaP database\\
{\bf http://www.ncbi.nlm.nih.gov/gap}. 
The same dataset has been analyzed by \ctn{Bhattacharya15} without considering the sex variable as the covariate,
and by \ctn{Bhattacharya16}, who incorporate the sex variable in their gene-environment interaction model. 
Although \ctn{Bhattacharya15} obtained significant genetic and gene-gene interaction effects, their later
study after considering sex as the environmental variable, revealed strong effects of the sex variable but 
no significant gene-gene interaction, although many of the genes turned out to be individually significant. 
%see \ctn{Bhattacharya16}. 
In our current HDP based analysis, we once again obtain strong effects of the sex variable,
but in contrast with \ctn{Bhattacharya16}, although we obtain significant genetic effects, none of the genes
turned out to be significant individually. Moreover, the subject-wise gene-gene interactions, although of small
magnitude, turned out to be significant in some cases, and interestingly (and apparently counter-intuitively) 
seem to be instrumental in counter-acting the disease rather than provoking it.

%MI (more commonly, heart attack), is a complex disease and is a leading cause
%of death and disability all over the world. Much investigation has been carried out for detecting the genetic causes
%of myocardial infarction, all of which are based on the assumption that 
%the main contributory factors for the disease are the mutations in the proteins associated with 
%the pathophysiology of atherosclerosis (see \ctn{Erdmann10}).

%Although the GWA studies have revealed a lot of genetic information regarding
%MI (an overview of the main results can be found in \ctn{Erdmann10}), only a very few of the detected genes are related to traditional
%risk factors (LDL-cholesterol, diabetes and LP[a] etc.), and the other genes increase the risk by pathogenetic mechanisms
%that are not yet properly understood. Despite much success in deciphering the marginal effects of many SNPs, not much has been achieved in the gene-gene interaction front. According to \ctn{Musameh15}, burden of multiple testing renders the standard
%GWAS samples underpowered to detect such effects, while \ctn{LucasG12} blame the complexity of the epistatic effects as a reason behind the difficulty in detecting them. 

\subsection{{\bf Data description}}
\label{subsec:myo_data}
We recall that the MI Gen data obtained from dbGaP consists of observations on presence/absence of
minor alleles at $727478$ SNP markers associated with 22 autosomes and the sex chromosomes of 
$2967$ cases of early-onset myocardial infarction, $3075$ age and sex matched controls. 
The average age at the time of MI was 41 years among the male cases and 47 years among the female cases. 
%The data also consists of the sex information of the individuals, which we incorporate in our Bayesian model. 
The data broadly represents a mixture of four sub-populations: Caucasian, Han Chinese,
Japanese and Yoruban. 
%Since the names of the genes were not provided in the dataset, SNPs were mapped on to the corresponding genes using the 
%Ensembl human genome database ({\bf http://www.ensembl.org/}). However, technical glitches prevented us from
%obtaining information on the genes associated with all the markers. 
Using the Ensembl human genome database ({\bf http://www.ensembl.org/}) we could categorize
$446765$ markers out of $727478$ with respect to $37233$ genes.

As in \ctn{Bhattacharya16} we considered 32 genes covering 1251 loci, for 200 individuals.
These 1251 loci include 33 SNPs that are believed to be associated with MI and also those that are believed to be associated 
with different cardiovascular end points like LDL cholesterol, smoking, blood pressure, body mass, etc. Other than the 33 SNPs, the remaining 1218 SNPs are
not known to be associated with the disease.
See \ctn{Bhattacharya16} for the details and the relevant references.

%in various GWA studies published in NHGRI catalogue and augmented this set further with another set of 
%SNPs found to be marginally associated with MI in the MIGen study (see \ctn{LucasG12}). Our study also includes 
%SNPs that are reported to be associated with MI in various other studies, see \ctn{Erdmann10}, \ctn{LuQi11} and \ctn{Wang04}. 
%In all, we obtained 271 SNPs.
%Unfortunately, only 33 of them turned out to be common to the SNPs of our original MI dataset on genotypes, 
%which has been mapped on to the genes using the Ensembl human genome database.
%However, we included in our study all the SNPs associated with the genes containing the 33 common SNPs. Specifically, our study involves the genotypic information on 32 genes covering 1251 loci, including
%the 33  previously identified loci for $200$ individuals. We chose this relatively small number of individuals
%to ensure computational feasibility. However, even this data set, along with our model and prior, yielded results
%that are compatible with, and complement the results established in the literature.

Since the four broad sub-populations are not unlikely to admit further genetic sub-divisions, it makes sense to
set the maximum number of mixture components, $M$, to a value much larger than 4. As before, we set $M=30$; 
we also set $\nu_1=\nu_2=1$,
so that $\tilde\bH$ is the uniform distribution on $[0,1]$. As in the simulation experiments, here also
the structures $\alpha_{G,ik}=0.1\times\exp\left(100+\mu_G+\beta_G E_{ik}\right)$, 
$\alpha_{G_0,k}=0.1\times\exp\left(100+\mu_{G_0}+\beta_{G_0} \bar E_{k}\right)$ and
$\alpha_H=0.1\times\exp\left(100+\mu_H+\beta_H \bar{\bar E}\right)$, where 
$\mu_G,\mu_{G_0},\mu_H\stackrel{iid}{\sim}U(0,1)$ and $\beta_G,\beta_{G_0},\beta_H\stackrel{iid}{\sim}U(-1,1)$,
ensured adequate number of sub-populations and satisfactory mixing of MCMC.
For the null data and model, we follow the same procedure as discussed in Section \ref{subsubsec:null_model_e_choice}.

\subsection{{\bf Remarks on model implementation}}
\label{subsec:myo_implementation}
%{\bf ** provide more details on computation cost, including CPU, memory, time, etc.**}

Our parallel MCMC algorithm detailed in Section \ref{sec:computation} takes about 7 days to generate
30,000 iterations on our VMware %consisting of $50$ double-threaded, $64$-bit physical cores, each running at $2493.990$ MHz.
consisting of 1 TB RAM, 60 double-threaded, 64-bit physical cores, each
running at 2.5 GHz; 50 such cores were available to us.
We discard the first $10,000$ iterations as burn-in, using the subsequent 20,000 iterations for our Bayesian inference.
Informal convergence diagnostics such as trace plots, although did not demosntarte excellent mixing properties, did not indicate evidence of non-convergence.
Some instances are provided in Section \ref{sec:simstudy} of the supplement.

\subsection{{\bf Results of the real data analysis}}
\label{subsec:realdata_results}

\subsubsection{{\bf Effect of the sex variable}}
\label{subsubsec:sex_effect}

We obtain $P(|\beta_{G}|<\varepsilon_{\beta_G}|\mbox{Data})\approx 0$, 
$P(|\beta_{G_0}|<\varepsilon_{\beta_{G_0}}|\mbox{Data})\approx 0$ and 
$P(|\beta_H|<\varepsilon_{\beta_H}|\mbox{Data})\approx 1$. In other words, although $\bar{\bar E}$ (here $E$ being
the sex variable) is insignificant, both $E_{ik}$ and $\bar E_k$ are very significant. Thus, in this study, 
sex seems to play an important role in influencing the genes.

\subsubsection{{\bf Roles of individual genes}}
\label{subsubsec:influential_genes}

\begin{figure}%[htp]
\centering
\subfigure[Posterior probability of no genetic effect with respect to clustering metric.]{ \label{fig:clustering_hypotheses}
\includegraphics[width=15cm,height=8cm]{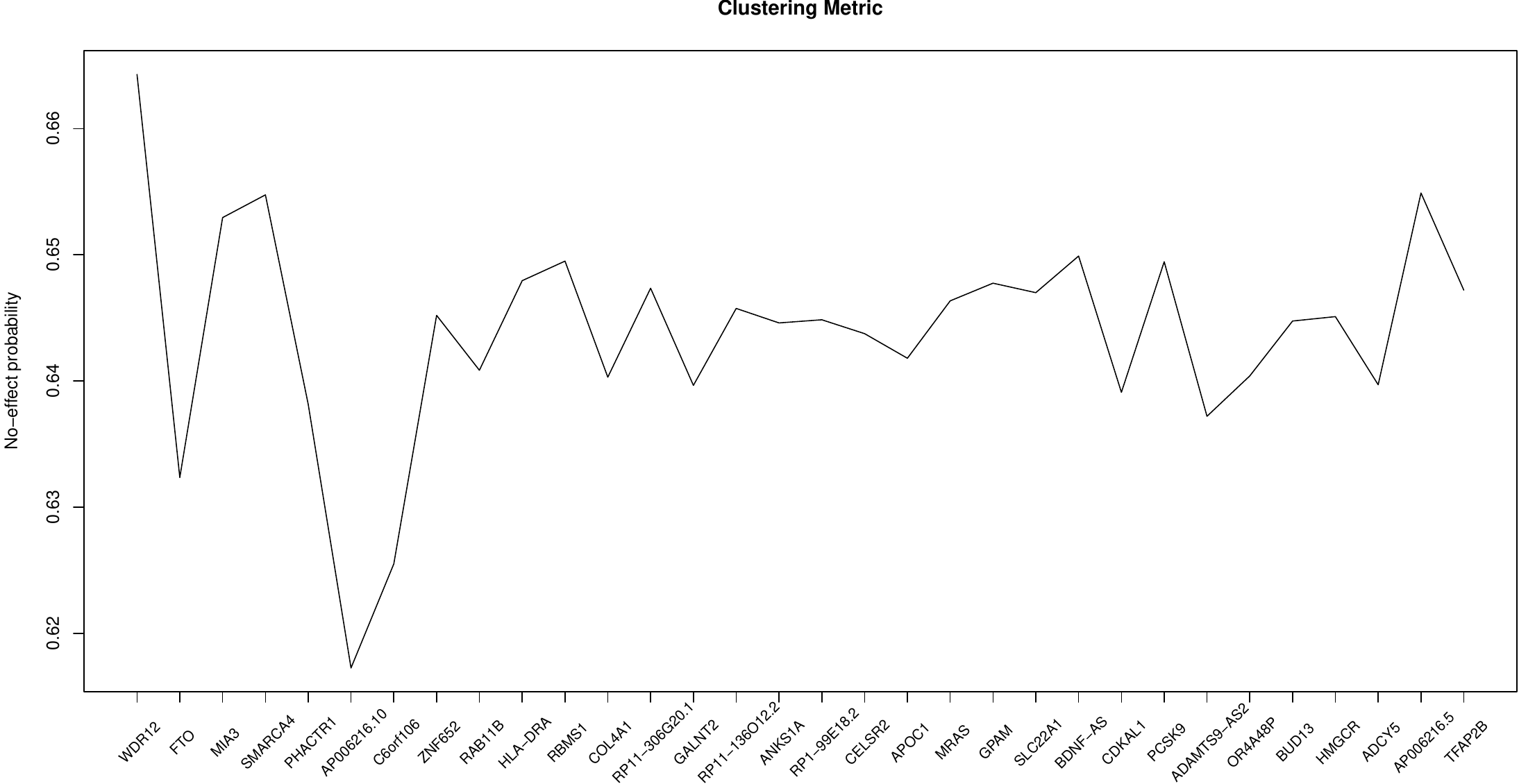}}\\
\vspace{4mm}
\subfigure[Posterior probability of no genetic effect with respect to Euclidean metric.]{ \label{fig:euclidean_hypotheses} 
\includegraphics[width=15cm,height=8cm]{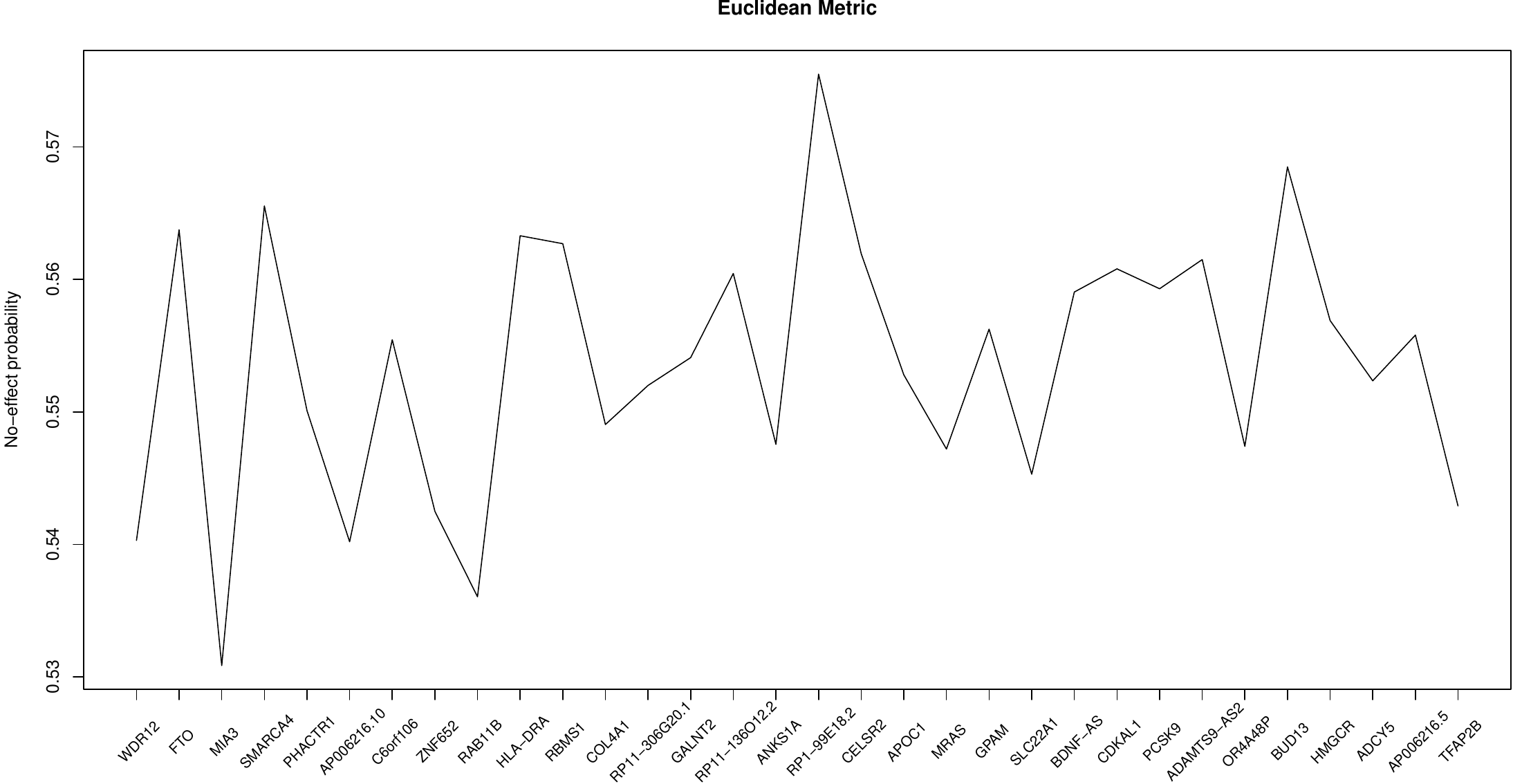}}
\caption{{\bf Posterior probabilities of no individual genetic influence:} 
Index plots of the posterior probabilities of the null hypotheses for (a) clustering metric
and (b) Euclidean metric, for the $32$ genes.}
\label{fig:null_hypotheses}
\end{figure}

With the clustering metric we obtained $P\left(d^*<\epsilon_1|\mbox{Data}\right)\approx 0.030$
while that with the Euclidean distance we obtained $P\left(d^*_E<\epsilon_2|\mbox{Data}\right)\approx 0.540$.
That is, the maximum of the gene-wise clustering metrics turns out to be significant, while
the maximum of the gene-wise Euclidean metrics is seen to be insignificant.
The same ambiguity was also obtained by \ctn{Bhattacharya16}. The tests of the marginal genes are expected
to shed some light regarding this dilemma.
The posterior probabilities of the null hypotheses (of no significant genetic influence) are shown in 
Figure \ref{fig:null_hypotheses}.
As is observed, none of the individual genes turned out to be significant, for either the clustering metric
or the Euclidean metric. Our result is not much different from that of \ctn{Bhattacharya16} who also note that
their marginal probabilities, at least for the clustering metric, are not significantly small to provide
strong enough evidences against the nulls.

Now, at least from the clustering metric perspective, it is necessary to explain the issue that
all the genes are insignificant individually but still the maximum of the gene-wise clustering metrics
is significant. The key to this issue seem to be the correlations between the distances, which are induced
by gene-gene interactions. We explain this phenomenon using a bivariate normal example.
Let $(X_1,X_2)$ have a bivariate normal distribution with means 0, variances 1, and correlation $\rho$.
\begin{figure}%[htp]
\centering
\includegraphics[width=9cm,height=6cm]{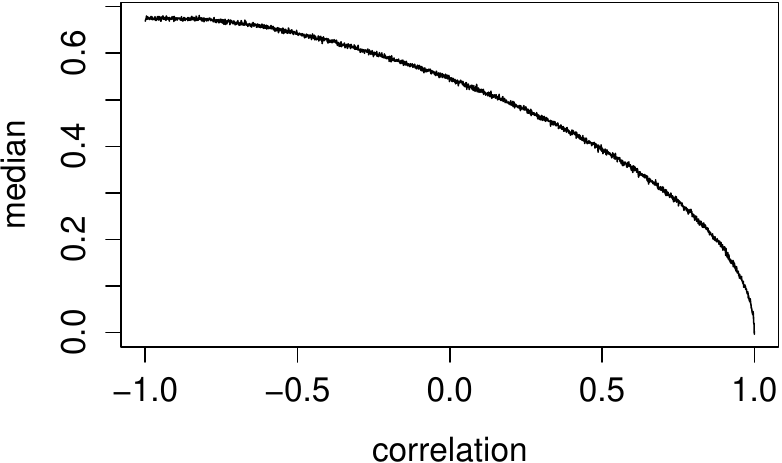}
\caption{{\bf Bivariate normal example:} Plot of the median of $\max\{X_1,X_2\}$ with respect to
the correlation $\rho$.}
\label{fig:medians_maxima_correlations}
\end{figure}
Figure \ref{fig:medians_maxima_correlations} depicts the median of $\max\{X_1,X_2\}$ as a function of $\rho$,
which is seen to be increasing as $\rho$ decreases from 1 to -1. On the other hand, the medians of the 
marginal distributions of $X_1$ and $X_2$ remain zero, irrespective of the value of $\rho$.
Thus, it seems that gene-gene interaction does have some role to play in this study.

\subsubsection{{\bf Gene-gene interactions}}
\label{subsubsec:ggi}

\begin{figure}%[htp]
\centering
\subfigure[Male case.]{ \label{fig:ggi_male1_case} 
\includegraphics[width=7.5cm,height=7.5cm]{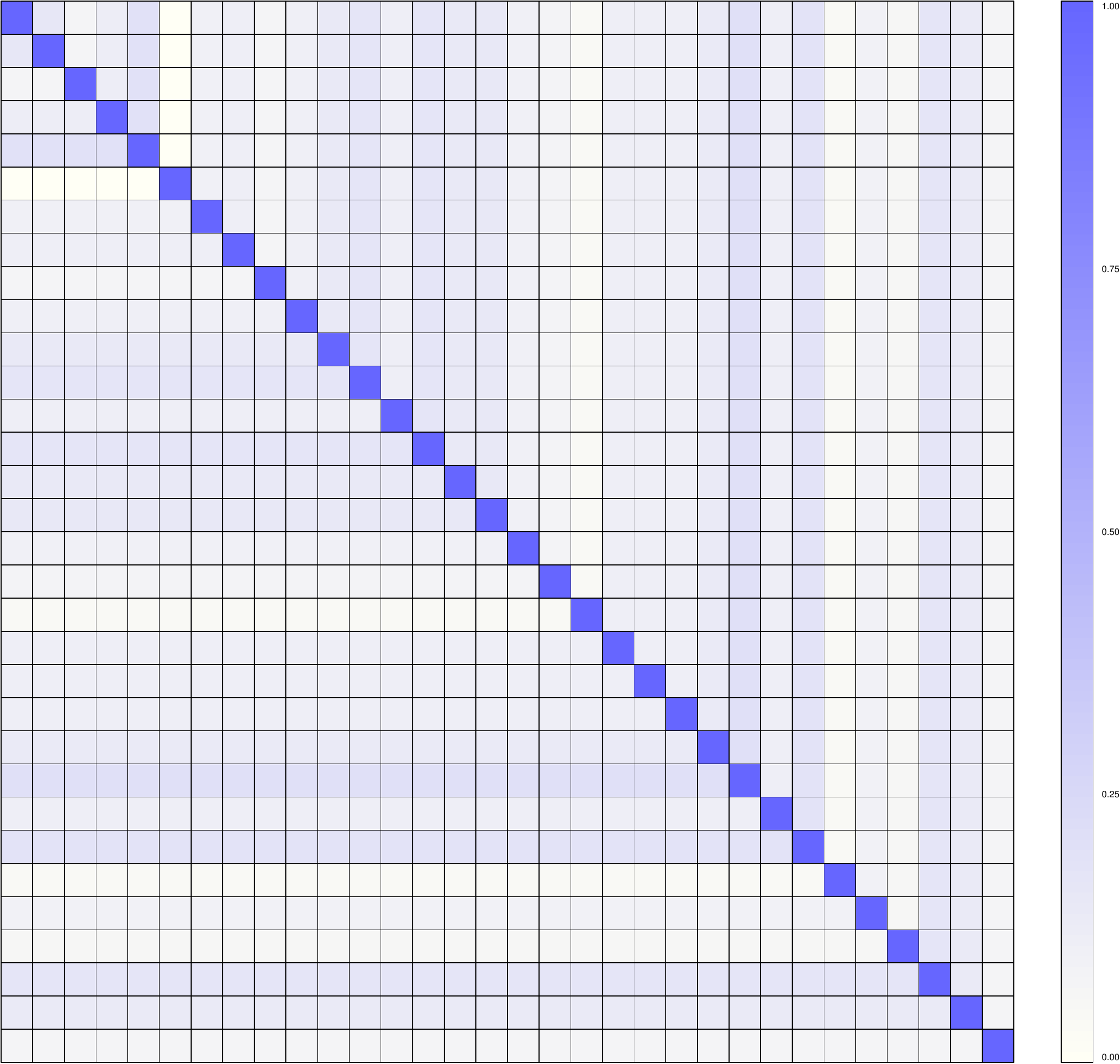}}
\hspace{2mm}
\subfigure[Male control.]{ \label{fig:ggi_male1_control} 
\includegraphics[width=7.5cm,height=7.5cm]{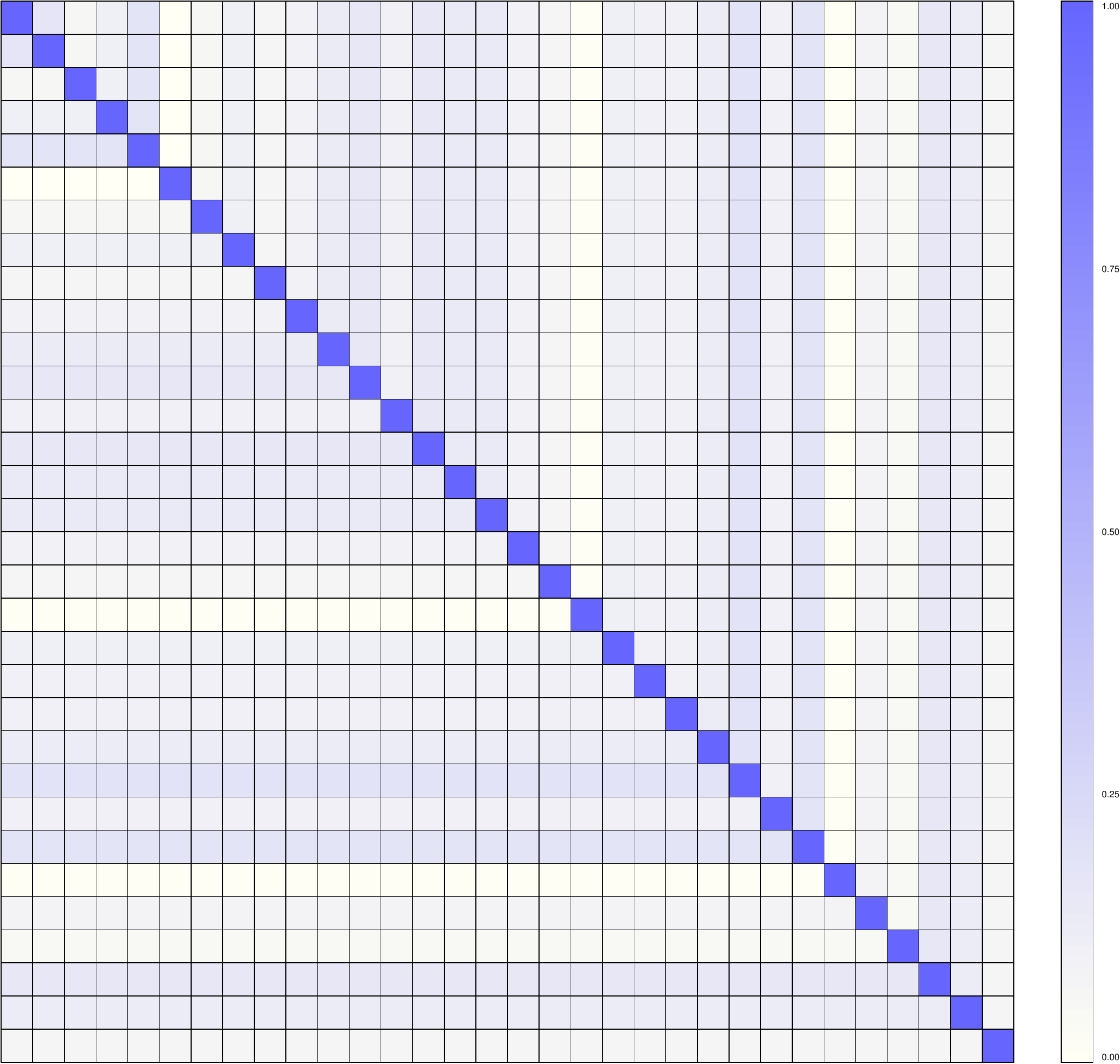}}
\\[2mm]
%\subfigure[Male case.]{ \label{fig:ggi_male46_case} 
%\includegraphics[width=7.5cm,height=7.5cm]{plots_realdata/ggi_nolabel_male46_case-crop.pdf}}
%\hspace{2mm}
%\subfigure[Male control.]{ \label{fig:ggi_male1_control} 
%\includegraphics[width=7.5cm,height=7.5cm]{plots_realdata/ggi_nolabel_male1_control-crop.pdf}}
\caption{{\bf Typical median gene-gene posterior correlation plot for male cases and male control.}}
\label{fig:ggi_plots_male}
\end{figure}

\begin{figure}%[htp]
\centering
\subfigure[Female case.]{ \label{fig:ggi_female48_case} 
\includegraphics[width=7.5cm,height=7.5cm]{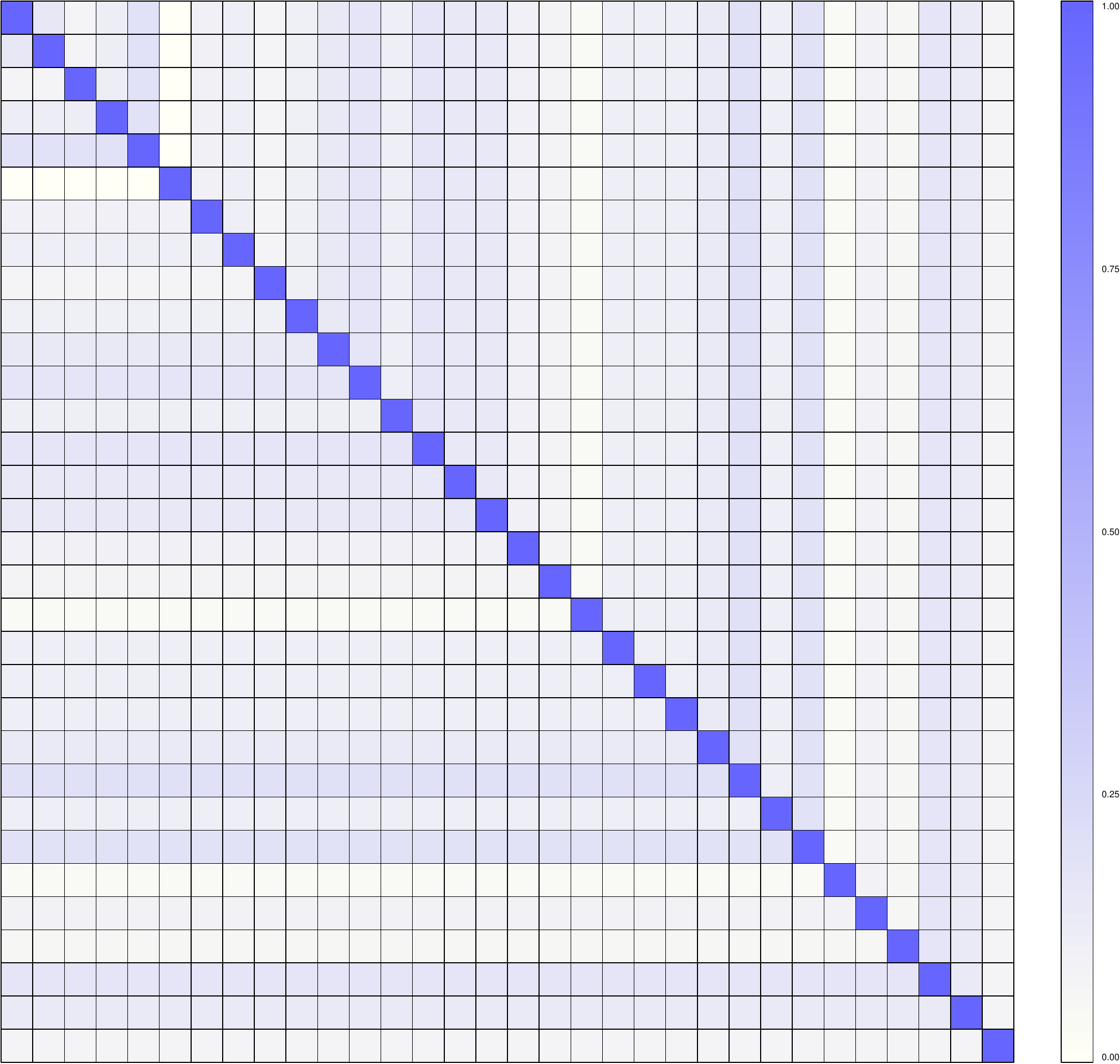}}
\hspace{2mm}
\subfigure[Female control.]{ \label{fig:ggi_female52_control} 
\includegraphics[width=7.5cm,height=7.5cm]{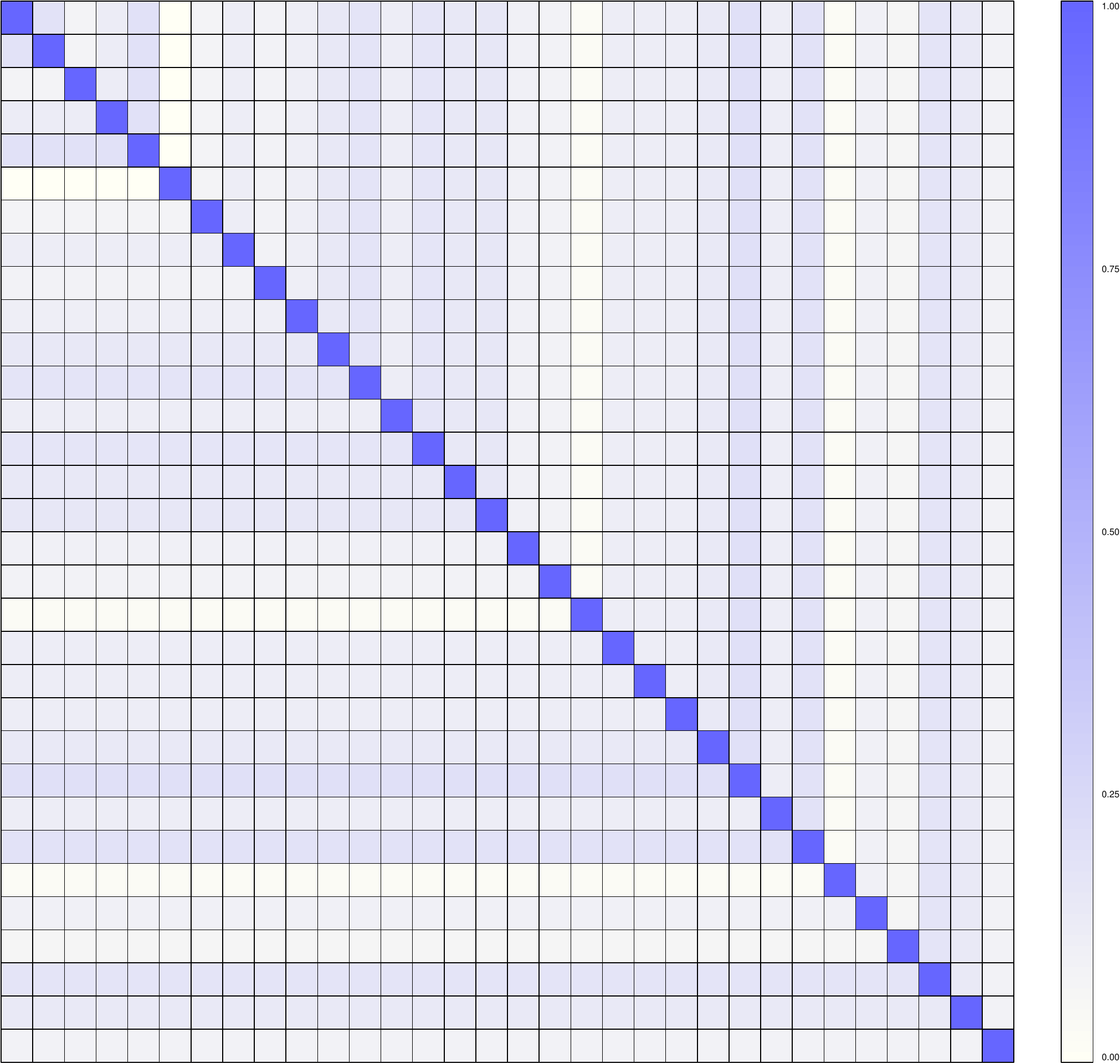}}
\\[2mm]
%\subfigure[Female control.]{ \label{fig:ggi_female55_control} 
%\includegraphics[width=7.5cm,height=7.5cm]{plots_realdata/ggi_nolabel_female55_control-crop.pdf}}
%\hspace{2mm}
%\subfigure[Female control.]{ \label{fig:ggi_female9_control} 
%\includegraphics[width=7.5cm,height=7.5cm]{plots_realdata/ggi_nolabel_female9_control-crop.pdf}}
\caption{{\bf Typical median gene-gene posterior correlation plot for female cases and female controls.}}
\label{fig:ggi_plots_female}
\end{figure}

\begin{figure}%[htp]
\centering
\subfigure[Male case.]{ \label{fig:ggi_interaction_male1_case} 
\includegraphics[width=7.5cm,height=7.5cm]{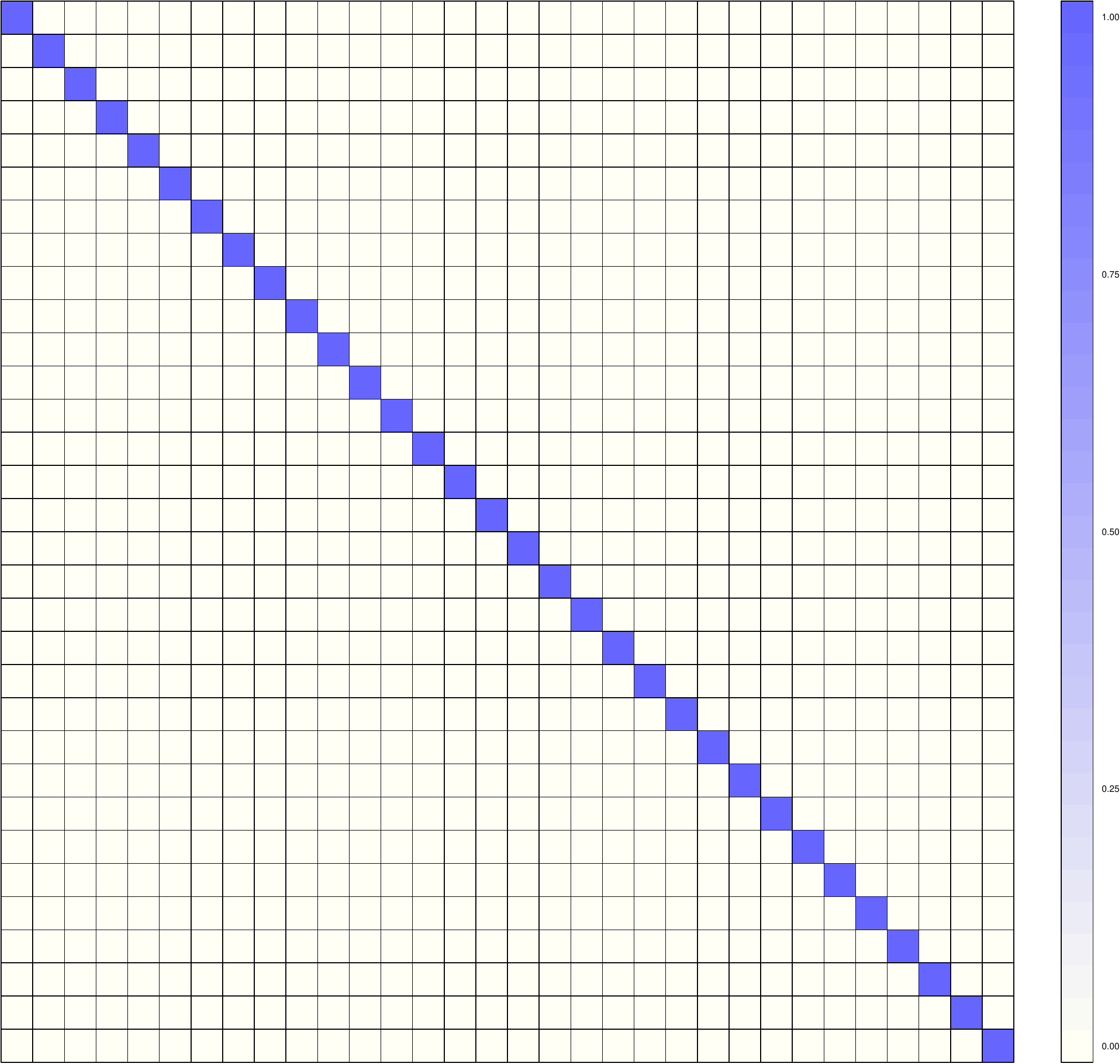}}
\hspace{2mm}
\subfigure[Male case.]{ \label{fig:ggi_interaction_male30_case} 
\includegraphics[width=7.5cm,height=7.5cm]{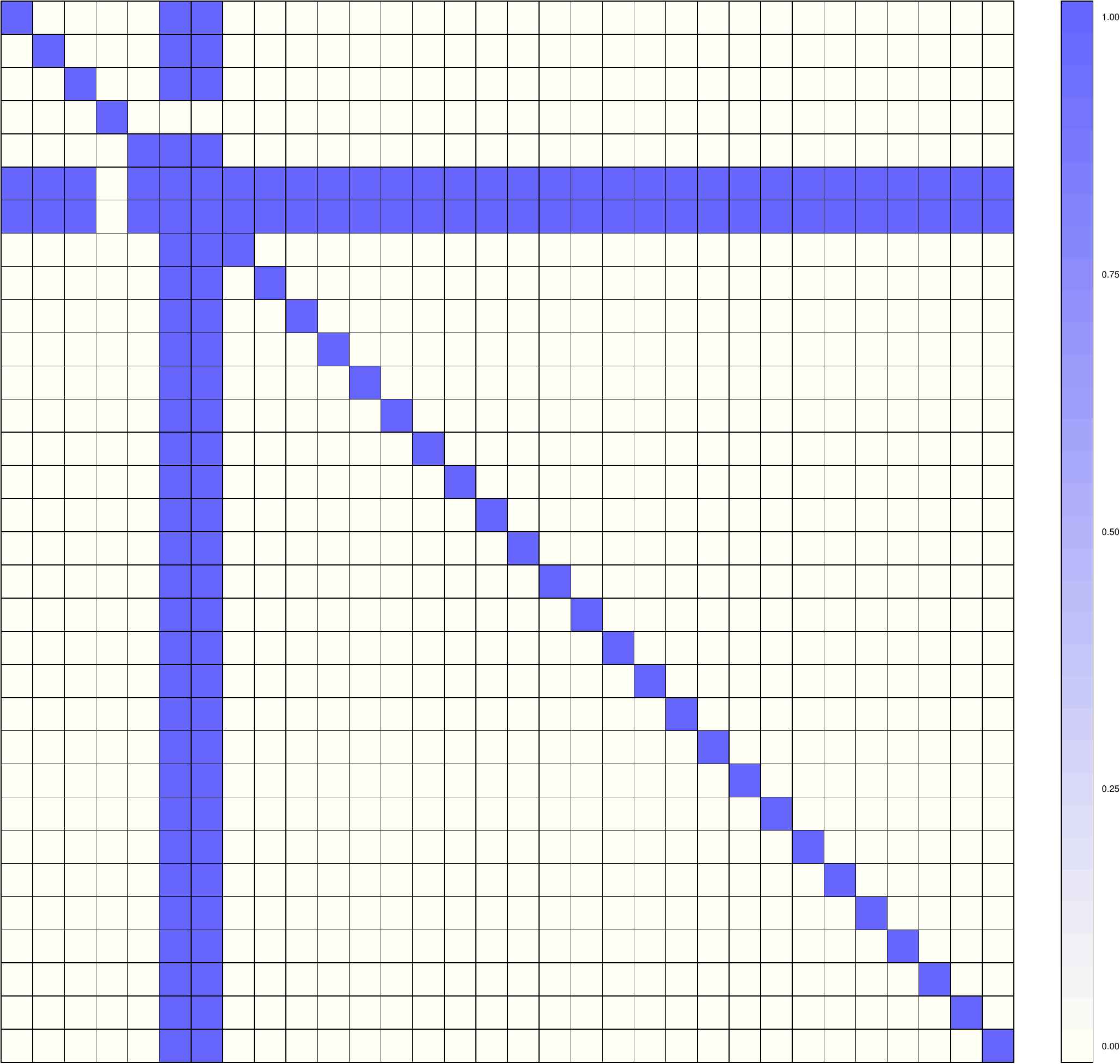}}
\\[2mm]
\subfigure[Male case.]{ \label{fig:ggi_interaction_male46_case} 
\includegraphics[width=7.5cm,height=7.5cm]{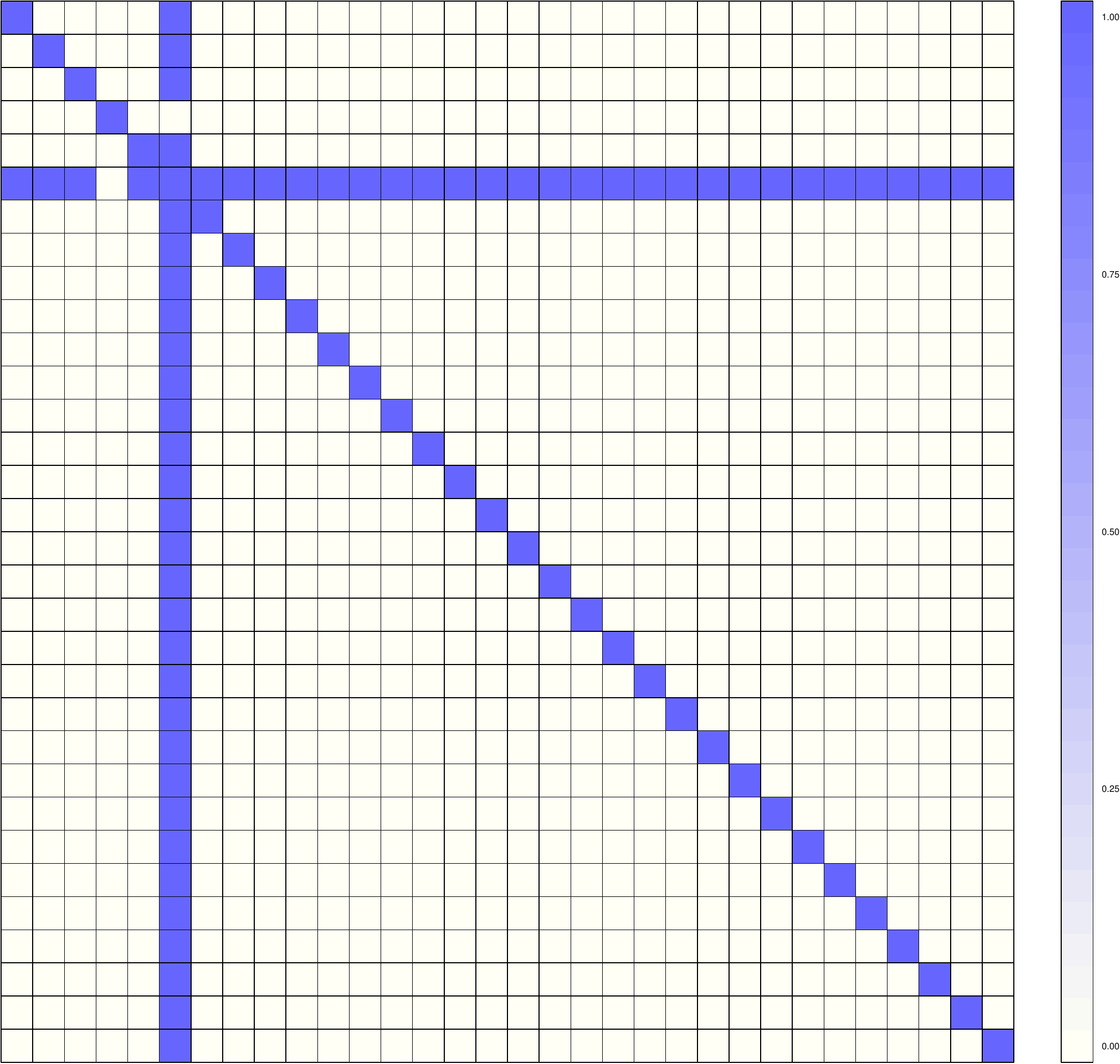}}
\hspace{2mm}
\subfigure[Male control.]{ \label{fig:ggi_interaction_male1_control} 
\includegraphics[width=7.5cm,height=7.5cm]{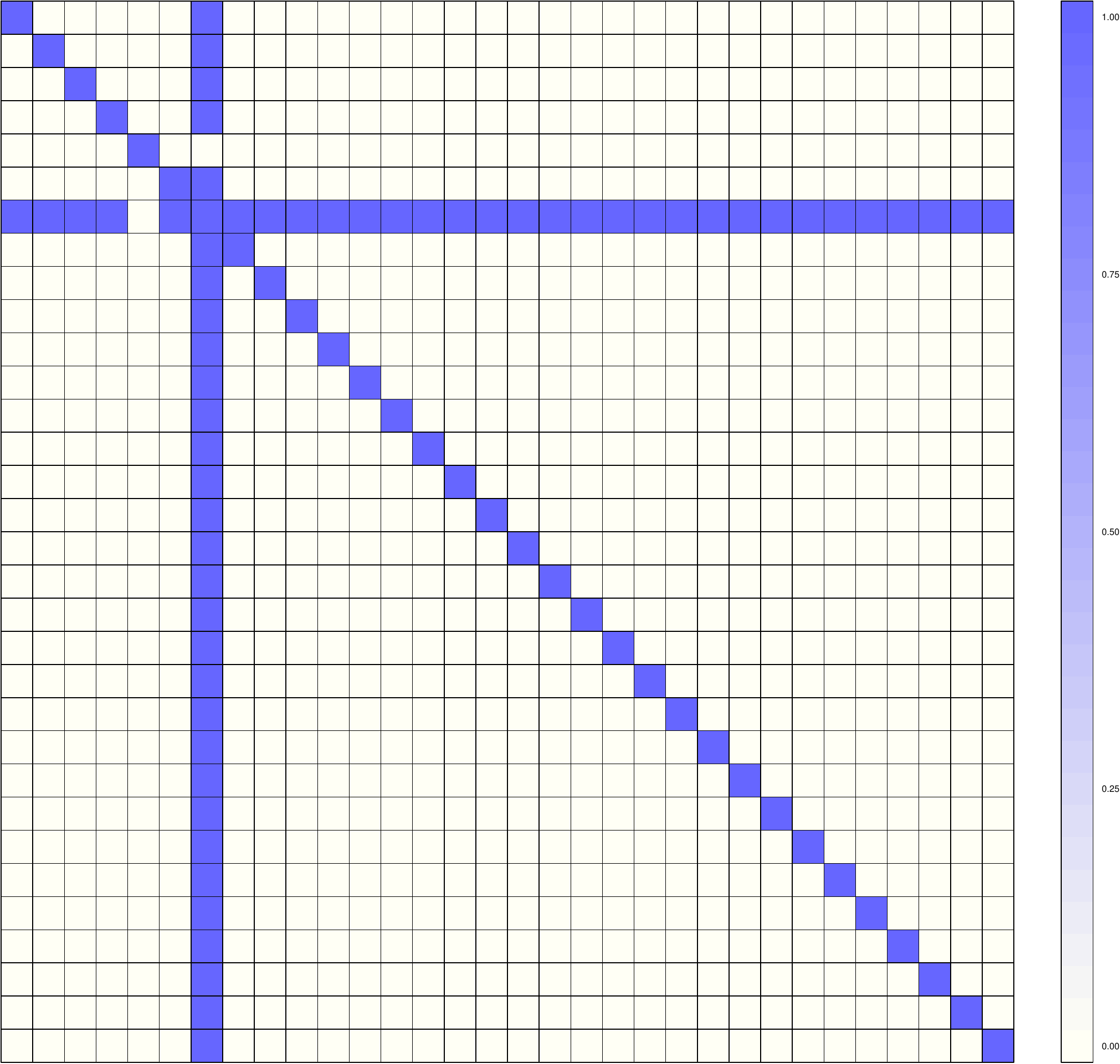}}
\caption{{\bf Presence/absence of gene-gene interactions for typical male cases and controls: Blue denotes presence and white represents
absence of gene-gene interaction.}}
\label{fig:ggi_interaction_plots_male}
\end{figure}

\begin{figure}%[htp]
\centering
\subfigure[Female case.]{ \label{fig:ggi_interaction_female48_case} 
\includegraphics[width=7.5cm,height=7.5cm]{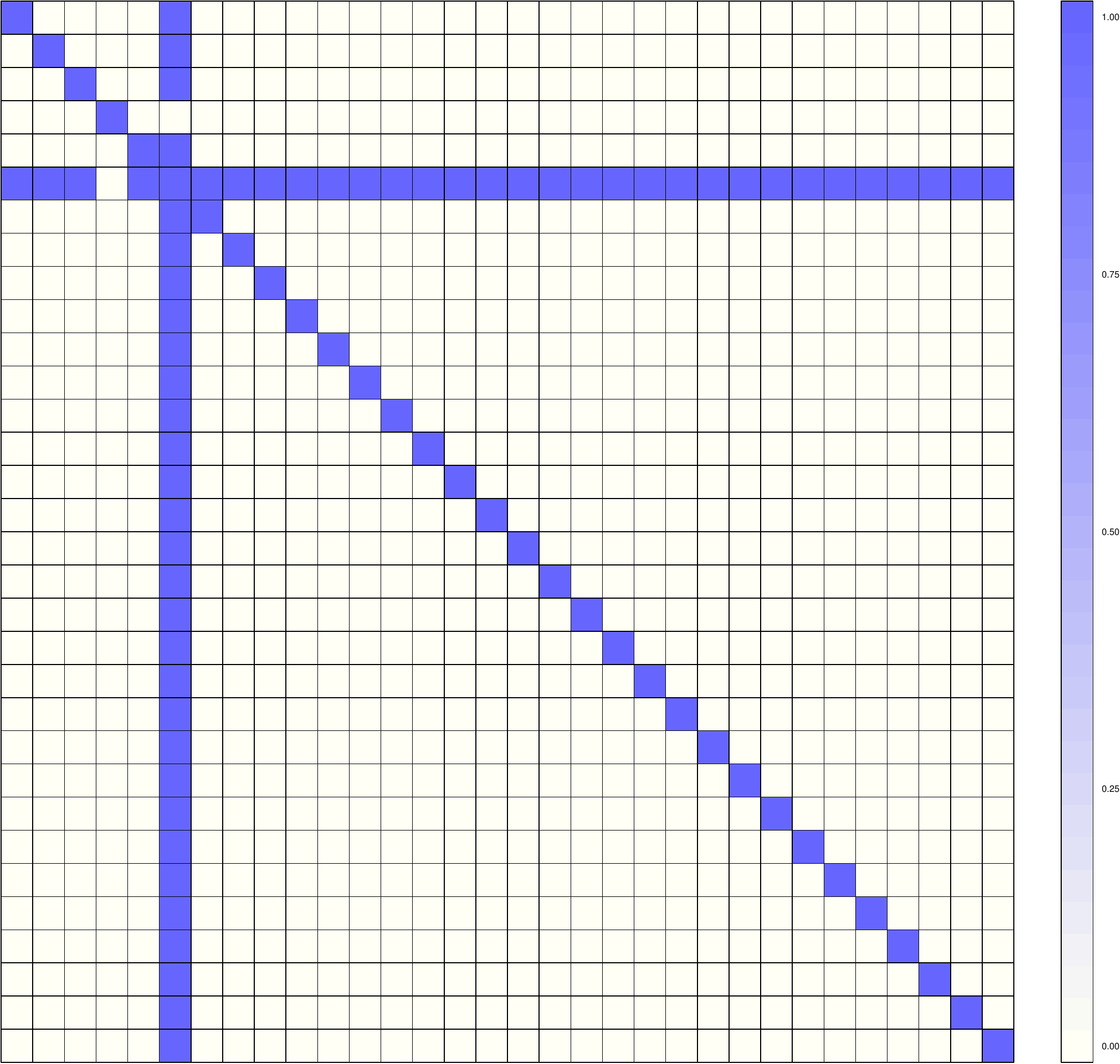}}
\hspace{2mm}
\subfigure[Female control.]{ \label{fig:ggi_interaction_female52_control} 
\includegraphics[width=7.5cm,height=7.5cm]{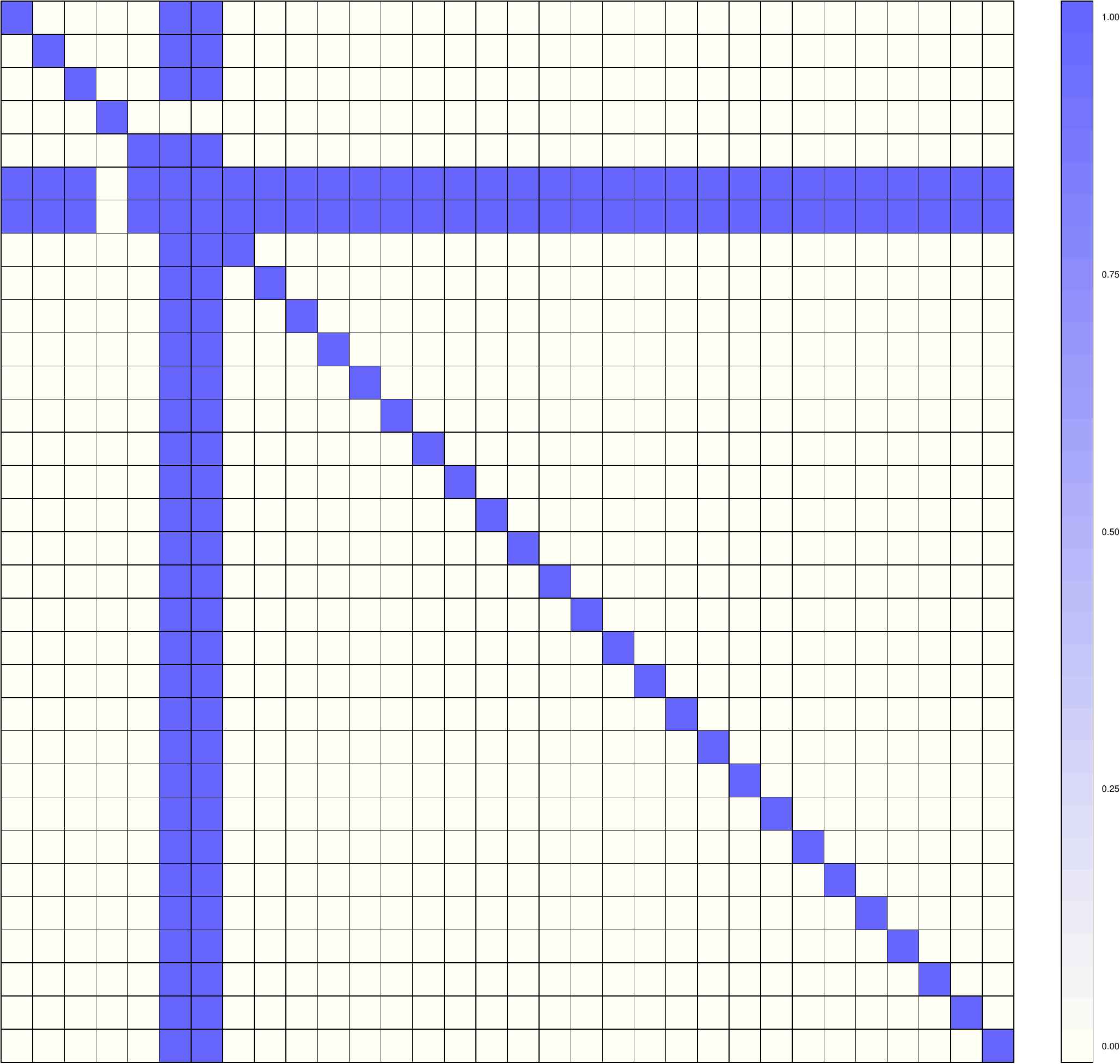}}
\\[2mm]
\subfigure[Female control.]{ \label{fig:ggi_interaction_female55_control} 
\includegraphics[width=7.5cm,height=7.5cm]{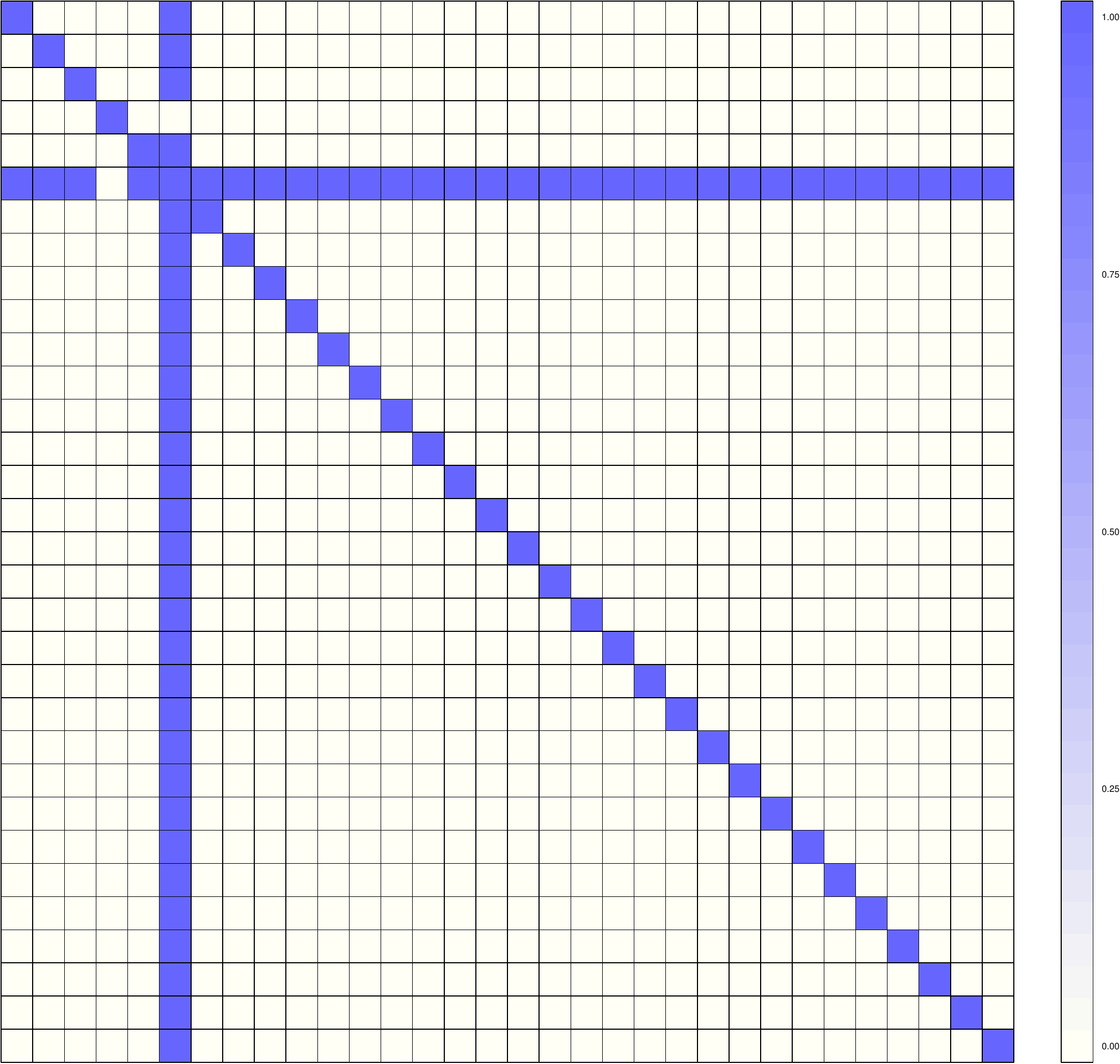}}
\hspace{2mm}
\subfigure[Female control.]{ \label{fig:ggi_interaction_female9_control} 
\includegraphics[width=7.5cm,height=7.5cm]{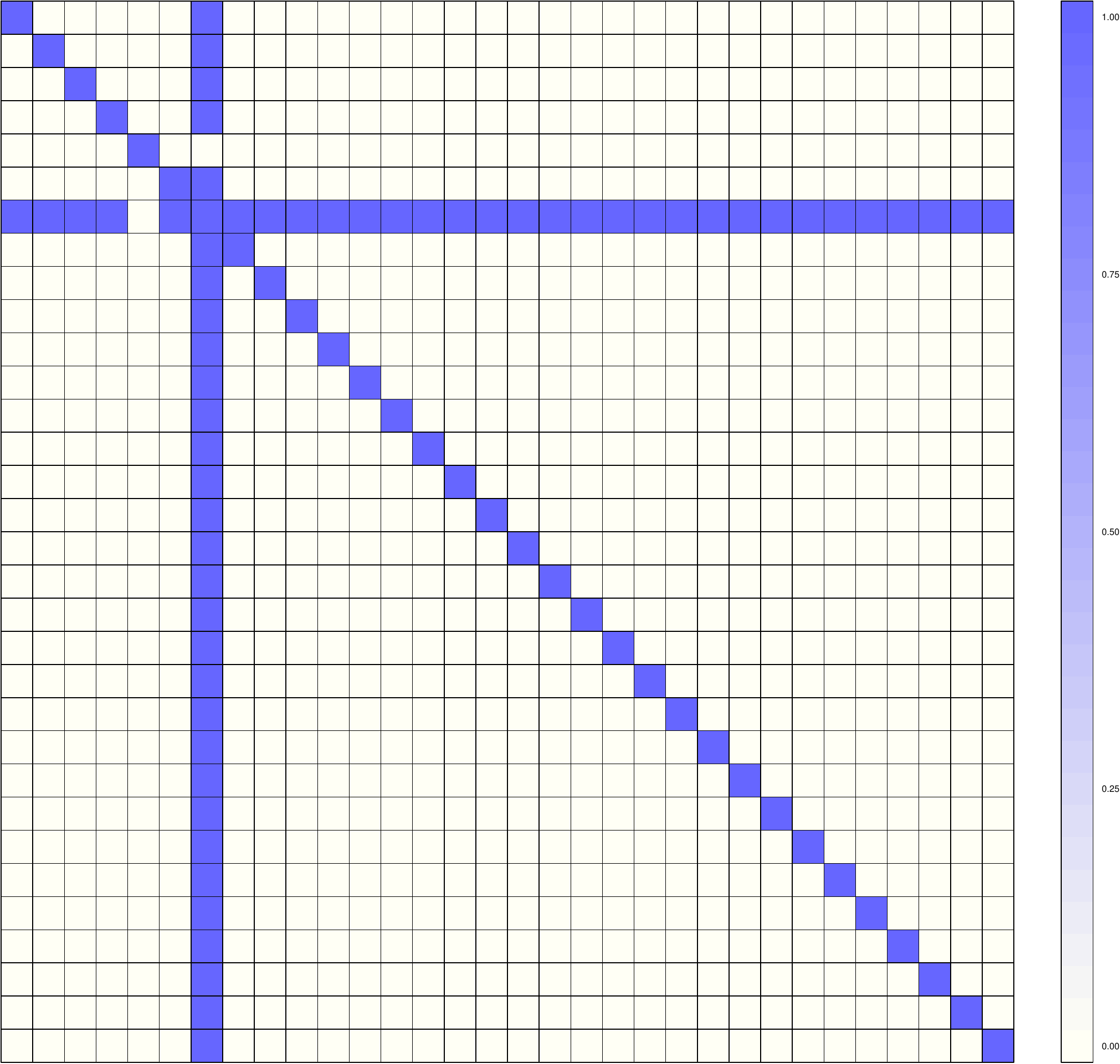}}
\caption{{\bf Presence/absence of gene-gene interactions for typical female cases and controls: Blue denotes presence and white represents
absence of gene-gene interaction.}}
\label{fig:ggi_interaction_plots_female}
\end{figure}

Unlike \ctn{Bhattacharya16}, where there is a single gene-gene correlation structure for all the individuals,
our current model has provision for subject-specific gene-gene correlations. Figures \ref{fig:ggi_plots_male} 
and \ref{fig:ggi_plots_female} show the typical gene-gene correlations representative of cases and controls in all males and 
females respectively. Essentially, the pictures represent the
gene-gene correlation patterns for all the subjects. The
color intensities correspond to the absolute values of the correlations.
Although the correlations are small in all the subjects,
the tests of hypotheses reveal some interesting structures. Figures \ref{fig:ggi_interaction_plots_male}
and \ref{fig:ggi_interaction_plots_female} represent the all possible interacting patterns found in the study. Panel (a) of Figure \ref{fig:ggi_interaction_plots_male} represents 9 male cases where no gene-gene
interaction is significant. Panel (b) shows the genetic interaction pattern in some male cases where $AP006216.10$ and $C6orf106$, interact with
all the other genes. Panel (c) shows the results of significance tests of gene-gene interactions
for some male cases, for whom only $AP006216.10$ interacts with all the other genes in the study. A representative interaction pattern for the male controls shown in panel (d), indicates that only $C6orf106$ or only 
$AP006216.10$ interacts with every gene, but in a few subjects both $AP006216.10$ and $C6orf106$ interact with all the genes.

Even for the females, the two genes, $AP006216.10$ and $C6orf106$, play significant role in gene-gene interactions. Indeed,
in our data, unlike the 9 male cases, there is no female for whom all gene-gene interactions 
are insignificant. The relevant representative plots for the females, given by Figure \ref{fig:ggi_interaction_plots_female},
shows that for all the female cases, only $AP006216.10$ interacts with the other genes. For the female controls,
either only $AP006216.10$ or only $C6orf106$ interacts with the other genes, or both $AP006216.10$ and $C6orf106$ interact significantly with the other genes included in the study.

The messages gained from our analysis seem to be somewhat counter-intuitive but perhaps quite insightful. 
Our tests indicate that the genes have insignificant marginal effect. Thus, some external or non-genetic factors might have some significant role to play. But for most of the subjects, at least one of the genes 
$AP006216.10$ and $C6orf106$ interact with every other
gene. The subjects, for whom no significant genetic interactions involving $AP006216.10$ and $C6orf106$ were detected, turned out to be male cases, indicating that the lack of genetic interaction
in these males failed to get them any preventive measure against MI. 
On the other hand, the interactions of the genes $AP006216.10$ and $C6orf106$
with all the genes seemed to reduce the risk of the disease for the other subjects. Thus, in this study,
the gene-gene interactions seem to have a beneficial effect on the subjects.
It also seems that only a small proportion of males are prone to the risk of having no beneficial gene-gene interactions.

Note that our results are broadly consistent with those obtained by \ctn{Bhattacharya16} but are more
precise and informative. Indeed, they also noted relatively small impact of the individual genes and
small gene-gene correlations. Our current ideas and analyses also support their conclusion that external factors 
(in particular, sex) are
perhaps playing a bigger role in explaining case-control with respect to MI. We recall (see \ctn{Bhattacharya16})
that with respect to the data that we used, the empirical conditional probability of a male given case is about $0.38$,
and that of a male given control is about $0.50$, so that females seem to be more at risk, given our data.
The inherent coherence of the Bayesian paradigm upholds the sex factor by attaching little importance to the individual
genes. However, in contrast with \ctn{Bhattacharya16} who found no interacting genes, here it turns out that
the genes $AP006216.10$ and $C6orf106$ in interaction with other genes generally lower the risk of the individuals with respect to MI. Importantly, each of the few males having no such interactions turned out to be a case. This seems to be roughly in accordance with the popular belief that males are
more susceptible to MI than females. Our Bayesian model coherently weaves together the prior and the data 
and brings out this information in spite of the data-driven information that females are more prone to MI than males.
We also note that \ctn{LucasG12}, who analyzed the same MI dataset using logistic regression, reached the conclusion
that there is no significant gene-gene interaction. Thus, their result completely supports that of \ctn{Bhattacharya16}
and are also very much in keeping with our current results.

\subsubsection{{\bf Posteriors of the number of sub-populations}}
\label{subsubsec:no_of_components}
\begin{figure}%[htp]
\centering
\subfigure[Male case.]{ \label{fig:comp_male1_case}
\includegraphics[width=7.5cm,height=7.5cm]{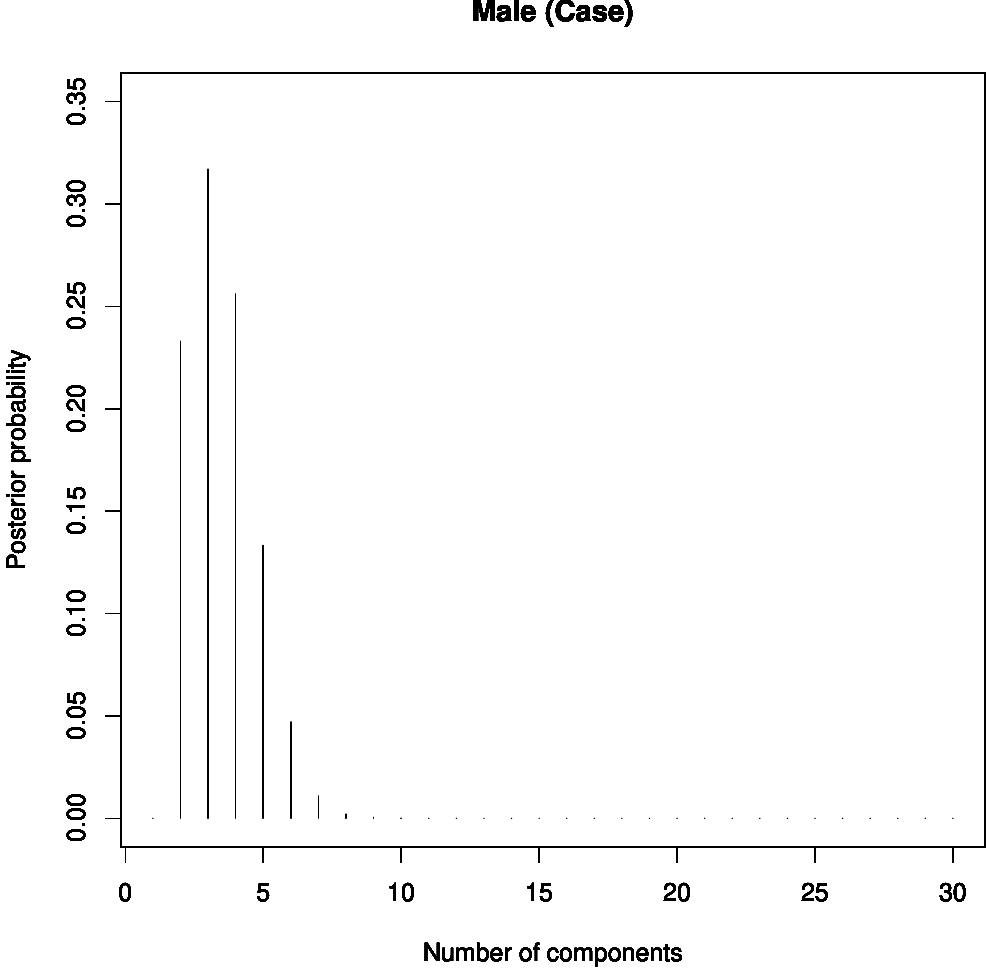}}
\hspace{2mm}
\subfigure[Male case.]{ \label{fig:comp_male30_case}
\includegraphics[width=7.5cm,height=7.5cm]{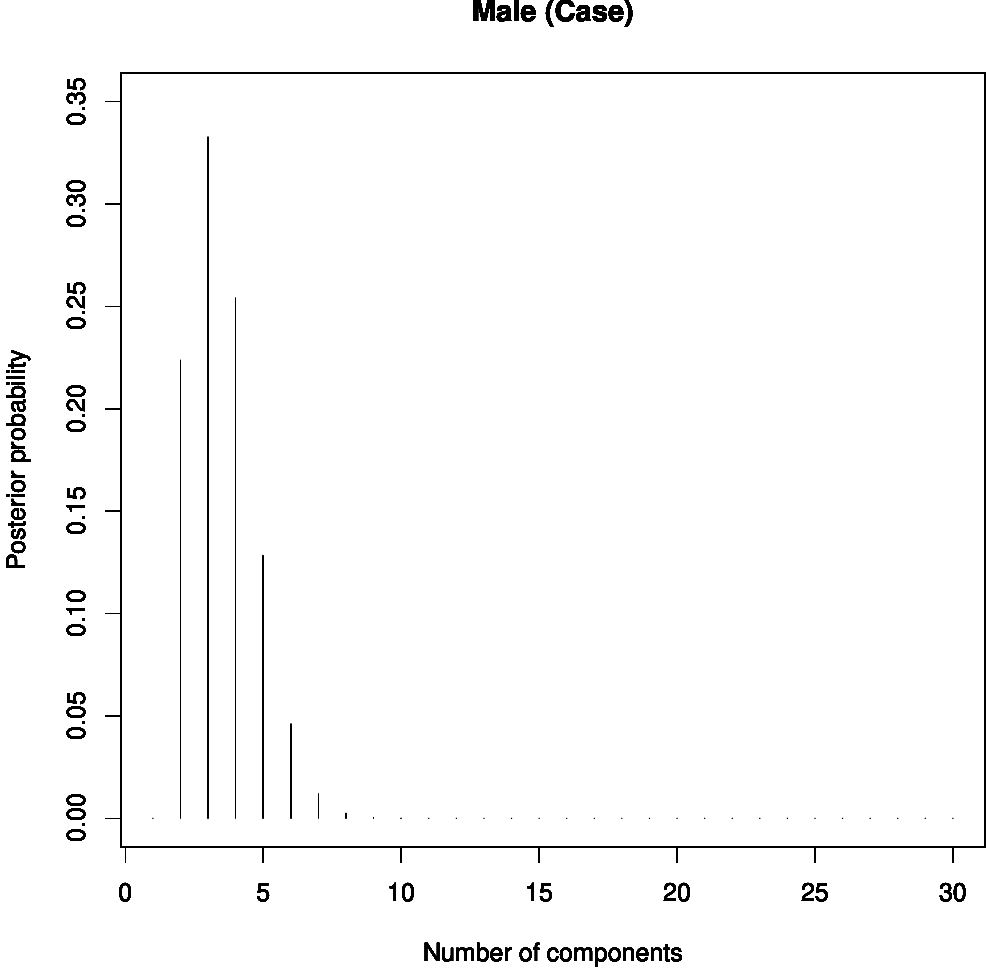}}\\
\vspace{2mm}
\subfigure[Male case.]{ \label{fig:comp_male46_case}
\includegraphics[width=7.5cm,height=7.5cm]{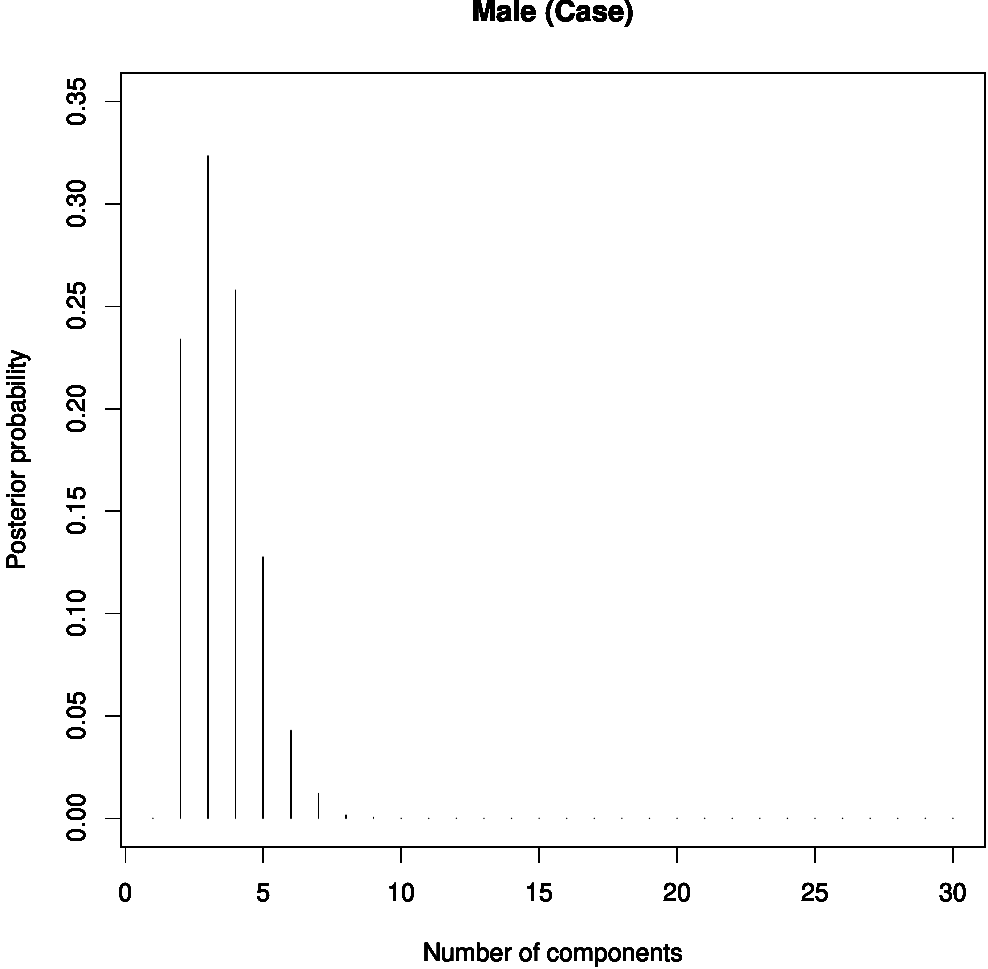}}
\hspace{2mm}
\subfigure[Male control.]{ \label{fig:comp_male1_control}
\includegraphics[width=7.5cm,height=7.5cm]{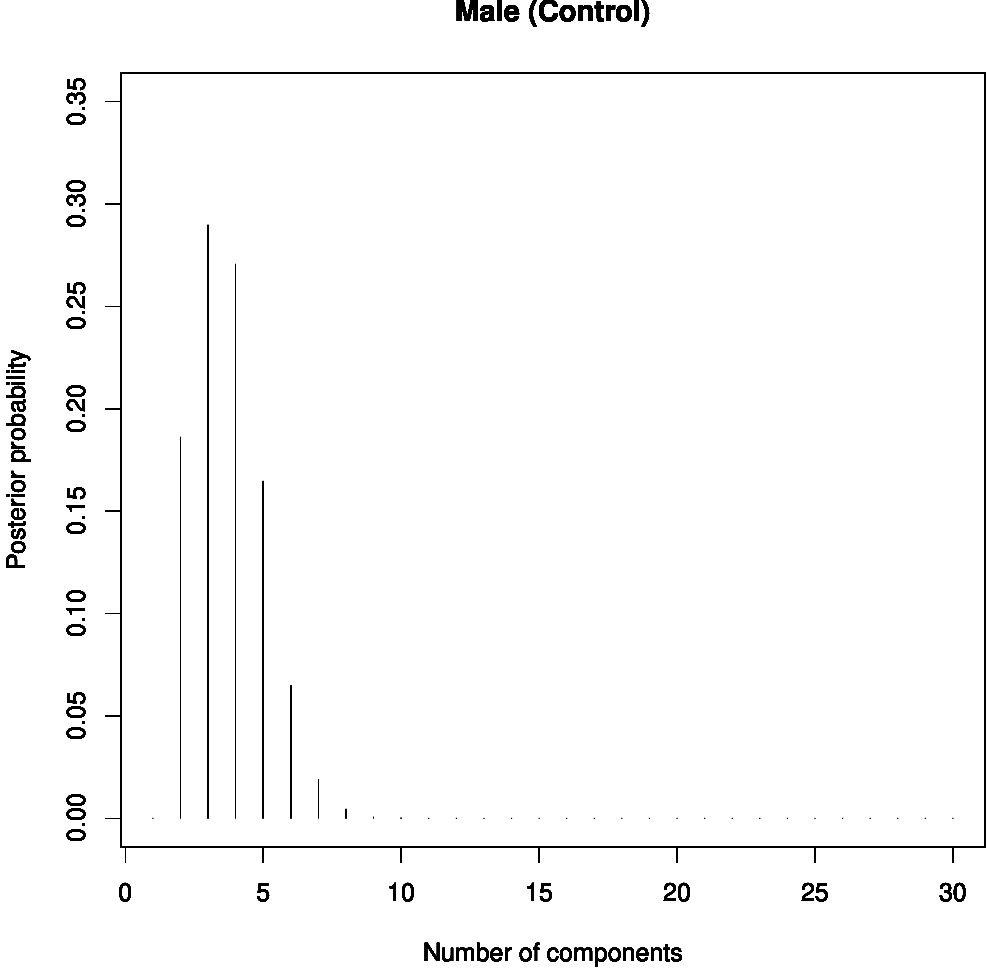}}
\caption{{\bf Posteriors of the number of components for some males.}}
\label{fig:ggi_comp_realdata1}
\end{figure}

\begin{figure}%[htp]
\centering
\subfigure[Female case.]{ \label{fig:comp_female48_case}
\includegraphics[width=7.5cm,height=7.5cm]{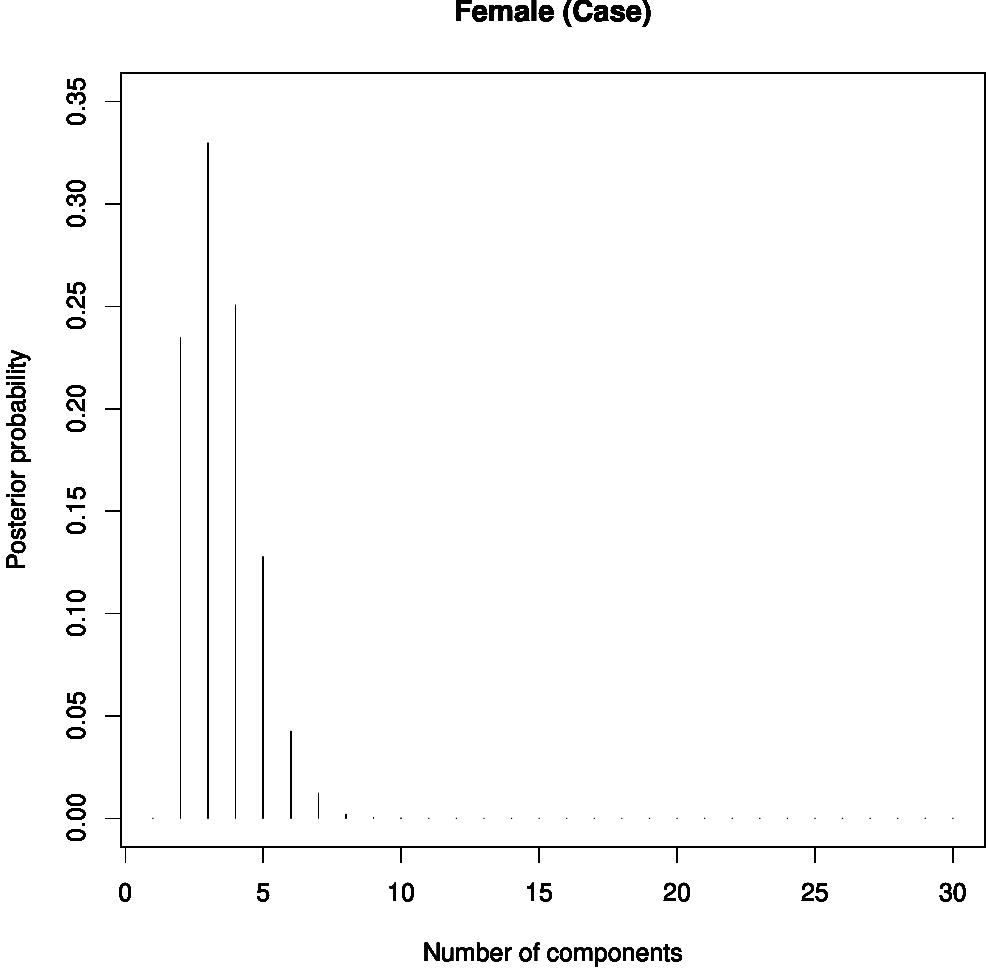}}
\hspace{2mm}
\subfigure[Female control.]{ \label{fig:comp_female52_control}
\includegraphics[width=7.5cm,height=7.5cm]{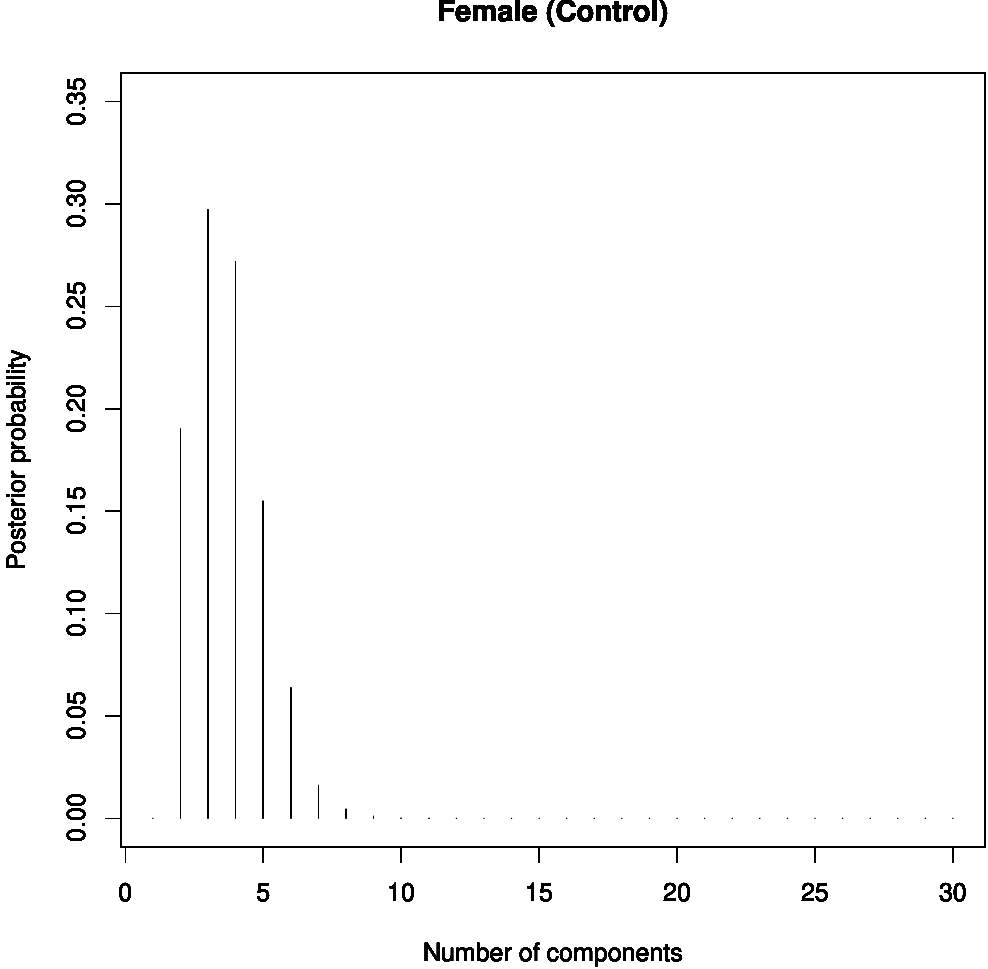}}\\
\vspace{2mm}
\subfigure[Female control.]{ \label{fig:comp_female55_control}
\includegraphics[width=7.5cm,height=7.5cm]{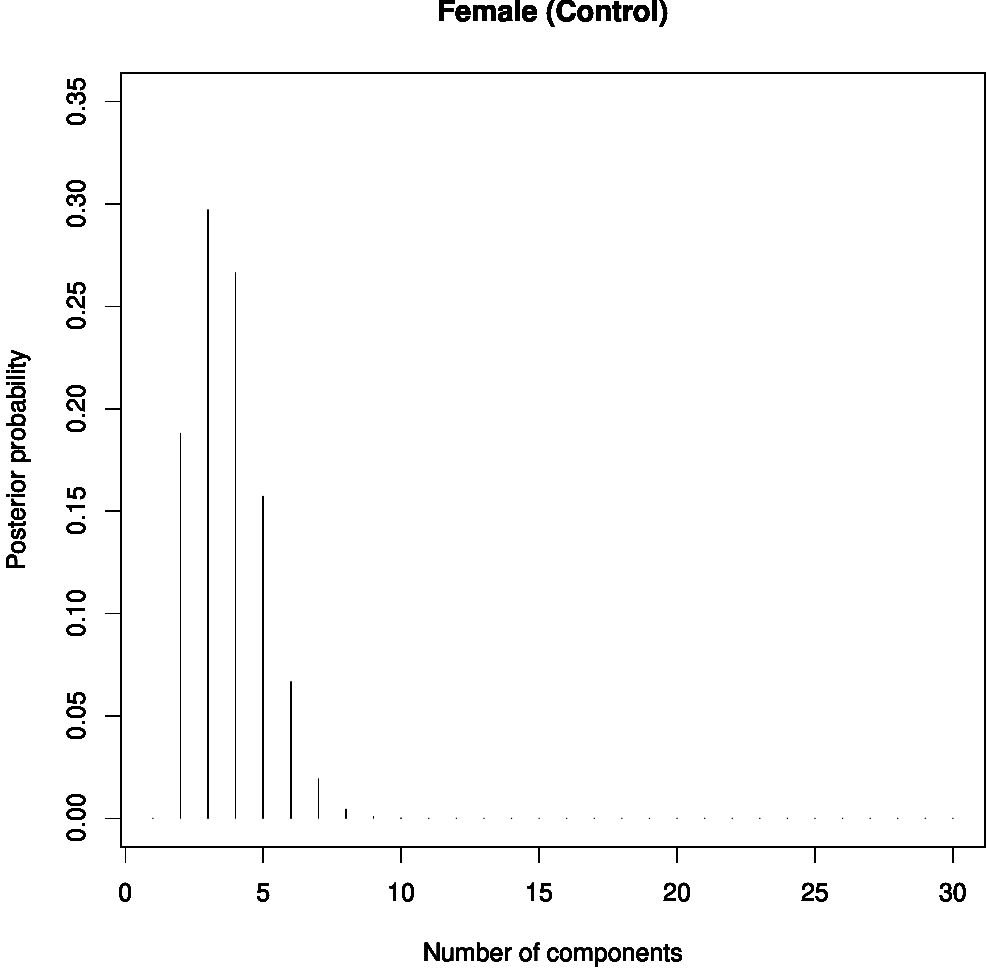}}
\hspace{2mm}
\subfigure[Female control.]{ \label{fig:comp_female9_control}
\includegraphics[width=7.5cm,height=7.5cm]{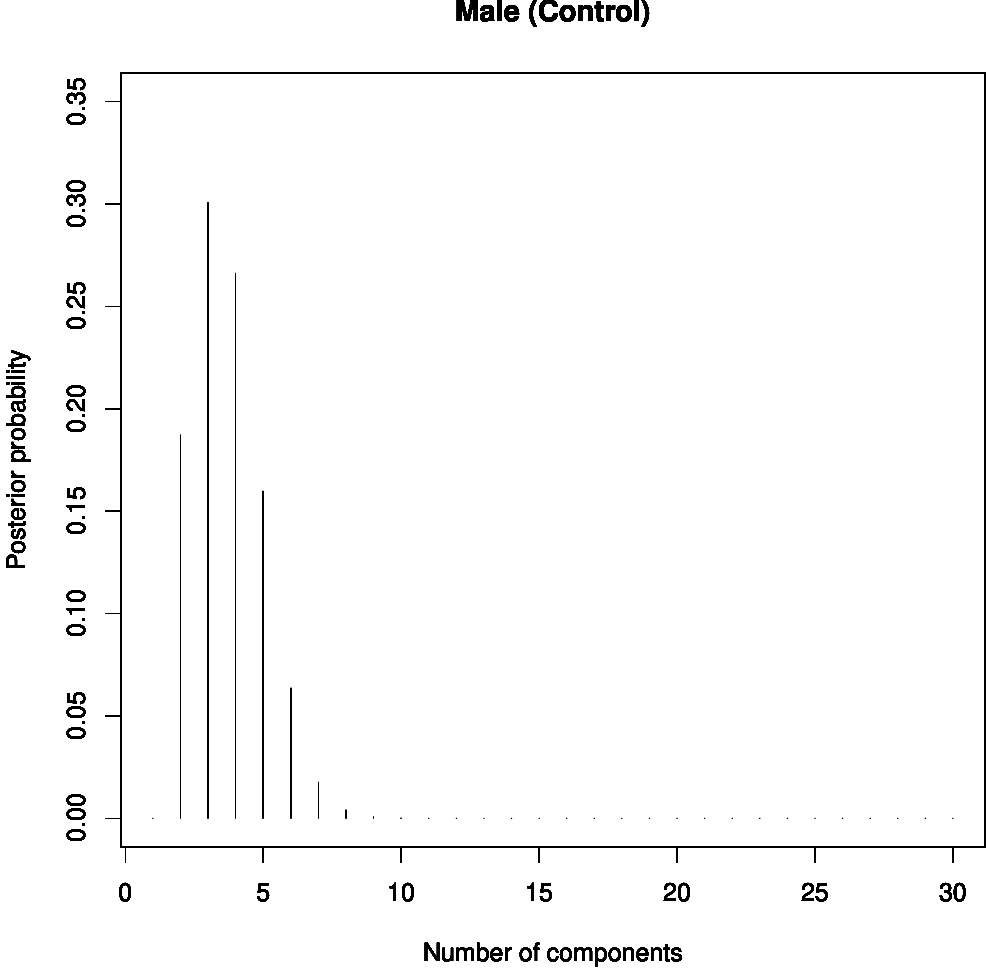}}
\caption{{\bf Posteriors of the number of components for some females.}}
\label{fig:ggi_comp_realdata2}
\end{figure}

Figures \ref{fig:ggi_comp_realdata1} and \ref{fig:ggi_comp_realdata2} show the posteriors of the number 
of sub-populations for the same males and females associated with Figures \ref{fig:ggi_interaction_plots_male}
and \ref{fig:ggi_interaction_plots_female}, respectively.
Observe that the posteriors are quite similar, with the mode at $3$ and $4$ components
receiving the next highest probability. Thus, the 4 sub-populations, irrespective of sex, are well-supported by our model,
showing that these can not be further sub-divided genetically. This is not unexpected, since the roles of the individual genes
are not significant in our study. 
Our result broadly agrees with \ctn{Bhattacharya16} who obtained for different genes, the modes at $5$ components,
with $4$ components receiving the next highest posterior mass. 
%Recall that BB obtained different posteriors for different genes,
%which was the case because the genes turned out to be very significant with significant interactions with each other.
%Since in our case genes seem to play very little role, such identical posteriors of the number of components
%are to be expected. Our results also consistent with the four broad sub-populations composed of Caucasians, Han Chinese,
%Japanese and Yoruban, and coherently asserts that these can not be further sub-divided in terms of genetic variations,
%given the insignificant roles of the genes, given the sex variable.

\section{{\bf Summary and conclusion}}
\label{sec:conclusion}
In this paper, we have proposed a novel Bayesian nonparametric gene-gene and gene-environment
interaction model based on hierarchies of Dirichlet processes. This model is a significant improvement
over that of \ctn{Bhattacharya16} in the sense of much clear interpretability and accounting for
subject-specific gene-gene interactions. Moreover, the interactions arise as natural by-products 
of our nonparametric structure based on HDP, and are not based on matrix normal distributions, as in
\ctn{Bhattacharya15} and \ctn{Bhattacharya16}, and hence, are more realistic. 
We propose a novel parallel MCMC algorithm to implement our model, that combines powerful technology with 
conditionally independent structures inherent within our HDP based model and efficient TMCMC methods. 
The Bayesian tests of hypotheses that we employ in this paper are appropriately modified versions
of those proposed in \ctn{Bhattacharya16}.

Applications of our ideas to biologically realistic datasets generated
under $5$ different set-ups characterized by different combinations and structures associated
with gene-gene and gene-environment interactions demonstrated encouraging performance of our model and methods.
Our analysis of the MI dataset showed strong impact of the sex variable, which is consistent with the results
of \ctn{Bhattacharya16}. Our tests showed no effect of the individual genes, which is also in keeping with
\ctn{Bhattacharya16} who obtained relatively weak marginal effects. But most interestingly, 
even though we obtained very weak gene-gene correlations in accordance with \ctn{Bhattacharya16} and \ctn{LucasG12}, 
our tests on gene-gene interaction showed that two genes, $AP006216.10$ and $C6orf106$, generally 
interact with all the other genes in a beneficial way so as to fight the disease. Moreover, the only situations
where all the gene-gene interactions turned out to be insignificant, were the male cases, showing that the usual belief
that males are more prone to heart attack than females may hold some value from this perspective.

Although many standard methods are commonly used in GWAS to identify the genetic and the environmental effects, there are several reasons that point towards the fact that our approach
is not comparable with the existing methods. Firstly, our main objective is to provide a very flexible Bayesian nonparametric approach
that adequately accounts for various levels of uncertainties, whereas the objective of
the existing approaches is to provide computationally feasible algorithms for analyzing
very large datasets. Since the goals are different, different performance criteria must
be devised to measure the performances of the approaches, and hence, they are not
comparable. Another point which renders our Bayesian approach incomparable with the existing approaches such as PLINK, Bayesian variable selection regression (BVSR) model and the Bayesian
epistasis association mapping (BEAM) etc, is that both the simulated and the real data sets are associated with multiple sub-populations and unlike our Bayesian approach none of the existing methods coherently accounts for the unknown
number of sub-populations. As we argued in our paper, for different sub-populations,
the genes and the loci may interact differently, and so it is of utmost importance to
account for such uncertainty. Indeed, Bhattacharjee et al. (2010) empirically demon-
strate that methods ignoring population sub-structures can incur severe bias leading
to large-scale false positives.

So far, due to insufficient computational resources, we are compelled to restrict focus on a relatively small portion
of the data.
For improving our computing infrastructure, we have already taken the initiative of procuring supercomputing facilities from the Govt. of India, a project led, on behalf of
Indian Statistical Institute, by the second author of this paper. With such a facility, we will be able to analyze the entire MI dataset with much ease.

\newpage

\input{supp2}

%\newpage
\pagebreak
\renewcommand\baselinestretch{1.3}
\normalsize
\bibliographystyle{ECA_jasa}
\bibliography{irmcmc}

\end{document}

%% file: supp2.tex
\renewcommand\thefigure{S-\arabic{figure}}
\renewcommand\thetable{S-\arabic{table}}
\renewcommand\thesection{S-\arabic{section}}

\setcounter{section}{0}
\setcounter{figure}{0}
\setcounter{table}{0}

\begin{center}
{\bf \LARGE Supplementary Material}
\end{center}

\section{{\bf An MCMC method using Gibbs sampling and TMCMC}}
\label{sec:inferential_procedure}

\subsection{{\bf Full conditionals}}
\label{subsec:fullcond}

\subsubsection{{\bf Full conditional of $\bH_k$}}
\label{subsubsec:H_fullcond}
First observe that %conditionally on $\bzeta_S$, the remaining unknowns and the data, 
for $k=0,1$, the full conditional of $\bH_k$ is given by
\begin{equation}
[\bH_k|\cdots]\sim DP\left(\alpha_H+n_{\cdot k},
\frac{\alpha_H \tilde\bH+\sum_{s=1}^Sn_{sk}\delta_{\boeta_s}}{\alpha_H+n_{\cdot k}}\right),
\label{eq:H_fullcond1}
\end{equation}
where $n_{sk}=\#\{r\in\{1,\ldots,R_k\}:\bxi_{rk}=\boeta_s\}$ and $n_{\cdot k}=\sum_{s=1}^Sn_{sk}$.

%Exact simulation from Dirichlet processes is possible using the retrospective
%method (see \ctn{Papas08}) in conjunction with Sethuraman's characterization of Dirichlet processes (see \ctn{Sethuraman94}). 

\subsubsection{{\bf Full conditional of $\bG_{0,jk}$}}
\label{subsubsec:G_fullcond}
Similarly, the full conditional of $\bG_{0,jk}$ is given, for $j=1,\ldots,J$ and $k=0,1$, by
%of the unknowns, and the data, $\bG_{0,jk}$ depends only upon $\bphi_{1jk},\ldots,\bphi_{T_{jk},jk}$, 
%so that its full conditional is given by
\begin{equation}
[\bG_{0,jk}|\cdots]\sim DP\left(\alpha_{G_0,k}+n_{\cdot jk},
\frac{\alpha_{G_0,k}\bH_k+\sum_{l=1}^{R_k}n_{ljk}\delta_{\bxi_{lk}}}{\alpha_{G_0,k}+n_{\cdot jk}}\right),
\label{eq:G_fullcond1}
\end{equation}
where $n_{ljk}=\#\{(t,i)\in\{1,\ldots,\tau_{ijk}\}\times\{1,\ldots,N_k\}:\bphi_{tijk}=\bxi_{lk}\}$
and $n_{\cdot jk}=\sum_{l=1}^{R_k}n_{ljk}$.
%where $M_{tijk}=\#\{m\in\{1,\ldots,M\}:\bp_{mijk}=\bphi_{tijk}\}$ and $M_{\cdot ijk}=\sum_{t=1}^{\tau_{jk}}M_{tijk}$.

The full conditionals of $\bH_k$ and $\bG_{0,jk}$ given by (\ref{eq:H_fullcond1}) and (\ref{eq:G_fullcond1}) indicate
generating the infinite-dimensional random probability measures using Sethuraman's characterization of Dirichlet processes 
(see \ctn{Sethuraman94}). However, in our case, forming the infinite-dimensional Sethuraman's construction
is not necessary; instead, it will be required to simulate from the random probability measures having distributions
(\ref{eq:H_fullcond1}) and (\ref{eq:G_fullcond1}). Such simulations are possible using the retrospective
method (see \ctn{Papas08}) which avoids dealing with infinitely many objects.

\subsubsection{{\bf Full conditional of $\bp_{mijk}$}}
\label{subsubsec:p_fullcond}

%$\bP_{Mijk}\backslash\left\{\bp_{mijk}\right\}$.
%, where, in our notation,
%Let $\bP_{Mijk}=\left\{\bp_{1ijk},\ldots,\bp_{Mijk}\right\}$. 
The associated Polya urn distribution of $\bp_{mijk}$ given $\bP_{Mijk}\backslash \{\bp_{mijk}\}$, 
derived by marginalizing over $\bG_{ijk}$, is the following:
\begin{align}
\left[\bp_{mijk}|\bP_{Mijk}\backslash \{\bp_{mijk}\}\right]
&=\frac{\alpha_{G,ik}}{\alpha_{G,ik}+M-1}\bG_{0,jk}\left(\bp_{mijk}\right)
+\frac{1}{\alpha_{G,ik}+M-1}\sum_{m'\neq m=1}^M\delta_{\bp_{m'ijk}}\left(\bp_{mijk}\right)
\label{eq:polya}\\
&=\frac{\alpha_{G,ik}}{\alpha_{G,ik}+M-1}\bG_{0,jk}\left(\bp_{mijk}\right)
+\frac{1}{\alpha_{G,ik}+M-1}\sum_{t=1}^{\tau_{ijk}}M_{tijk}\delta_{\bphi_{tijk}}\left(\bp_{mijk}\right).
\label{eq:polya2}
\end{align}
where $M_{tijk}=\#\{m'\in\{1,\ldots,M\}\backslash\{m\}:\bp_{m'ijk}=\bphi_{tijk}\}$ %and $M_{\cdot ijk}=\sum_{t=1}^{\tau_{jk}}M_{tijk}$.
and $\delta_{\bphi_{tijk}}(\cdot)$ denotes point mass at $\bphi_{tijk}$.

Given $z_{ijk}=m$, on combining the Polya urn distribution with the likelihood 
$\prod_{r=1}^{L}f(x_{ijkr}|p_{mijkr})$ we obtain the following full conditional of $\bp_{mijk}$:
\begin{align}
\left[\bp_{mijk}|\cdots\right]
&\propto \alpha_{G,ik}\prod_{r=1}^{L}f(x_{ijkr}|p_{mijkr})\bG_{0,jk}\left(\bp_{mijk}\right)
+\sum_{t=1}^{\tau_{ijk}}M_{tijk}\prod_{r=1}^{L}f(x_{ijkr}|\phi_{tijkr})\delta_{\bphi_{tijk}}\left(\bp_{mijk}\right).
\label{eq:polya3}
\end{align}
Note that in (\ref{eq:polya3}), $\bG_{0,jk}$, drawn from (\ref{eq:G_fullcond1}), 
is not available in closed form and only admits the form dictated by Sethuraman's construction, given, almost surely, by
\begin{equation}
\bG_{0,jk}=\sum_{l=1}^{\infty}\tilde p_l\delta_{\tilde\bxi_{ljk}},
\label{eq:sethuraman_G}
\end{equation}
where $\tilde p_1=V_1$, $\tilde p_l=V_l\prod_{s<l}(1-V_s)$, for $l\geq 2$, 
with $V_1,V_2,\ldots\stackrel{iid}{\sim}\mbox{Beta}\left(\alpha_{G_0,k}+n_{\cdot jk},1\right)$,
and for $l=1,2,\ldots$,
$\tilde\bxi_{ljk}\stackrel{iid}{\sim}
\frac{\alpha_{G_0,k}\bH_k+\sum_{l=1}^{R_k}n_{ljk}\delta_{\bxi_{lk}}}{\alpha_{G_0,k}+n_{\cdot jk}}$.

In (\ref{eq:polya3}), the posterior proportional to 
$\prod_{r=1}^{L}f(x_{ijkr}|p_{mijkr})\bG_{0,jk}\left(\bp_{mijk}\right)$, which we denote by
$[\bG_{0,jk}|\bX_{ijk}]$, is the
discrete distribution that puts mass $C_{ijk}\tilde p_t\prod_{r=1}^{L}f(x_{ijkr}|\tilde\xi_{tjkr})$
to the point $\tilde\bxi_{tjk}$,
for $t=1,2,\ldots$, where  
\begin{equation}
C_{ijk}=\left(\sum_{t=1}^{\infty}\tilde p_t\prod_{r=1}^{L}f(x_{ijkr}|\tilde\xi_{tjkr})\right)^{-1}
\label{eq:Cjk}
\end{equation}
is the normalizing constant.
Combining these with (\ref{eq:polya3}) it follows that
\begin{align}
\left[\bp_{mijk}|\cdots\right]
&=\alpha_{G,ik}\bar C C^{-1}_{ijk}[\bG_{0,jk}\left(\bp_{mijk}\right)|\bX_{ijk}]
+\bar C\sum_{t=1}^{\tau_{ijk}}M_{tijk}\prod_{r=1}^{L}f(x_{ijkr}|\phi_{tijkr})\delta_{\bphi_{tijk}}\left(\bp_{mijk}\right),
\label{eq:polya4}
\end{align}
where
\begin{equation*}
\bar C=\left[\alpha_{G,ik}C^{-1}_{jk}+\sum_{t=1}^{\tau_{ijk}}M_{tijk}\prod_{r=1}^{L_j}f(x_{ijkr}|\phi_{tijkr})\right]^{-1}
\end{equation*}
is the normalizing constant of $\left[\bp_{mijk}|\cdots\right]$.

\subsection{{\bf Retrospective method for simulating from $\left[\bp_{mijk}|\cdots\right]$}}
\label{subsec:retro1}

From (\ref{eq:polya4}) it follows that, to draw from $\left[\bp_{mijk}|\cdots\right]$, it is required to simulate
from $[\bG_{0,jk}\left(\bp_{mijk}\right)|\bX_{ijk}]$ with probability proportional to $C^{-1}_{ijk}$.
However, since $C_{ijk}$ involves an infinite series, its calculation is infeasible. 
The same issue also prevents the traditional simulation methods to draw from the discrete distribution $[\bG_{0,jk}|\bX_{ijk}]$. 
In this case, the retrospective sampling method proposed in Section 3.5 of \ctn{Papas08} is the appropriate
method for our purpose. We first briefly discuss the role of such method in
simulating from $[\bG_{0,jk}|\bX_{ijk}]$, and then argue that a by-product of the method can be used
to estimate $C_{ijk}$ arbitrarily accurately.

\subsubsection{{\bf Retrospective method to draw from $[\bG_{0,jk}\left(\bp_{mijk}\right)|\bX_{ijk}]$}}
\label{subsubsec:retro2}

Note that the retrospective method requires $\prod_{r=1}^{L}f(x_{ijkr}|\phi_{tijkr})$
in our case to be uniformly bounded for all $\bphi_{tijk}$, which holds in our case, as 
$f(x_{ijkr}|\phi_{tjkr})$ represents the Bernoulli distribution, which is bounded above by 1. We briefly
describe the method as follows. Let
\begin{equation}
c_{\ell}(K)=\sum_{a=1}^K\tilde p_a\prod_{r=1}^{L}f(x_{ijkr}|\tilde\xi_{ajkr})
\label{eq:c_l}
\end{equation}
and
\begin{equation}
c_{u}(K)=c_{\ell}(K)+(1-\sum_{a=1}^K\tilde p_a). 
\label{eq:c_u}
\end{equation}
Let us also
define 
$\breve p_{\ell,a}(K)=\tilde p_a\prod_{r=1}^{L}f(x_{ijkr}|\tilde\xi_{ajkr})/c_{\ell}(K)$ and
$\breve p_{u,a}(K)=\tilde p_a\prod_{r=1}^{L}f(x_{ijkr}|\tilde\xi_{ajkr})/c_{u}(K)$. To simulate
from $[\bG_{0,jk}|\bX_{ijk}]$ we first generate $U\sim \mbox{Uniform}(0,1)$, and given $U$,
choose $\tilde\bxi_{tjk}$ when 
\begin{equation}
\sum_{a=1}^{t-1}\breve p_{u,a}(K)\leq U\leq\sum_{a=1}^{t}\breve p_{\ell,a}(K).
\label{eq:retro1}
\end{equation}
In fact, $K$ needs to be increased and $\tilde p_t$ and $\prod_{r=1}^{L}f(x_{ijkr}|\tilde\xi_{ajkr})$ simulated
retrospectively, till (\ref{eq:retro1}) is satisfied for some $t\leq K$. 

%where $\tilde p_1=V_1$, $\tilde p_l=V_l\prod_{s<l}(1-V_s)$, for $l\geq 2$, 
%with $V_1,V_2,\ldots\stackrel{iid}{\sim}Beta\left(\alpha_{G_0,k}+n_{\cdot,jk},1\right)$,
% 

%Observe that, following our recommended procedure of generation from the full conditional of $[p_{mijk}|\cdots]$,
%separate simulation from the full conditionals of $\bH_k$ and $\bG_{0,jk}$, given by (\ref{eq:H_fullcond1})
%and (\ref{eq:G_fullcond1}), are rendered unnecessary. However, for the purpose
%of updating the parameters %$\alpha_G$, $\bbeta_G$, 
%$\mu_{G_0}$, $\bbeta_{G_0}$, $\mu_H$
%and $\bbeta_H$ involved in %$\alpha_{G,ik}$, 
%$\alpha_{G_0,k}$ and $\alpha_H$, we still need to continue augmenting $p^*_l$'s such that for any given $\epsilon>0$,
%$\sum_{l=1}^Kp^*_l>1-\epsilon$, for $K\geq K_0$, for some $K_0\geq 1$. We also need to keep track of
%the probabilities $\tilde p^*_1=\tilde V^*_1$ and 
%$\tilde p^*_l=\tilde V^*_l\prod_{s<l}(1-\tilde V^*_s)$, for $l\geq 2$, associated with $\boeta^*_1,\boeta^*_2,\ldots$, where 
%$\tilde V^*_l\stackrel{iid}{\sim}\mbox{Beta}\left(\alpha_H+n_{\cdot k},1\right)$, for $l\geq 1$. In other words,
%we need to ensure that for any given $\epsilon>0$, $\sum_{l=1}^K\tilde p^*_l>1-\epsilon$, for $K\geq K_0$, for some $K_0\geq 1$.

\subsubsection{{\bf Retrospective method for estimating $C_{ijk}$ arbitrarily accurately}}
\label{subsubsec:retro4}
By choosing $K$ to be large enough, the quantities $c_{\ell}(K)$ and $c_{u}(K)$ given by
(\ref{eq:c_l}) and (\ref{eq:c_u}), respectively, can be made arbitrarily close. In other words, for
any $\epsilon>0$, there exists $K_0\geq 1$ such that $|c_{\ell}(K)-c_{u}(K)|<\epsilon$, for $K\geq K_0$.
Thus, for any such $K\geq K_0$, one may approximate $C_{ijk}$ with $\left[c_{\ell}(K)\right]^{-1}$. 
In practice, it is only required to simulate $\tilde U\sim \mbox{Uniform}(0,1)$ and simulate from
$[\bG_{0,jk}\left(\bp_{mijk}\right)|\bX_{ijk}]$ if $\tilde U\leq \bar C C^{-1}_{ijk}$. 
For sufficiently small $\epsilon$ and for finite number of simulations, it will generally hold
that $\tilde U\leq \bar C C^{-1}_{ijk}$ if and only if $\tilde U\leq \bar C_{\epsilon} c_{\ell}(K)$, for $K\geq K_0$,
where 
\begin{equation*}
\bar C_{\epsilon}=\left[c^{-1}_{\ell}(K)+\sum_{t=1}^{\tau_{ijk}}M_{tijk}\prod_{r=1}^{L}f(x_{ijkr}|\phi_{tijkr})\right]^{-1}.
\end{equation*}

\subsubsection{{\bf Retrospective method to simulate from 
$\frac{\alpha_{G_0,k}\bH_k+\sum_{l=1}^{R_k}n_{ljk}\delta_{\bxi_{lk}}}{\alpha_{G_0,k}+n_{\cdot jk}}$}}
\label{subsubsec:retro3}
Note that the retrospective simulation method requires simulation of
$\tilde\bxi_{ljk}\stackrel{iid}{\sim}
\frac{\alpha_{G_0,k}\bH_k+\sum_{l=1}^{R_k}n_{ljk}\delta_{\bxi_{lk}}}{\alpha_{G_0,k}+n_{\cdot jk}}$,
for $l=1,2,\ldots$. This requires simulation from $\bH_k$ with probability proportional to
$\alpha_{G_0,k}$. For this, we first simulate $U\sim\mbox{Uniform}(0,1)$. We then simulate 
%$\boeta^*_{1k}\sim\bH_k$ 
a realization from $\bH_k$ 
after generating $\bH_k$ from the Dirichlet process given by (\ref{eq:H_fullcond1}).
Note that we do not have to generate the entire random probability measure $\bH_k$ for this; we only need
to generate as many realizations $\boeta^*_{lk}$'s from 
$\frac{\alpha_H\tilde\bH+\sum_{s=1}^Sn_{sk}\delta_{\boeta_s}}{\alpha_H+n_{\cdot k}}$ 
and as many $p^*_{lk}=V^*_{lk}\prod_{s<l}(1-V^*_{lk})$; $l=1,2\ldots$,
with $p^*_{1k}=V^*_{1k}$, with $V^*_{lk}\stackrel{iid}{\sim}\mbox{Beta}\left(\alpha_{H}+n_{\cdot k},1\right)$, as required
to satisfy $\sum_{l=1}^{t-1}p^*_{lk}< U\leq \sum_{l=1}^{t}p^*_{lk}$, for some $t\geq 1$ (with $p^*_0=0$).
We then report $\tilde\bxi_{1jk}=\boeta^*_{tk}$ with probability proportional to $\alpha_{G_0,k}$
and $\tilde\bxi_{1jk}=\bxi_{\tilde ljk}$ with probability proportional to $n_{\tilde l jk}$, for $\tilde l\in\{1,\ldots,R_k\}$. 
We repeat this procedure for generating $\bxi_{ljk}$; $l\geq 2$,
by sequentially augmenting the existing simulations of $\boeta^*_{lk}$'s and $p^*_{lk}$'s
with new draws from $\tilde\bH$ and $\mbox{Beta}\left(\alpha_{H}+n_{\cdot k},1\right)$, if needed. 
Indeed, note that for augmentation of $p^*_{lk}$'s, only extra $V^*_{lk}$'s need to be generated 
from $\mbox{Beta}\left(\alpha_{H}+n_{\cdot k},1\right)$.

\subsection{{\bf Updating procedure of $z_{ijk}$ and $\bp_{mijk}$}}
\label{subsec:update_z_p}
The full conditional of $z_{ijk}$ is given by the following:
\begin{equation}
[z_{ijk}=m|\cdots]\propto\pi_{mijk}\prod_{r=1}^{L_j}f\left(\bx_{ijkr}|p_{mijkr}\right);
%{\sum_{m'=1}^M\pi_{m'jk}\prod_{r=1}^{L_j}f\left(\bx_{ijkr}|p_{m'jkr}\right)}
\label{eq:fullcond_z}
\end{equation}
for $m=1,\ldots,M$.

In Section \ref{subsec:retro1} we have devised a method of simulating from the full conditional of $\bp_{mijk}$
given the data and the remaining variables. For our convenience, we re-formulate the full conditional
in terms of the dishes $\bphi_{tjk}$ and the indicators of the dishes, which we denote by $t_{mijk}$, where
$t_{mijk}=t$ if and only if $p_{mijk}=\phi_{tijk}$; $t=1,\ldots,\tau_{ijk}$.

%Let $\left\{\bp^*_{1ijk},\ldots,\bp^*_{\tau_{ijk} ijk}\right\}$ denote the distinct elements
%in $\bP_{Mijk}=\left\{\bp_{1ijk},\ldots,\bp_{Mijk}\right\}$. Also let $\bC_{ijk}=\left\{c_{1ijk},\ldots,c_{Mijk}\right\}$
%denote the configuration vector, where $c_{mijk}=\ell$ if and only if $\bp_{mijk}=\bp^*_{\ell ijk}$.

Now let $\tau^{(m)}_{ijk}$ denote the number of elements in 
%$\bP_{-Mijkm}$.
%=
$\bP_{Mijk}\backslash\left\{\bp_{mijk}\right\}$ that arose from $[\bG_{0,jk}|\bX_{ijk}]$.
%, where, in our notation,
%$\bP_{Mijk}=\left\{\bp_{1ijk},\ldots,\bp_{Mijk}\right\}$. 
Also let ${\bphi^m}^*_{tijk}=\left\{{\phi^m}^*_{tijkr};~r=1,\ldots,L\right\};~t=1,\ldots,\tau^{(m)}_{ijk}$ 
denote the parameter vectors arising from $[\bG_{0,jk}|\bX_{ijk}]$. Further, let ${\bphi^m}^*_{tijk}$
occur $M_{mtijk}$ times. 

Then we update $t_{mijk}$ using Gibbs steps, where the full conditional distribution of $t_{mijk}$ is given by
\begin{equation}
[t_{mijk}=t|\cdots]\propto\left\{\begin{array}{ccc}q^*_{t, mijk} & \mbox{if} & t=1,\ldots,\tau^{(m)}_{ijk};\\
q_{0,mijk} & \mbox{if} & t=\tau^{(m)}_{ijk}+1,\end{array}\right.
\label{eq:fullcond_c}
\end{equation}
where
\begin{align}
q_{0,mijk} &=\alpha_{G,ik}C^{-1}_{ijk};
\label{eq:q0}\\
q^*_{t, mijk} &=M_{mtijk}\prod_{r=1}^{L_j}\left\{{\phi^{m}}^*_{tijkr}\right\}^{n_{1mijkr}}
\left\{1-{\phi^{m}}^*_{tijkr}\right\}^{n_{2mijkr}}.
\label{eq:q1}
\end{align}
In (\ref{eq:q0}) and (\ref{eq:q1}), $n_{1mijkr}$ and $n_{2mijkr}$ denote the number of $``a"$ and $``A"$ alleles,
respectively, at the $r$-th locus of the $j$-th gene of the $i$-th individual, associated with the $m$-th mixture component.
In other words, 
%$n_{1mjr}=\sum_{i:z_{ijk}=m}\left(x^1_{ijkr}+x^2_{ijkr}\right)$ and
%$n_{2mjr}=\sum_{i:z_{ijk}=m}\left\{2-\left(x^1_{ijkr}+x^2_{ijkr}\right)\right\}$.
$n_{1mijkr}=x^1_{ijkr}+x^2_{ijkr}$ and
$n_{2mijkr}=2-\left(x^1_{ijkr}+x^2_{ijkr}\right)$.
%The function $\beta(\cdot,\cdot)$ in the above equations is the Beta function such that for any $s_1>0, s_2>0$,
%$\beta(s_1,s_2)=\frac{\Gamma(s_1)\Gamma(s_2)}{\Gamma(s_1+s_2)}$; $\Gamma(\cdot)$ being the Gamma function.

Let ${n}^*_{1tijkr}=\sum_{m:t_{mijk}=t}n_{1mijkr}$ and  ${n}^*_{2tijkr}=\sum_{m:t_{mijk}=t}n_{2mijkr}$.
Then, for $t=1,\ldots,\tau_{ijk}$; $r=1,\ldots,L_j$; $j=1,\ldots,J$ and $k=0,1$, 
update ${\bphi}^*_{tijk}$ by simulating from its full conditional distribution,
given by
\begin{equation}
[{\bphi}^*_{tijk}|\cdots]\sim [\bG_{0,jk}|\bX_{ijk}].
%\mbox{Beta}\left({n}^*_{1\ell ijkr}+\nu_{1ijkr},{n}^*_{2\ell ijkr}+\nu_{2ijkr}\right).
\label{eq:fullcond_p}
\end{equation}
The above simulation from $[{\bphi}^*_{tijk}|\cdots]$ is to be carried out by the retrospective 
method discussed in Sections \ref{subsubsec:retro2} and \ref{subsubsec:retro3}.

\subsection{{\bf Updating the missing data $\tilde{\bY}_{ijk}$}}
\label{subsec:update_missing_data}

From (\ref{eq:x_given_z}) it follows that 
\begin{equation}
[\tilde\bY_{ijk}|z_{ijk}]=\prod_{r=L_j+1}^{L}f\left(\by_{ijkr}\vert p_{z_{ijk} ijkr}\right).
\label{eq:y_given_z}
\end{equation}
Hence, given the other unknowns, $\tilde\bY_{ijk}$ can be updated by simply simulating from
the Bernoulli distributions given by (\ref{eq:y_given_z}). 

\subsection{{\bf Updating $\mu_{G}$, $\bbeta_{G}$, $\mu_{G_0}$, $\bbeta_{G_0}$, $\mu_H$
and $\bbeta_H$ using TMCMC}}
\label{subsec:update_other_parameters}

\subsubsection{{\bf Relevant factors for updating $\mu_G$ and $\bbeta_G$}}
\label{subsubsec:factors_mu_beta_G}
Let $$\mathcal L_G(\mu_G,\bbeta_G)
=\prod_{k=0}^1\prod_{i=1}^{N_k}\prod_{j=1}^J\prod_{m=2}^M[\bp_{mijk}|\bp_{lijk};l<m],$$ 
where $[\bp_{mijk}|\bp_{lijk};l<m]$ is given by (\ref{eq:polya_urn}).
%\begin{equation}
%[\bp_{mijk}|\bp_{lijk};l<m]=\frac{\alpha_{G,ik}}{\alpha_{G,ik}+m-1}\bG_{0,jk}\left(\bp_{mijk}\right)
%+\frac{1}{\alpha_{G,ik}+m-1}\sum_{l=1}^{m-1}\delta_{\bp_{lijk}}\left(\bp_{mijk}\right).
%\label{eq:polya_urn1}
%\end{equation}
Let $\pi_G(\mu_G,\bbeta_G)$ denote the prior on $(\mu_G,\bbeta_G)$. 
%We shall choose this prior to be 
%the product of two $iid$ normal distributions with mean $0$ and variance $100$.
Note that $\pi_G(\mu_G,\bbeta_G)\mathcal L_G(\mu_G,\bbeta_G)$ is the product of the only factors 
in the joint model consisting of $\mu_G$ and $\bbeta_G$.

\subsubsection{{\bf Relevant factors for updating $\mu_{G_0}$ and $\bbeta_{G_0}$}}
\label{subsubsec:factors_mu_beta_G0}
Now let $$\mathcal L_{G_0}(\mu_{G_0},\bbeta_{G_0})=
=\prod_{k=0}^1\prod_{i=1}^{N_k}\prod_{j=1}^J\prod_{t=2}^{\tau_{ijk}}[\bphi_{tijk}|\bphi_{lijk};l<t],$$ 
where $[\bphi_{tijk}|\bphi_{lijk};l<t]$ is given by (\ref{eq:polya_urn_phi}).

%\left(\alpha_{G_0,k}+n_{\cdot jk}\right)^K\prod_{l=1}^K\left(V^*_l\right)^{\alpha_{G_0,k}+n_{\cdot jk}},$$ and
Let $\pi_{G_0}(\mu_{G_0},\bbeta_{G_0})$ denote the prior on $(\mu_{G_0},\bbeta_{G_0})$. 
Then $\pi_{G_0}(\mu_{G_0},\bbeta_{G_0})\mathcal L_{G_0}(\mu_{G_0},\bbeta_{G_0})$ is the functional form
associated with $\mu_{G_0}$ and $\bbeta_{G_0}$. 
%As before, we shall
%fix the prior $\pi_{G_0}$ to be the product of two $iid$ normal distributions with mean $0$ and variance $100$.

\subsubsection{{\bf Relevant factors for updating $\mu_{H}$ and $\bbeta_{H}$}}
\label{subsubsec:factors_mu_beta_H}
Finally, we let $$\mathcal L_{H}(\mu_{H},\bbeta_{H})
=\prod_{k=0}^1\prod_{s=2}^{R_{k}}[\bxi_{sk}|\bxi_{lk};l<s],$$ 
where $[\bxi_{sk}|\bxi_{lk};l<s]$ is given y (\ref{eq:polya_urn_xi}).

%\left(\alpha_{H}+n_{\cdot k}\right)^K\prod_{l=1}^K\left(\tilde V^*_l\right)^{\alpha_{H}+n_{\cdot k}},$$ and
Let $\pi_{H}(\mu_{H},\bbeta_{H})$ be the prior on $(\mu_{H},\bbeta_{H})$. 
Then $\pi_{H}(\mu_{H},\bbeta_{H})\mathcal L_{H}(\mu_{H},\bbeta_{H})$ is the functional form
to be considered for updating $\mu_{H}$ and $\bbeta_{H}$. 
%Again, we shall
%set the prior $\pi_{H}$ to be the product of two $iid$ normal distributions with mean $0$ and variance $100$.

\subsubsection{{\bf Mixture of additive and multiplicative TMCMC for updating 
$\mu_{G}$, $\bbeta_{G}$, $\mu_{G_0}$, $\bbeta_{G_0}$, $\mu_H$
and $\bbeta_H$ in a single block}}
\label{subsubsec:additive_TMCMC}
We shall update all the parameters $\mu_{G}$, $\bbeta_{G}$, $\mu_{G_0}$, $\bbeta_{G_0}$, $\mu_H$ and 
$\bbeta_H$ using a mixture of additive and multiplicative TMCMC, where all the aforementioned
parameters are given either the additive move or the 
multiplicative move with equal probability, and where the acceptance ratio will be calculated by evaluating
the functional form
$$\pi_G(\mu_G,\bbeta_G)\mathcal L_G(\mu_G,\bbeta_G)\times 
\pi_{G_0}(\mu_{G_0},\bbeta_{G_0})\mathcal L_{G_0}(\mu_{G_0},\bbeta_{G_0})
\times \pi_{H}(\mu_{H},\bbeta_{H})\mathcal L_{H}(\mu_{H},\bbeta_{H})$$
at the numerator and the denominator corresponding to the proposed and the current values 
of $\mu_{G}$, $\bbeta_{G}$, $\mu_{G_0}$, $\bbeta_{G_0}$, $\mu_H$ and $\bbeta_H$, with all other unknowns
held fixed at their current values, multiplied by an appropriate Jacobian whenever the multiplicative
move is chosen. For details regarding mixture of additive and multiplicative
TMCMC, see \ctn{Dey14}.

\section{{\bf A parallel algorithm for implementing our MCMC procedure}}
\label{sec:computation}

Recall that the mixtures associated with gene $j\in\{1,\ldots,J\}$, and individual
$i\in\{1,\ldots,N_k\}$ and case-control status $k\in\{0,1\}$,
are conditionally independent of each other, given the interaction parameters.
This allows us to update the mixture components in separate parallel processors, conditionally on the
interaction parameters. Once the mixture components are updated, we update the interaction parameters 
using a specialized form of TMCMC, in a single processor. 
Furthermore, the parameters of the HDP are also amenable to efficient parallelization.
The details are as follows.

\begin{itemize}

%\subsection{{\bf Updating the allocation variables using Gibbs steps}}
%\label{subsec:fullcond_z}

\item[(1)] 
\begin{enumerate}
\item[(a)] In processes numbered $0$ and $1$, simultaneously obtain the set of distinct elements $\bXi_{R_k,k}$; $k=0,1$, from 
$\{\phi_{tijk};t=1,\ldots,\tau_{ijk};i=1,\ldots,N_k;j=1,\ldots,J\}$; $k=0,1$. 
\item[(b)] Communicate $\bXi_{R_k,k}$; $k=0,1$, to all the processes.
\end{enumerate}

\item[(2)] 
\begin{enumerate}
\item[(a)] In process $0$, obtain the set of distinct elements $\bzeta_S$ from $\left\{\bXi_{R_k,k}; k=0,1\right\}$.
\item[(b)] Communicate $\bzeta_S$ to all the processes.
\end{enumerate}

\item[(3)] In processes numbered $0$ and $1$, do the following in parallel for $k=0,1$:  
\begin{enumerate}
\item[(a)] Simulate, following the retrospective method detailed in Section \ref{subsubsec:retro3},\\
$\boeta^*_{lk}\stackrel{iid}{\sim}\frac{\alpha_H\tilde\bH+\sum_{s=1}^Sn_{sk}\delta_{\boeta_s}}{\alpha_H+n_{\cdot k}}$;
$l=1,2,\ldots,\mathcal L$, for sufficiently large $\mathcal L$.
%\item[(b)] Form $p^*_{lk}=V^*_{lk}\prod_{s<l}(1-V^*_{lk})$; $l=2,3,\ldots,\mathcal L$,
%with $p^*_{1k}=V^*_{1k}$, where, for $l=1,2,\ldots,\mathcal L$, 
%$V^*_{lk}\stackrel{iid}{\sim}\mbox{Beta}\left(\alpha_{H}+n_{\cdot k},1\right)$. 
\item[(b)] Communicate the simulated values to all the processes.
\end{enumerate}

\item[(3)] Split $\left\{(j,k):~j=1,\ldots,J;~k=0,1\right\}$ in the available parallel processes.
\begin{enumerate}
\item[(a)] For each $(j,k)$, simulate, following the retrospective method detailed in Section \ref{subsubsec:retro3},
$\tilde\bxi_{ljk}\stackrel{iid}{\sim}
\frac{\alpha_{G_0,k}\bH_k+\sum_{l=1}^{R_k}n_{ljk}\delta_{\bxi_{lk}}}{\alpha_{G_0,k}+n_{\cdot jk}}$;
$l=1,2,\ldots,\mathcal L$.
\item[(b)] Communicate the simulated values to all the processes.
\end{enumerate}

\item[(4)] 
%Split the triplets $\left\{(i,j,k):~i=1,\ldots,N_k;~j=1,\ldots,J;~k=0,1\right\}$ in the available
%parallel processes. For our convenience, we 
\begin{enumerate}
\item[(a)] Split the triplets $\left\{(i,j,k):~i=1,\ldots,N_k;~j=1,\ldots,J;~k=0,1\right\}$ 
in the available parallel processes sequentially into 
$$\mathcal T_1=\left\{(i,j,0):~i=1,\ldots,N_0;~j=1,\ldots,J\right\}$$ and 
$$\mathcal T_2=\left\{(i,j,1):~i=1,\ldots,N_1;~j=1,\ldots,J\right\}.$$
\item[(b)]Then parallelise updation of the mixtures associated with $\mathcal T_1$, followed by those of
$\mathcal T_2$.
\item[(c)] If, for any $(i,j,k)$, retrospective simulation from $[\bG_{0,jk}|\bX_{ijk}]$ requires more than $\mathcal L$
simulations of $\tilde\bxi_{ljk}$ in step (3) (a), then increase $\mathcal L$ to $\mathcal L^*$, and
\begin{enumerate}
\item[(i)] For $k=0,1$, augment the simulations of 
$\left\{\boeta^*_{lk}; l=1,\ldots,\mathcal L\right\}$ with new simulations 
$\left\{\boeta^*_{lk}; l=\mathcal L+1,\ldots,\mathcal L^*\right\}$.
\item[(ii)] For $j=1,\ldots,J$ and for $k=0,1$, augment the simulations of 
$\left\{\tilde\bxi_{ljk}; l=1,\ldots,\mathcal L\right\}$ with new simulations
$\left\{\tilde\bxi_{ljk}; l=\mathcal L+1,\ldots,\mathcal L^*\right\}$.
\item[(iii)] Repeat (4) (a) and (4) (b).
\end{enumerate}
%\item[(d)] Communicate the results of updation to all the processes.
\end{enumerate}

\item[(5)] During each MCMC iteration, for each $(i,j,k)$ 
in each available parallel processor, update the allocation variables $z_{ijk}$, the proportions $\bp_{mijk}$; $m=1,\ldots,M$, 
and the missing data $\tilde\bY_{ijk}$, using the methods proposed in Sections \ref{subsec:update_z_p}
and \ref{subsec:update_missing_data}. 

\item[(6)] Communicate the results of updating in (4) and (5) to all the processes.

%Perform the necessary book-keeping detailed in  
%Section \ref{subsubsec:book_keeping}, and communicate the results to a the processor where the parameters
%$\mu_{G}$, $\bbeta_{G}$, $\mu_{G_0}$, $\bbeta_{G_0}$, $\mu_H$ and $\bbeta_H$ will be updated.

\item[(7)] 
\begin{enumerate}
\item[(a)]During each MCMC iteration, update the parameters 
$\mu_{G}$, $\bbeta_{G}$, $\mu_{G_0}$, $\bbeta_{G_0}$, $\mu_H$ and 
$\bbeta_H$ using additive TMCMC in a single block, as proposed in Section \ref{subsec:update_other_parameters},
in process number $0$.
\item[(b)] Communicate the updated results to all the processes.
\end{enumerate}
\end{itemize}

\section{{\bf Simulation studies}}
\label{sec:simstudy}

For simulation studies, we first generate realistic biological data for stratified population with known
gene-environment interaction from the GENS2 software of \ctn{Pinelli12}.  
To this data, we then apply our model and methodologies in an effort to detect gene-environment
interaction effects that are present in the data. We consider simulation studies 
in $5$ different true model set-ups: (a) presence of gene-gene and gene-environment interaction,
(b) absence of genetic or gene-environmental interaction effect,
(c) absence of genetic and gene-gene interaction effects but presence of environmental effect,
(d) presence of genetic and gene-gene interaction effects but absence of environmental effect,
and 
(e) independent and additive genetic and environmental effects.

\begin{figure}
\centering
\subfigure []{ \label{fig:exp1}
\includegraphics[width=6cm,height=5cm]{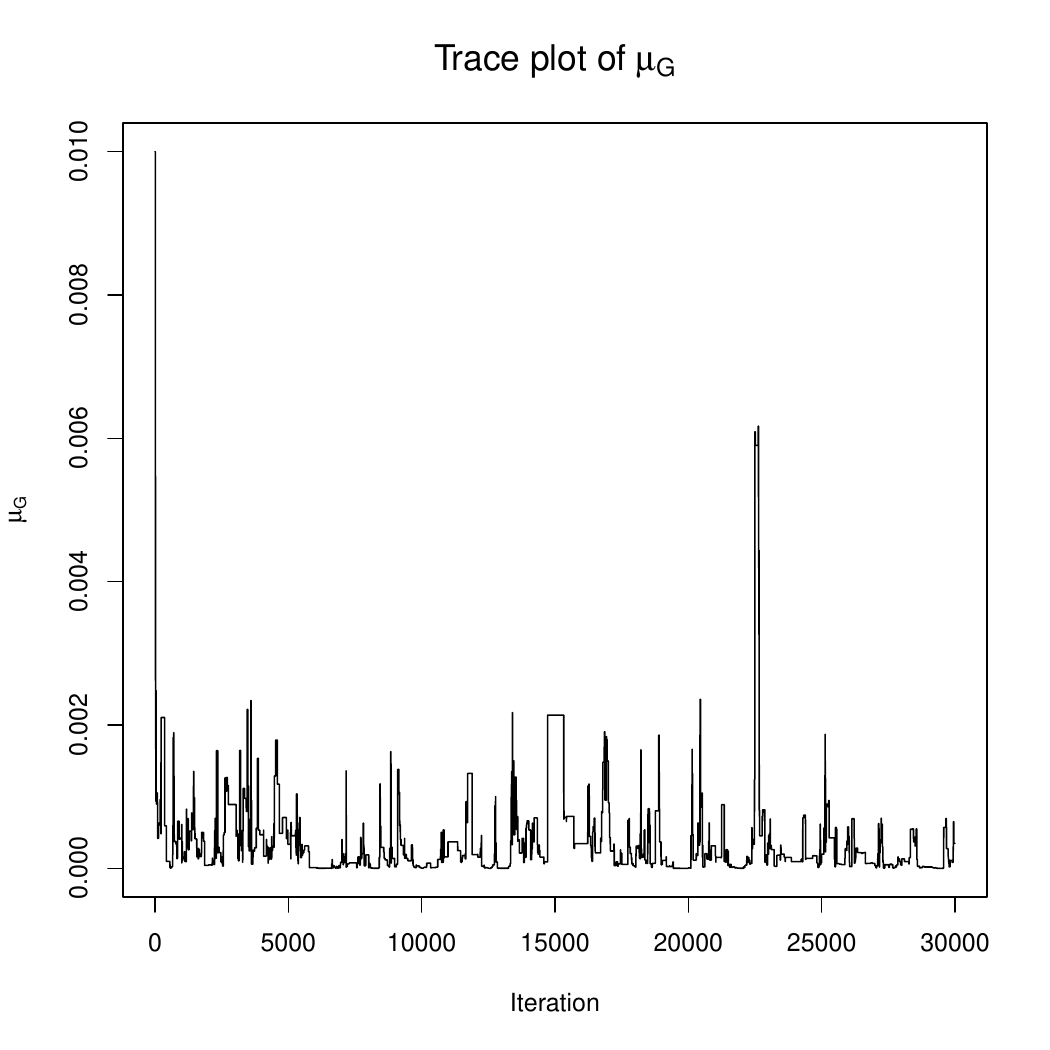}}
\hspace{2mm}
\subfigure []{ \label{fig:exp2}
\includegraphics[width=6cm,height=5cm]{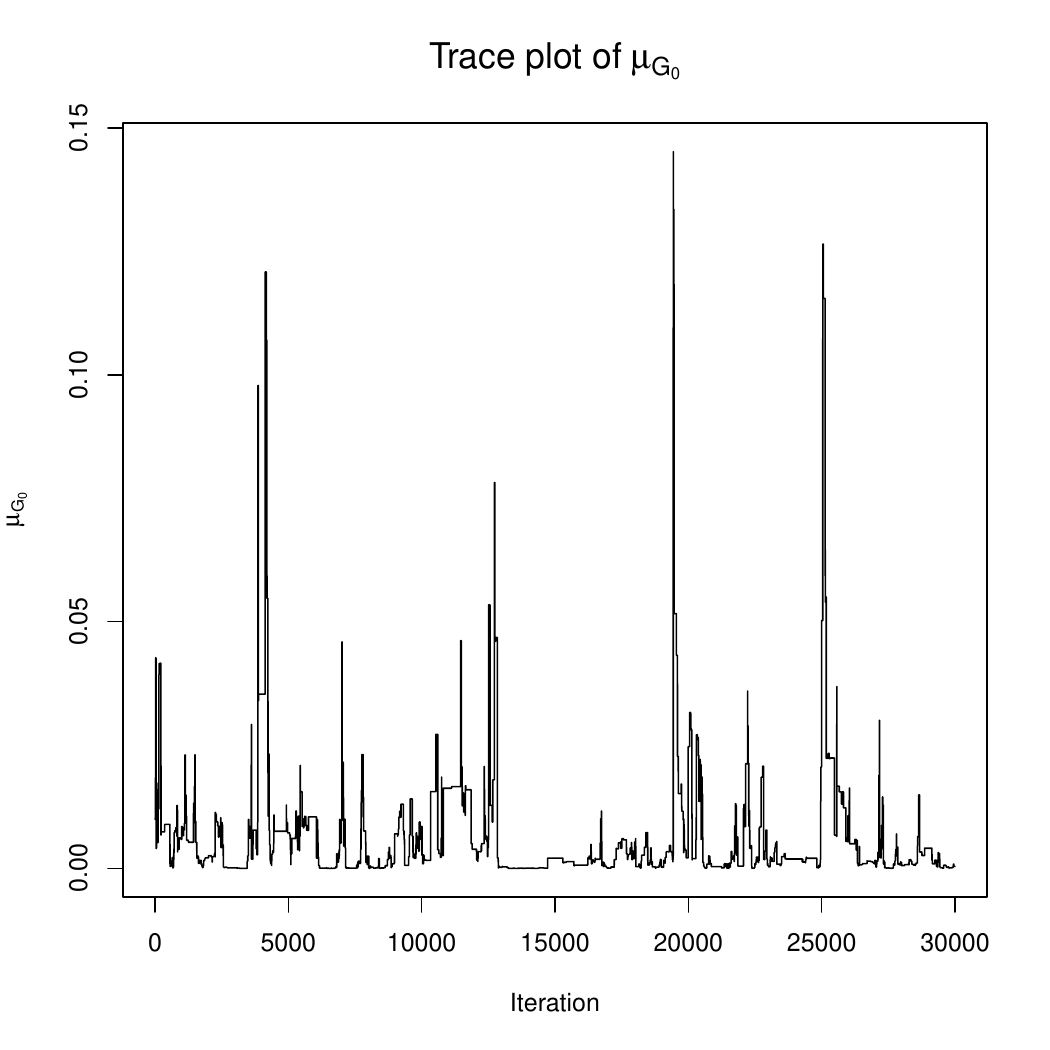}}\\
\vspace{2mm}
\subfigure []{ \label{fig:exp3}
\includegraphics[width=6cm,height=5cm]{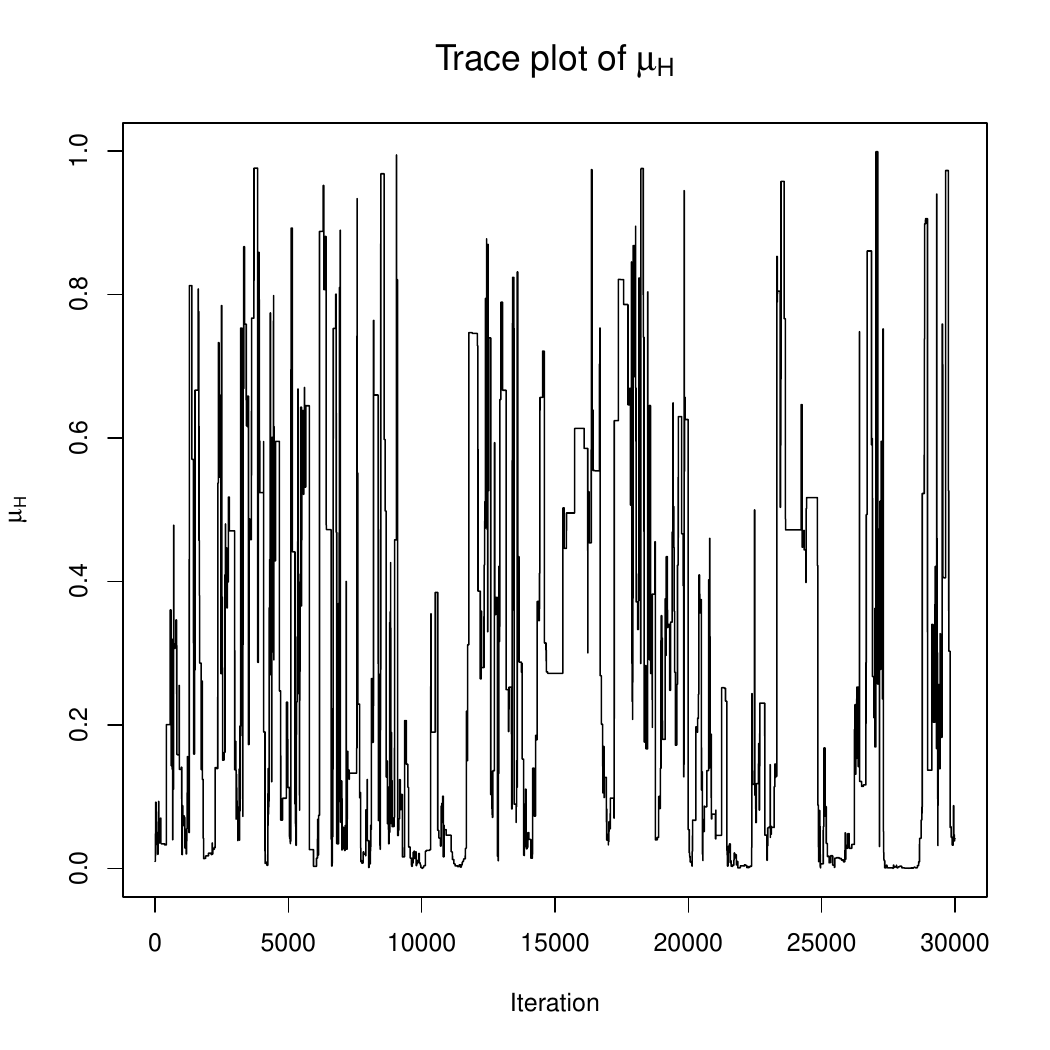}}
\hspace{2mm}
\subfigure []{ \label{fig:exp4}
\includegraphics[width=6cm,height=5cm]{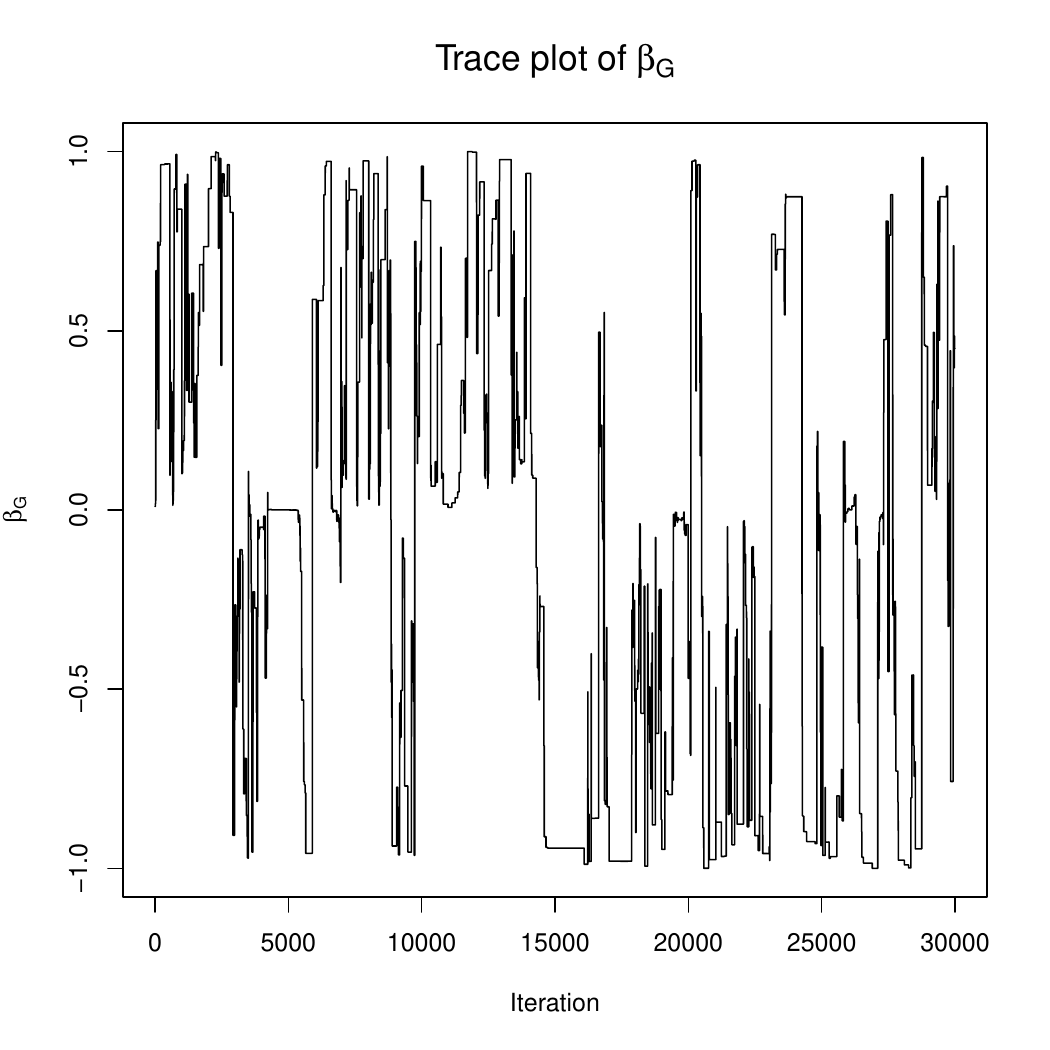}}\\
\vspace{2mm}
\subfigure []{ \label{fig:exp5}
\includegraphics[width=6cm,height=5cm]{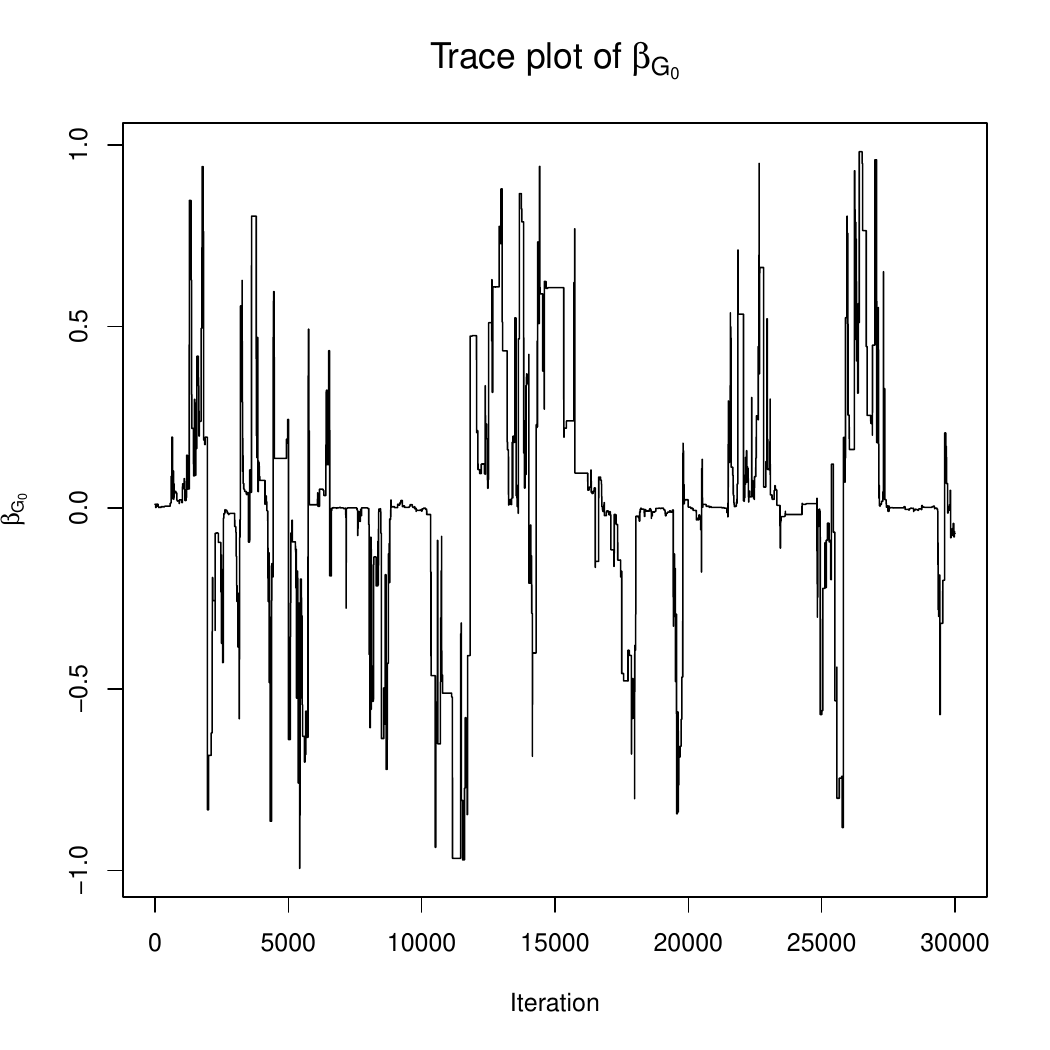}}
\hspace{2mm}
\subfigure []{ \label{fig:exp6}
\includegraphics[width=6cm,height=5cm]{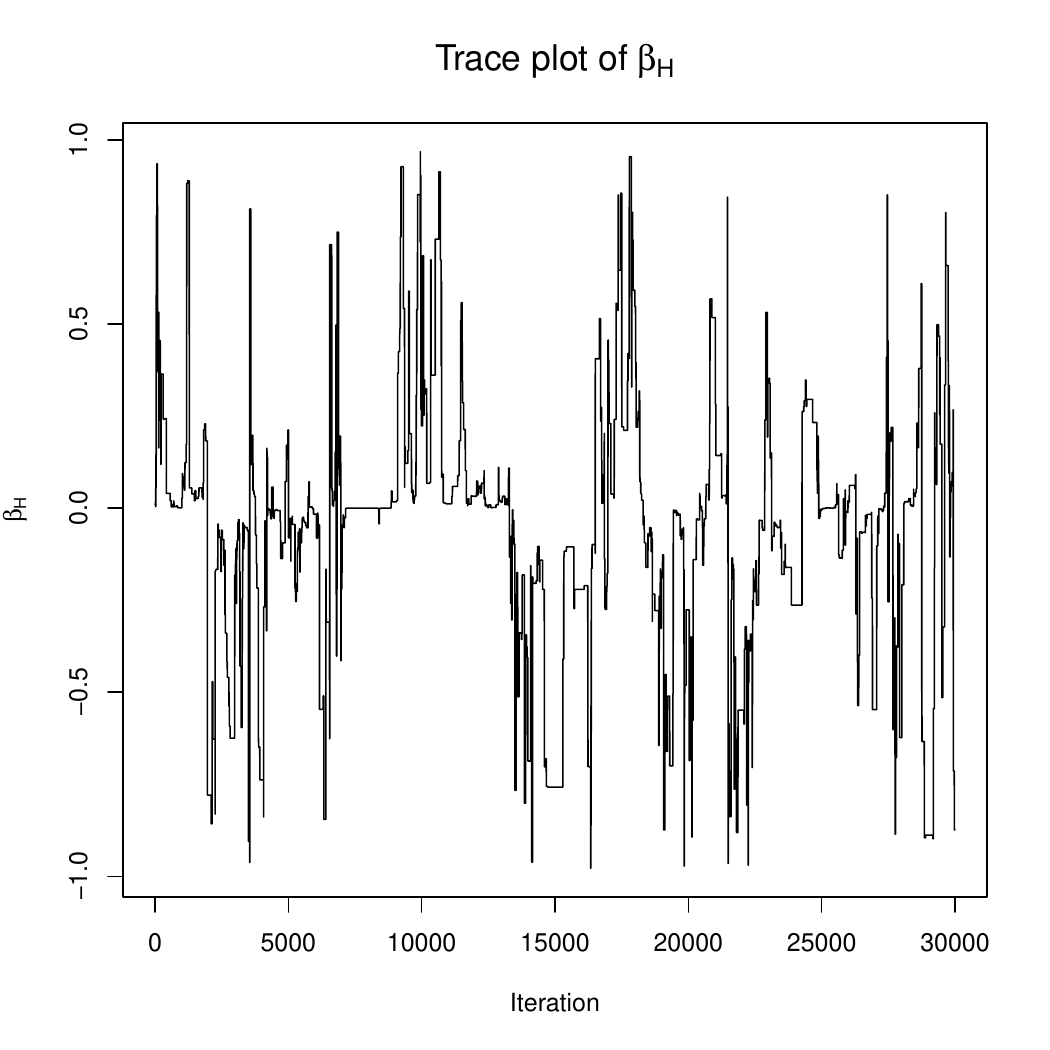}}
\caption{Trace plots for the HDP model.}
\label{fig:fig1}
\end{figure}

As we demonstrate, our model and methodologies successfully identify the effects of the
individual genes, gene-gene and gene-environment interactions, and the number of sub-populations. 
In all our applications, we set $M=30$, $\nu_1=\nu_2=1$, so that $\tilde\bH$ is the uniform distribution on $[0,1]$.
We set $\alpha_{G,ik}=0.1\times\exp\left(100+\mu_G+\beta_G E_{ik}\right)$, 
$\alpha_{G_0,k}=0.1\times\exp\left(100+\mu_{G_0}+\beta_{G_0} \bar E_{k}\right)$ and
$\alpha_H=0.1\times\exp\left(100+\mu_H+\beta_H \bar{\bar E}\right)$, where we assumed
$\mu_G,\mu_{G_0},\mu_H\stackrel{iid}{\sim}U(0,1)$ and $\beta_G,\beta_{G_0},\beta_H\stackrel{iid}{\sim}U(-1,1)$.
This structure ensured adequate number of sub-populations and reasonable mixing of MCMC.
Figure \ref{fig:fig1} shows some typical trace plots associated with
our HDP model. Although the mixing is not excellent, there is no evidence of non-convergence.
%In a second simulation study we generate a dataset with stratified population structure 
%from GENS2 where there is no gene-gene or gene-environment
%interaction, and then fit our nonparametric gene-gene and gene-environment interaction model to it.
%Once again we demonstrate the ability of our model and methods to correctly identify the number
%of sub-populations and the absence of the
%interactions.

\subsection{{\bf First simulation study: presence of gene-gene and gene-environment interaction}}
\label{subsec:first_simulation_study}

\subsubsection{{\bf Data description}}
\label{subsubsec:data_description}
As in \ctn{Bhattacharya16} we consider two genetic factors as allowed by GENS2
and simulated 5 data sets with gene-gene and gene-environment interaction with a one-dimensional environmental
variable, associated with 5 sub-populations. 
%The data sets consist of disease status, gender, environmental exposures and genotypes for each individual.
One of the genes consists of 1084 SNPs and another has 1206 SNPs, with 
one disease pre-disposing locus (DPL) at each gene. There are 113 individuals in each of the 5 data sets, from which
we selected a total of 100 individuals without replacement with probabilities 
assigned to the 5 data sets being $(0.1, 0.4, 0.2, 0.15, 0.15)$. 
%That is, we chose one of the 
%5 data sets with these probabilities and selected a row randomly from the chosen data set; we repeated
%this procedure 100 times without replacing the rows. In our final data set thus obtained, there
Our final dataset consists of 46 cases and 54 controls.
Since, in our case, the environmental variable is one-dimensional, $d=1$.

\subsubsection{{\bf Model implementation}}
\label{subsubsec:model_implementation}
We implemented our parallel MCMC algorithm on 50 cores in a 64-bit VMware with 64-bit physical cores, each
running at 2793.269 MHz.
Our code is written in C in conjunction with the Message Passing Interface (MPI) protocol for parallelisation.

The total time taken to implement $30,000$ MCMC iterations, where the first $10,000$ are discarded as burn-in,
is approximately 20 hours. We assessed convergence informally with trace plots, which indicated 
adequate mixing properties of our algorithm.

\subsubsection{{\bf Specifications of the thresholds $\varepsilon$'s using null distributions}}
\label{subsubsec:threshold}

Following the method outlined in Section \ref{subsubsec:null_model_e_choice} and setting $M$ to be 30, we obtain
$\varepsilon_{d^*}=0.200$, $\varepsilon_{\hat d_1}=0.167$, $\varepsilon_{\hat d_2}=0.167$,
$\varepsilon_{d^*_E}=0.250$, $\varepsilon_{d^*_{E,1}}=0.185$, $\varepsilon_{d^*_{E,2}}=0.173$,
$\varepsilon_{\beta_{G}}=0.874$, $\varepsilon_{\beta_{G_0}}=0.128$, $\varepsilon_{\beta_{H}}=0.219$.

\subsubsection{{\bf Results of fitting our model}}
\label{subsubsec:results_first_simulation_study}

%Figure \ref{fig:ggi_metric_plots} displays the posterior distributions of 
%$d^*=\underset{j=1,2}{\max}~\hat d\left(\bP_{30,j,0},\bP_{30,j,1}\right)$, 
%$\hat d_1=\hat d\left(\bP_{30,1,0},\bP_{30,1,1}\right)$ and 
%$\hat d_2=\hat d\left(\bP_{30,2,0},\bP_{30,2,1}\right)$, respectively.
%The diagrams show that in all the three cases, regions that are significantly bounded away from zero
%have high posterior probabilities compared to those closer to zero.
%%the region to the right of $0.2$ has significantly higher posterior 
%%probability compared to the left of $0.2$. 
%%, while the maximum value is around $0.6$. 
%For the purpose of formal Bayesian
%hypothesis, following the discussion in Section \ref{subsubsec:threshold}, we
%set $\varepsilon=0.233$. 
%%to be $1/3$ of the maximum value $0.6$, that is,
%%we specify $\varepsilon=0.2$. 

The posterior probabilities
$P\left(d^*<\varepsilon_{d^*}|\mbox{Data}\right)$, $P\left(\hat d_1<\varepsilon_{\hat d_1}|\mbox{Data}\right)$
and $P\left(\hat d_2<\varepsilon_{\hat d_2}|\mbox{Data}\right)$
empirically obtained from $20,000$ MCMC samples,
turned out to be $0.378$, $0.317$ and $0.324$, respectively. 
Hence, $H_{0,d^*}$, $H_{0,\hat d_1}$ and $H_{0,\hat d_2}$ are rejected, suggesting
the influence of significant genetic effects in the case-control study. 

However, 
$P\left(d^*_E<\varepsilon_{d^*_E}|\mbox{Data}\right)$, $P\left(\hat d_{E,1}<\varepsilon_{\hat d_{E,1}}|\mbox{Data}\right)$
and $P\left(\hat d_{E,2}<\varepsilon_{\hat d_{E,2}}|\mbox{Data}\right)$ are given, approximately, by
$0.558$, $0.561$ and $0.550$, respectively, which seem to contradict the results of the clustering based hypothesis tests.
This can be explained as follows. Since $\bG_{0,jk}$ are discrete, the parameters $\bp_{mijk}$, even if
generated from $\bG_{0,jk}$, coincide with positive probability, so that the effective dimensionalities of 
$\mbox{logit}\left(\bar P_{Mi_0jk=0}\right)$ and $\mbox{logit}\left(\bar P_{Mi_1jk=1}\right)$ are drastically
reduced, so that the Euclidean distance between these two vectors is substantially small. As such, the Euclidean 
distance fails to reject the null even if it is false. As noted in \ctn{Bhattacharya16}, even the clustering metric in this
scenario is not completely satisfactory since this involves clustering distance between two empirically obtained 
central clusterings which may not be very accurate unless the sample sizes for case and control are very large.
However, compared to the Euclidean distance, the clustering metric turns out to be far more reliable.

\begin{figure}%[htp]
\centering
\subfigure[Posterior probability of no gene-gene interactions in control subjects.]{ \label{fig:control_ggi}
\includegraphics[width=15cm,height=8cm]{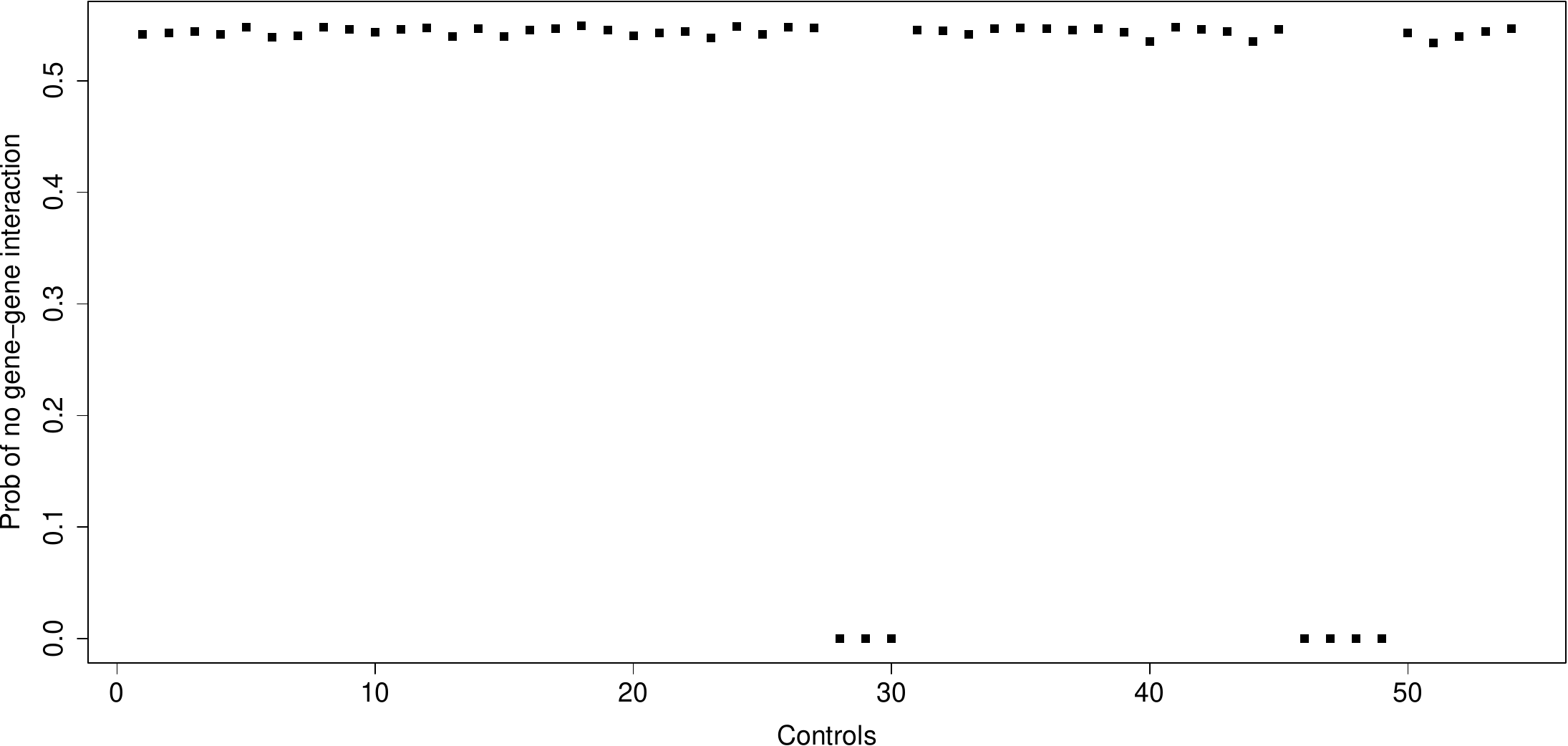}}\\
\vspace{4mm}
\subfigure[Posterior probability of no genetic effect with respect to cases.]{ \label{fig:case_ggi}
\includegraphics[width=15cm,height=8cm]{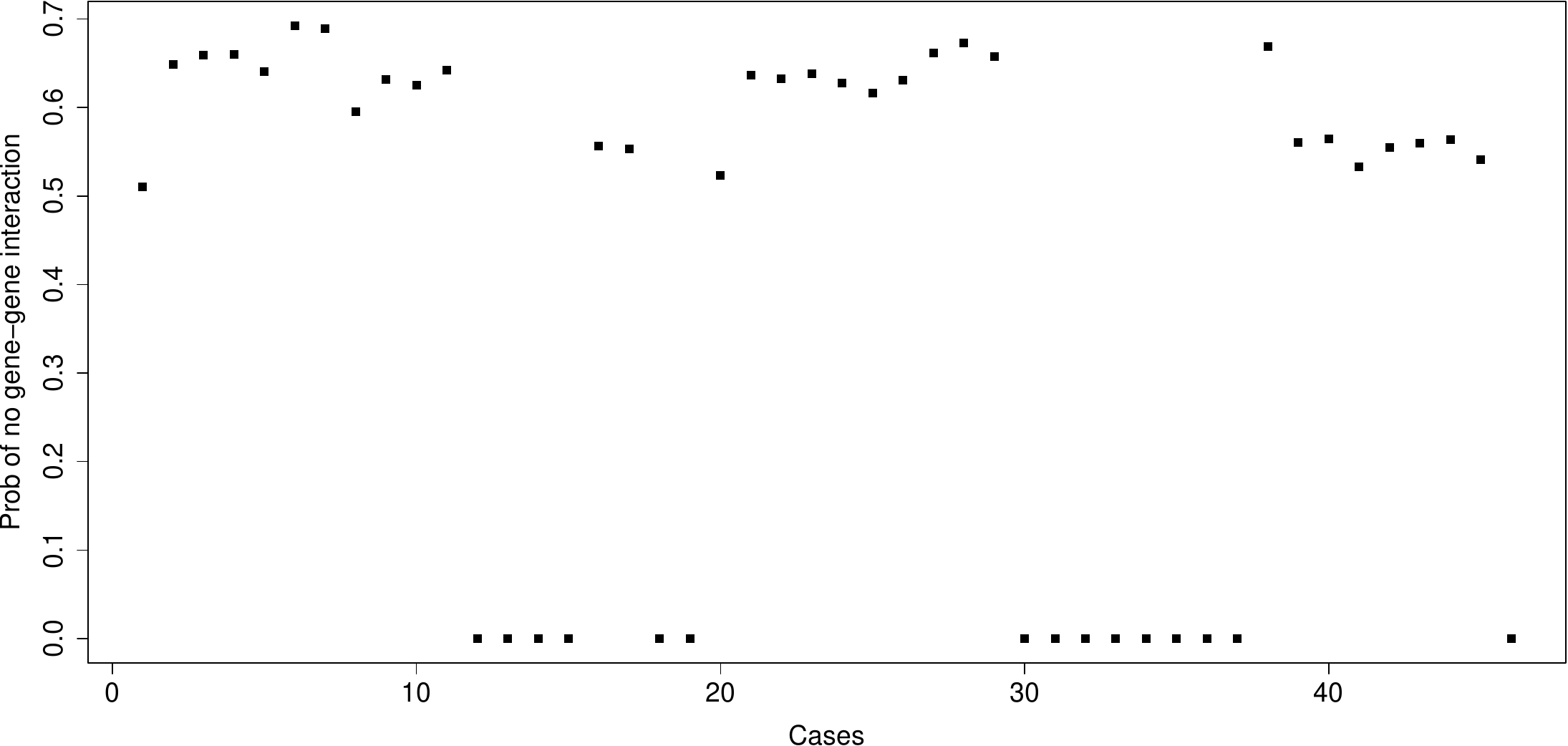}}
\caption{{\bf Presence of gene-gene and gene-environment interaction:}
Index plots of the posterior probabilities of no gene-gene interactions in (a) controls
and (b) cases, with respect to the two genes.}
\label{fig:prob_no_ggi_1}
\end{figure}

To check the influence of the environmental 
variable on the genes we compute the posterior probabilities 
$P\left(|\beta_{G}|<\varepsilon_{\beta_{G}}|\mbox{Data}\right)$, 
$P\left(|\beta_{G_0}|<\varepsilon_{\beta_{G_0}}|\mbox{Data}\right)$
and $P\left(|\beta_{H}|<\varepsilon_{\beta_{H}}|\mbox{Data}\right)$.
The probabilities turned out to be $0.544$, $0.550$ and $0.191$, respectively, showing that
$\beta_{H}$ is very significant. That is, the environmental variable has a significant overall effect on the genes. 
%Now if gene-gene interaction is found to be significant, then the interaction of the environment and gene 1 
%would seem to have affected gene 2 as well, so that both $H_{0,\hat d_1}$ and $H_{0,\hat d_2}$ are rejected. 
%Hence, we now investigate the significance of gene-gene interaction.
Figure \ref{fig:prob_no_ggi_1} depicts the posterior probabilities of no gene-gene interactions for
the controls and cases, showing the prominence of several gene-gene interactions in both control and case groups. 
As to be expected, in the case group, more instances of gene-gene interactions 
turned out to be significant compared to the control group.

\begin{figure}%[htp]
\centering
\subfigure[Posterior of $\tau_{110}$.]{ \label{fig:ggi_comp_prob1}
\includegraphics[width=7cm,height=7cm]{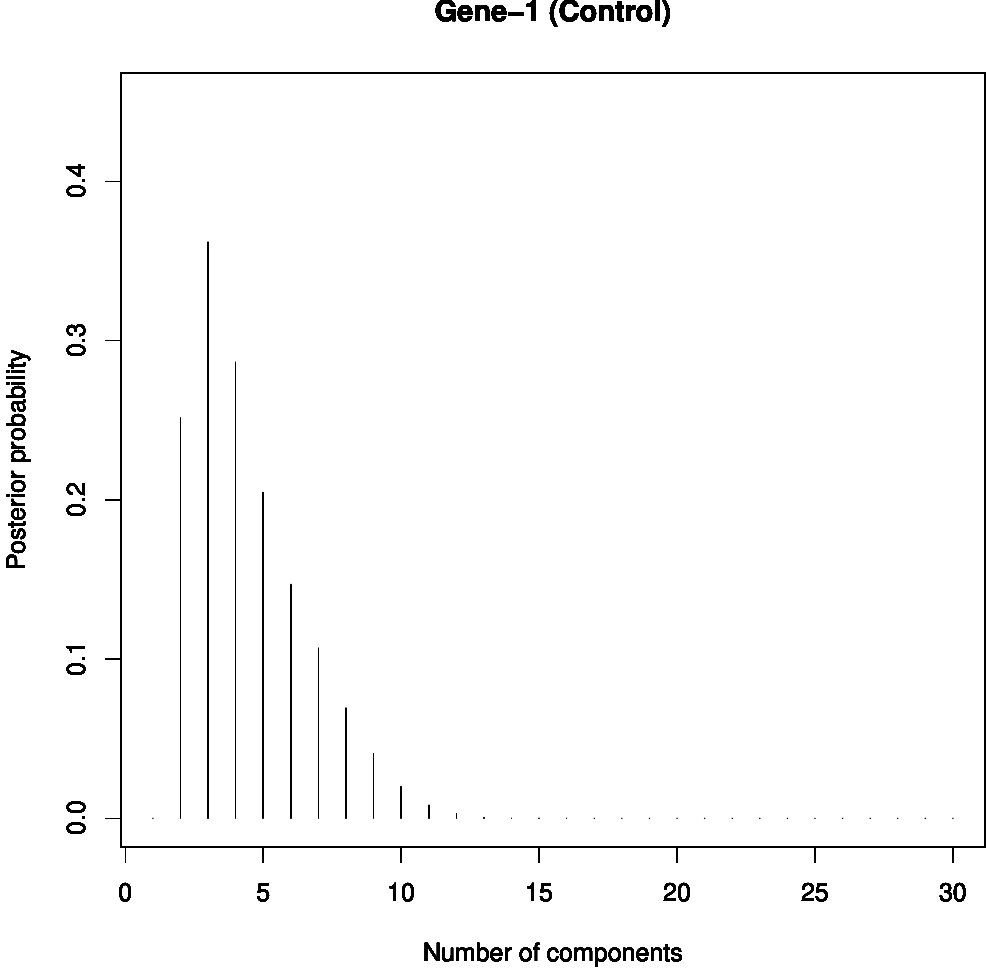}}
\hspace{2mm}
\subfigure[Posterior of $\tau_{211}$.]{ \label{fig:ggi_comp_prob2}
\includegraphics[width=7cm,height=7cm]{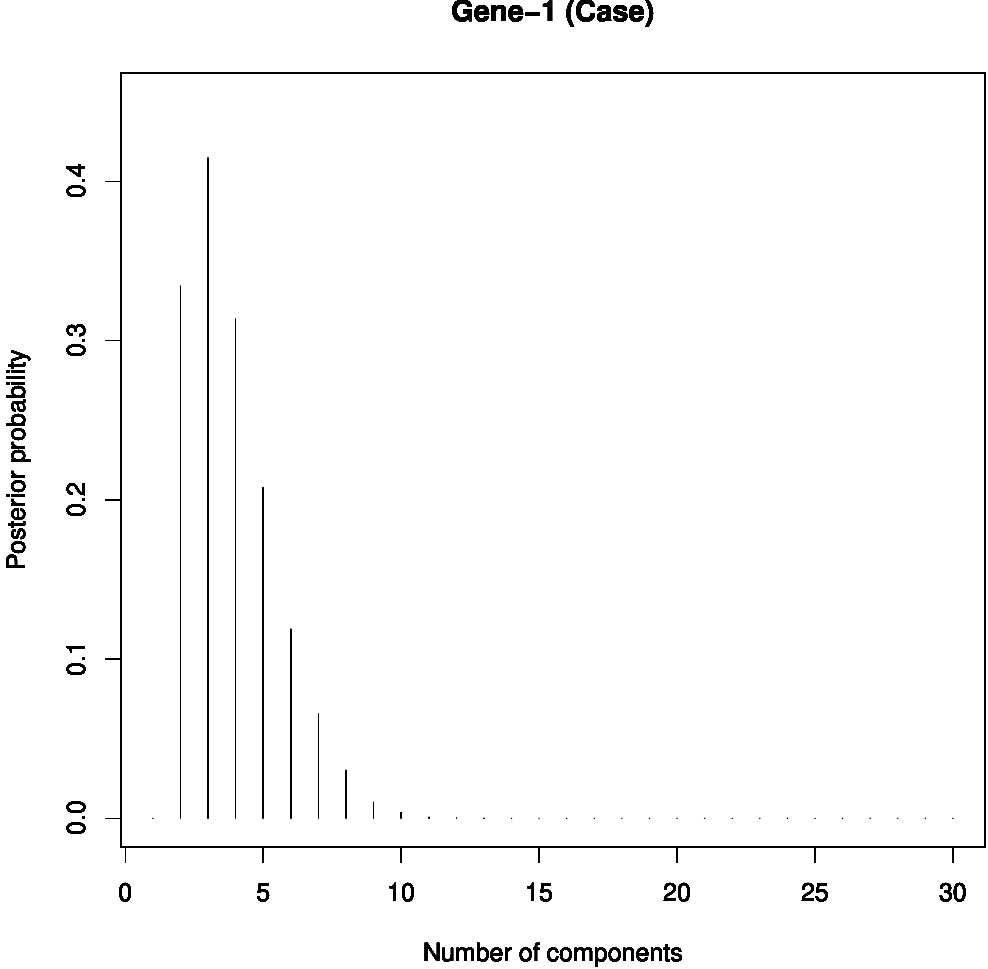}}
\hspace{2mm}
\subfigure[Posterior of $\tau_{120}$.]{ \label{fig:ggi_comp_prob3}
\includegraphics[width=7cm,height=7cm]{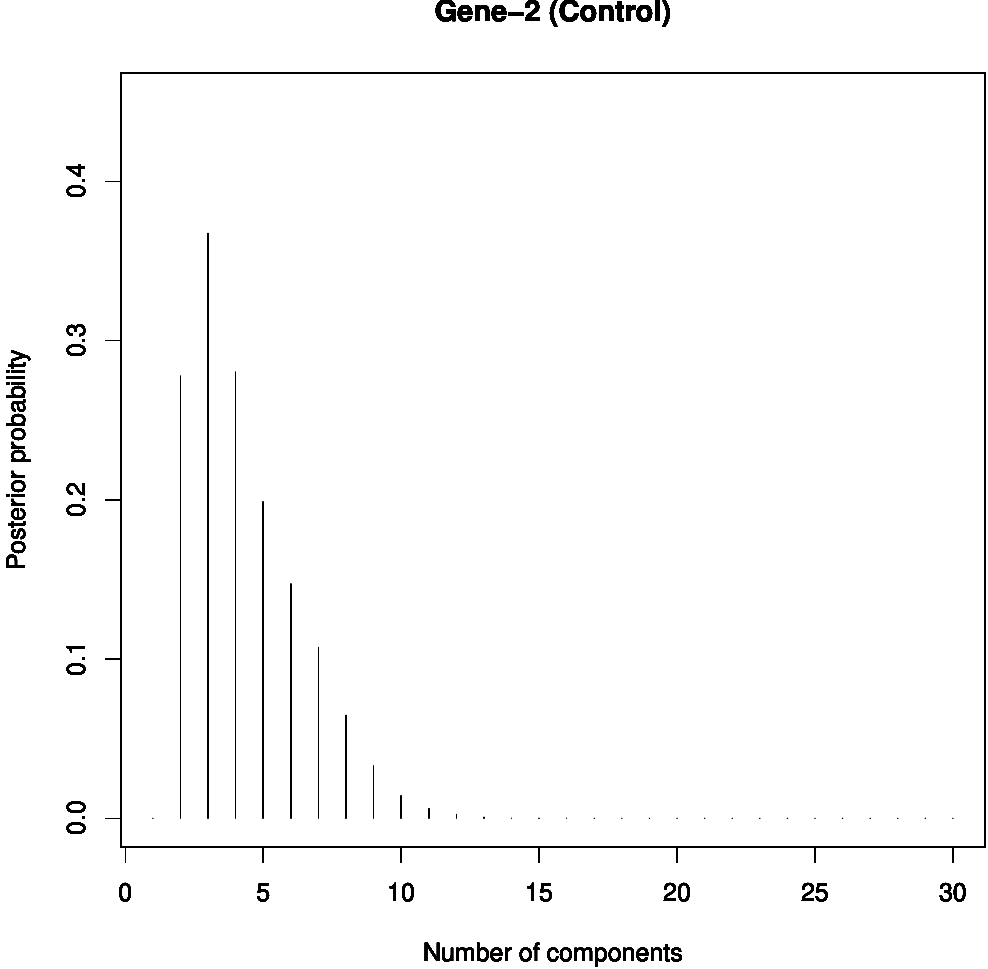}}
\hspace{2mm}
\subfigure[Posterior of $\tau_{221}$.]{ \label{fig:ggi_comp_prob4}
\includegraphics[width=7cm,height=7cm]{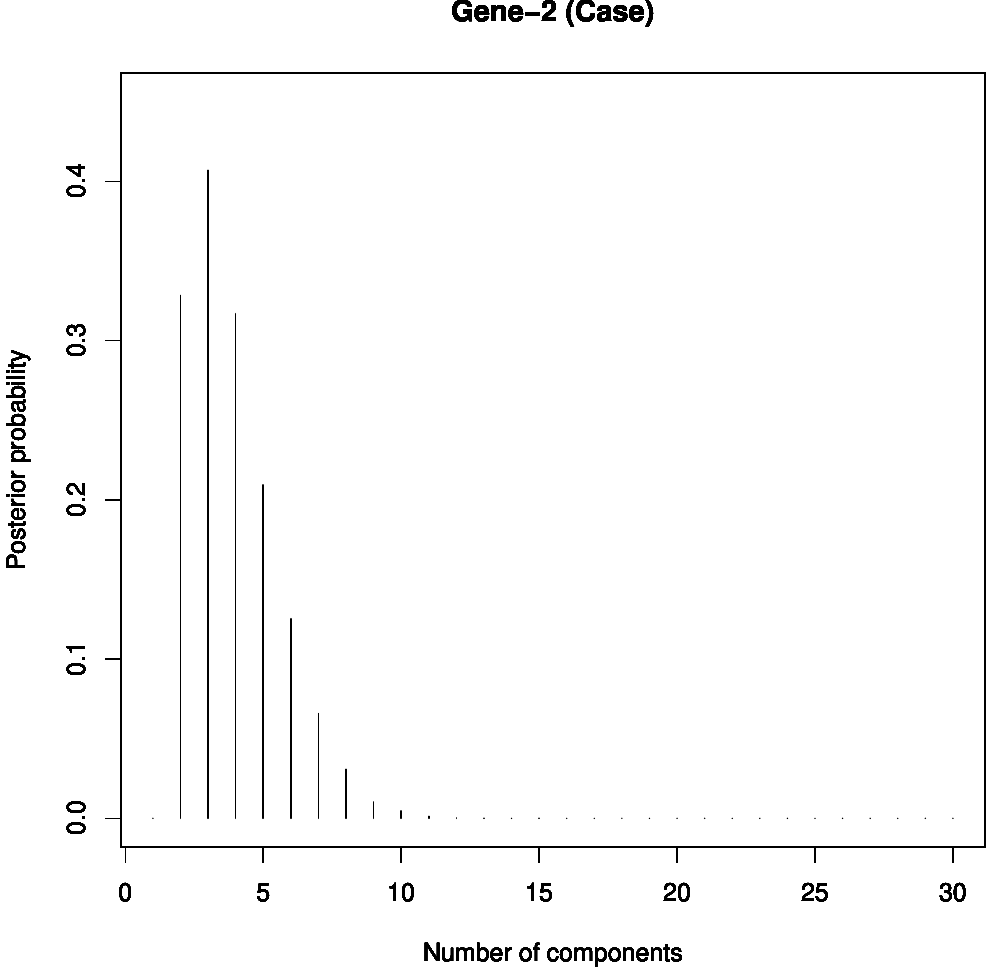}}
\caption{{\bf Gene-gene and gene-environment interaction:} Posterior distributions of the number of sub-populations.}
%for each pair ($j,k$); $j=1,2$; $k=0,1$.}
\label{fig:ggi_comp}
\end{figure}

The posteriors of the number of sub-populations, some of which are shown in Figure \ref{fig:ggi_comp}, give high probabilities
to $5$, the true number of sub-populations.
%The posteriors also closely match those associated with the null model, as depicted in Figure \ref{fig:ggnull_comp}.

%\begin{figure}%[htp]
%\centering
%\subfigure[Null posterior of $\tau_{110}$.]{ \label{fig:ggnull_comp_prob1}
%\includegraphics[width=7cm,height=7cm]{plots/nullcomp_prob1-crop.pdf}}
%\hspace{2mm}
%\subfigure[Null posterior of $\tau_{211}$.]{ \label{fig:ggnull_comp_prob2}
%\includegraphics[width=7cm,height=7cm]{plots/nullcomp_prob2-crop.pdf}}
%\hspace{2mm}
%\subfigure[Null posterior of $\tau_{120}$.]{ \label{fig:ggnull_comp_prob3}
%\includegraphics[width=7cm,height=7cm]{plots/nullcomp_prob3-crop.pdf}}
%\hspace{2mm}
%\subfigure[Null posterior of $\tau_{221}$.]{ \label{fig:ggnull_comp_prob4}
%\includegraphics[width=7cm,height=7cm]{plots/nullcomp_prob4-crop.pdf}}
%\caption{{\bf Gene-Gene and gene-environment Interaction:} Posterior distributions of the number of distinct components
%associated with the null model.}
%%for each pair ($j,k$); $j=1,2$; $k=0,1$.}
%\label{fig:ggnull_comp}
%\end{figure}

\subsubsection{{\bf Detection of DPL}}
\label{subsubsec:dpl}

The correct positions of the DPL, provided by GENS2, are $rs13266634$ and $rs7903146$, 
for the first and second gene respectively. 
Due to the LD effects implied by the highly correlated structure of our 
current HDP based model, the actual DPL are difficult to locate. Notably, our model is considerably more structured 
than those of \ctn{Bhattacharya15} and \ctn{Bhattacharya16}, and any inappropriate dependence structure would
render the task of DPL finding far more difficult than our previous models. Nevertheless, we demonstrate
that our HDP model can detect DPLs with more precision compared to our previous matrix-normal-inverse-Wishart
model for gene-environment interactions.

%and hence, unlike those works, 
%not much success is expected of our model regarding accurately locating the DPLs.

Following \ctn{Bhattacharya15} and \ctn{Bhattacharya16}, and writing $\bp^r_{ijk}=\left\{p_{mijkr}:m=1,\ldots,M\right\}$, we  
declare the $r$-th locus of the $j$-th gene as disease pre-disposing if, for the $r$-th locus, the Euclidean distance
$d^r_j\left(\mbox{logit}\left(\bp^r_{i_0jk=0}\right),\mbox{logit}\left(\bp^r_{i_1jk=1}\right)\right)$, 
between $\mbox{logit}\left(\bp^r_{i_0jk=0}\right)$ and $\mbox{logit}\left(\bp^r_{i_1jk=1}\right)$, is significantly larger than 
$d^{r'}_j\left(\bp^{r'}_{i_0jk=0},\bp^{r'}_{i_1jk=1}\right)$, for $r'\neq r$.  
We adopt the graphical method as in our previous works.
\begin{figure}%[htp]
\centering
\subfigure[Index plot for the first gene]{ \label{fig:index_plot_gene1}
\includegraphics[width=8cm,height=8cm]{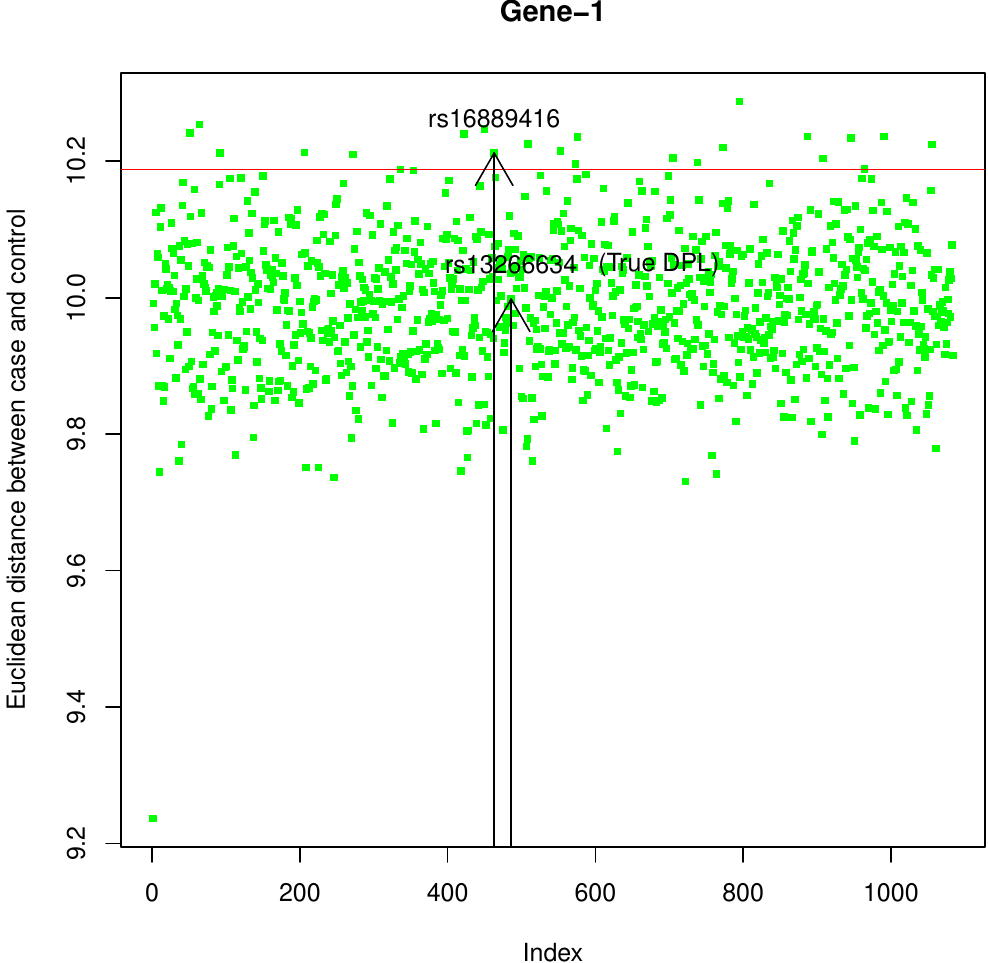}}
%\vspace{2mm}
\subfigure[Index plot for the second gene.]{ \label{fig:index_plot_gene2}
\includegraphics[width=8cm,height=8cm]{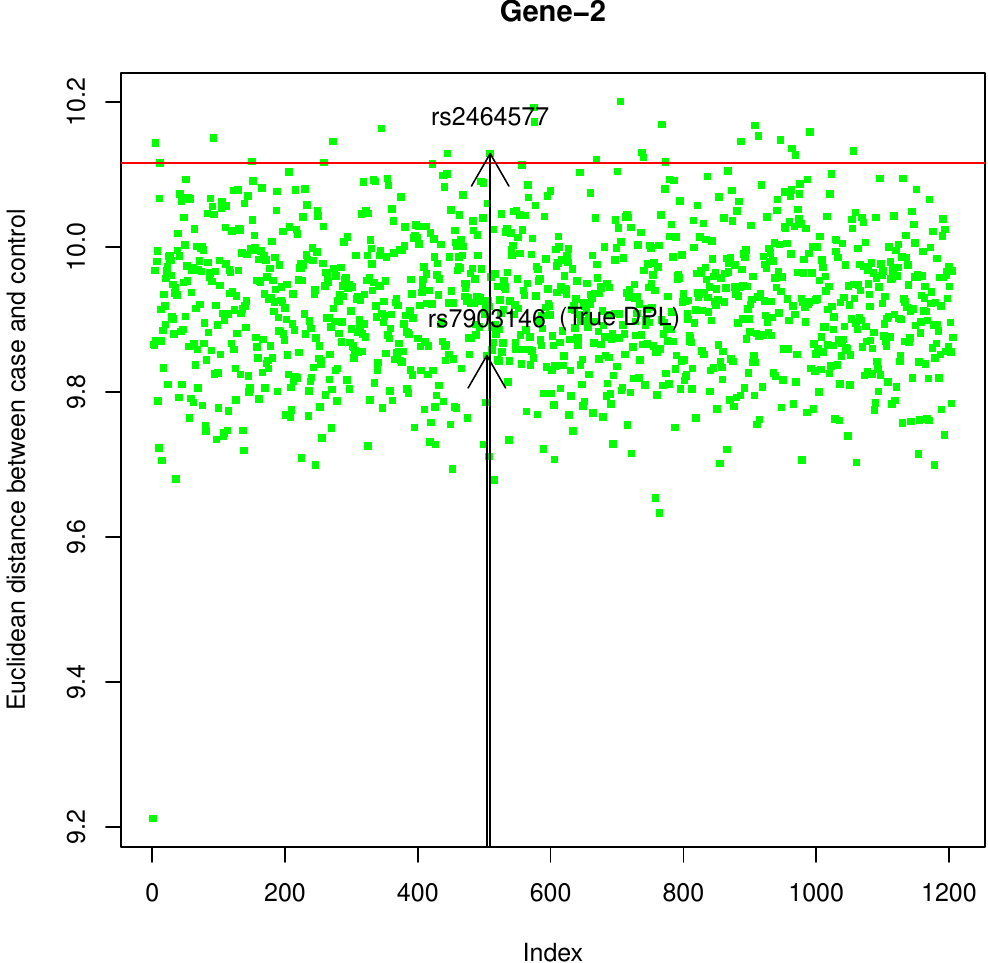}}
\caption{{\bf Presence of gene-gene and gene-environment interaction:} Plots of the Euclidean distances 
$\left\{d^r_j\left(\mbox{logit}\left(\bp^r_{i_0jk=0}\right),\mbox{logit}\left(\bp^r_{i_1jk=1}\right)\right);
~r=1,\ldots,L_j\right\}$
against the indices of the loci, for $j=1$ (panel (a)) and $j=2$ (panel (b)).}
\label{fig:index_plots}
\end{figure}
The red, horizontal lines in the panels of Figure \ref{fig:index_plots} 
represent the cut-off value such that the points above the
horizontal line are those with the highest $2\%$ Euclidean distances. 
The actual DPLs of the two genes, as well as their nearest neighbours with Euclidean distances
on or above the red, horizontal lines, are shown in the figures. That even such small sets of
SNPs with highest 2\% Euclidean distances consist of close neighbours of the true DPLs, is quite encouraging.
Observe that the DPL detection is more precise for the second gene in the sense that the closest neighbour
of the actual DPL above the red, horizontal line is closer to the true DPL than for the first gene.

The above results on DPL detection is also a significant improvement over \ctn{Bhattacharya16} 
where highest 10\% Euclidean distances
were considered, suggesting that our current HDP based model is more appropriate compared to our
previous matrix-normal-inverse-Wishart model for gene-environment interaction. 

%Although
%in this case, the DPLs were not detected with appreciable precision, we shall demonstrate with subsequent
%simulation studies that greater precision can be achieved even with our HDP-based model when the data arises
%from models with simpler dependence structures which does not involve environmental effects on the genes.

\subsection{{\bf Second simulation study: no genetic or environmental effect}}
\label{subsec:second_simulation_study}

Here we use the same case-control genotype data set as used by \ctn{Bhattacharya15} in their second simulation study
where genetic effects are absent, 
consisting of 49 cases and 51 controls and 5 sub-populations with the mixing proportions
$(0.1, 0.4, 0.2, 0.15, 0.15)$. We use the same environmental data set generated in our first simulation study described in
Section \ref{subsec:first_simulation_study}, which is unrelated to this genotype data.

Here we obtain $P\left(d^*<\varepsilon_{d^*}|\mbox{Data}\right)\approx 0.407$. Although this does not cross the
$0.5$ benchmark, there is significant evidence in favour of the null, and falling short of $0.5$ can be attributed
to the slight deficiency of the distance between the two approximate central clusterings associated with case and control,
as already discussed in the context of the first simulation study. 

Also, in this study, 
$P\left(|\beta_{G}|<\varepsilon_{\beta_{G}}|\mbox{Data}\right)$, 
$P\left(|\beta_{G_0}|<\varepsilon_{\beta_{G_0}}|\mbox{Data}\right)$
and $P\left(|\beta_{H}|<\varepsilon_{\beta_{H}}|\mbox{Data}\right)$ 
are given by $0.549$, $0.550$ and $0.649$, respectively, suggesting insignificance of the effect of the environmental 
variable on gene-gene interaction. 
As noted in \ctn{Bhattacharya16}, however, it is not straightforward to test whether or not
the environment is responsible for the case-control status. This is because we have modeled the genotype data conditionally
on case-control instead of modeling the case-control status conditionally on the environmental variable. 
\ctn{Bhattacharya16} use significance testing in a simple logistic regression framework to show insignificance 
of the environmental variable. 

As before, our model assigned high posterior probability to 5 sub-populations.

Note that since there is no genetic effect in this study, the question of detecting DPLs does not arise here.
%Figure \ref{fig:index_plots_nge} shows the plots of Euclidean distances between cases and controls for the loci
%of the two genes. Although for Gene-1 the Euclidean distance for the actual disease predisposing locus falls on the cut-off line,
%for the second gene it falls much short of the red, horizontal line. This can be explained as follows.  
%\begin{figure}%[htp]
%\centering
%\subfigure[Index plot for the first gene]{ \label{fig:index_plot_gene1_nge}
%\includegraphics[width=8cm,height=8cm]{plots/plot_loci_gene1_nge-crop.pdf}}
%%\vspace{2mm}
%\subfigure[Index plot for the second gene.]{ \label{fig:index_plot_gene2_nge}
%\includegraphics[width=8cm,height=8cm]{plots/plot_loci_gene2_nge-crop.pdf}}
%\caption{{\bf No genetic or environmental effect:} Plots of the Euclidean distances 
%$\left\{d^r_j\left(\mbox{logit}\left(\bp^r_{i_0jk=0}\right),\mbox{logit}\left(\bp^r_{i_1jk=1}\right)\right);
%~r=1,\ldots,L_j\right\}$
%against the indices of the loci, for $j=1$ (panel (a)) and $j=2$ (panel (b)).}
%\label{fig:index_plots_nge}
%\end{figure}

\subsection{{\bf Third simulation study: absence of genetic and gene-gene interaction effects
but presence of environmental effect}}
\label{subsec:third_simulation_study}

In this study we consider a case-control genotype data set simulated from GENS2 where case-control status depends
only upon the environmental data. The number of cases here is 47 and the number of controls is 53.
This is the same case-control genotype data set as used by \ctn{Bhattacharya16} in their third 
simulation study.

In this case, we find that 
$P\left(d^*<\varepsilon_{d^*}|\mbox{Data}\right)\approx 0.400$, which provides reasonable evidence in favour
of the null, even though the 0.5 benchmark is not crossed. Moreover,
$P\left(|\beta_{G}|<\varepsilon_{\beta_{G}}|\mbox{Data}\right)\approx 0.536$, 
$P\left(|\beta_{G_0}|<\varepsilon_{\beta_{G_0}}|\mbox{Data}\right)\approx 0.518$
and $P\left(|\beta_{H}|<\varepsilon_{\beta_{H}}|\mbox{Data}\right)\approx 0.504$, suggesting that the environmental
variable does not affect the genetic structure.
\ctn{Bhattacharya16} show by means AIC, %of Akaike Information Criterion (AIC), 
in the context of simple logistic regression, that the best model consists of the marginal effects of the second gene and the environment. In conjunction
with our HDP-based model which produces reasonable evidence in favour of accepting the hypothesis of no genetic effect, 
it may be possible to conclude that the environmental variable is responsible for the case-control status.

As before, 5 subpopulations get significant weight by our posterior distribution, and again, the question of
DPL detection is irrelevant here since there is no genetic effect.
%\begin{figure}%[htp]
%\centering
%\subfigure[Index plot for the first gene]{ \label{fig:index_plot_gene1_only_env}
%\includegraphics[width=8cm,height=8cm]{plots/plot_loci_gene1_only_env-crop.pdf}}
%%\vspace{2mm}
%\subfigure[Index plot for the second gene.]{ \label{fig:index_plot_gene2_only_env}
%\includegraphics[width=8cm,height=8cm]{plots/plot_loci_gene2_only_env-crop.pdf}}
%\caption{{\bf Absence of genetic and gene-gene interaction effects
%but presence of environmental effect:} Plots of the Euclidean distances 
%$\left\{d^r_j\left(\mbox{logit}\left(\bp^r_{i_0jk=0}\right),\mbox{logit}\left(\bp^r_{i_1jk=1}\right)\right);
%~r=1,\ldots,L_j\right\}$
%against the indices of the loci, for $j=1$ (panel (a)) and $j=2$ (panel (b)).}
%\label{fig:index_plots_only_env}
%\end{figure}

\subsection{{\bf Fourth simulation study: presence of genetic and gene-gene interaction effects
but absence of environmental effect}}
\label{subsec:fourth_simulation_study}

Here we use the same genotype data set as used by \ctn{Bhattacharya15} in their first simulation study
associated with genetic and gene-gene interaction effects, 
consisting of 41 cases and 59 controls and 5 sub-populations with the mixing proportions
$(0.1, 0.4, 0.2, 0.15, 0.15)$. We use the same environmental data set generated in our first simulation study described in
Section \ref{subsec:first_simulation_study}, which is unrelated to this case-control genotype data. 

\begin{figure}%[htp]
\centering
\subfigure[Posterior probability of no gene-gene interactions in control subjects.]{ \label{fig:control_ggi_2}
\includegraphics[width=15cm,height=8cm]{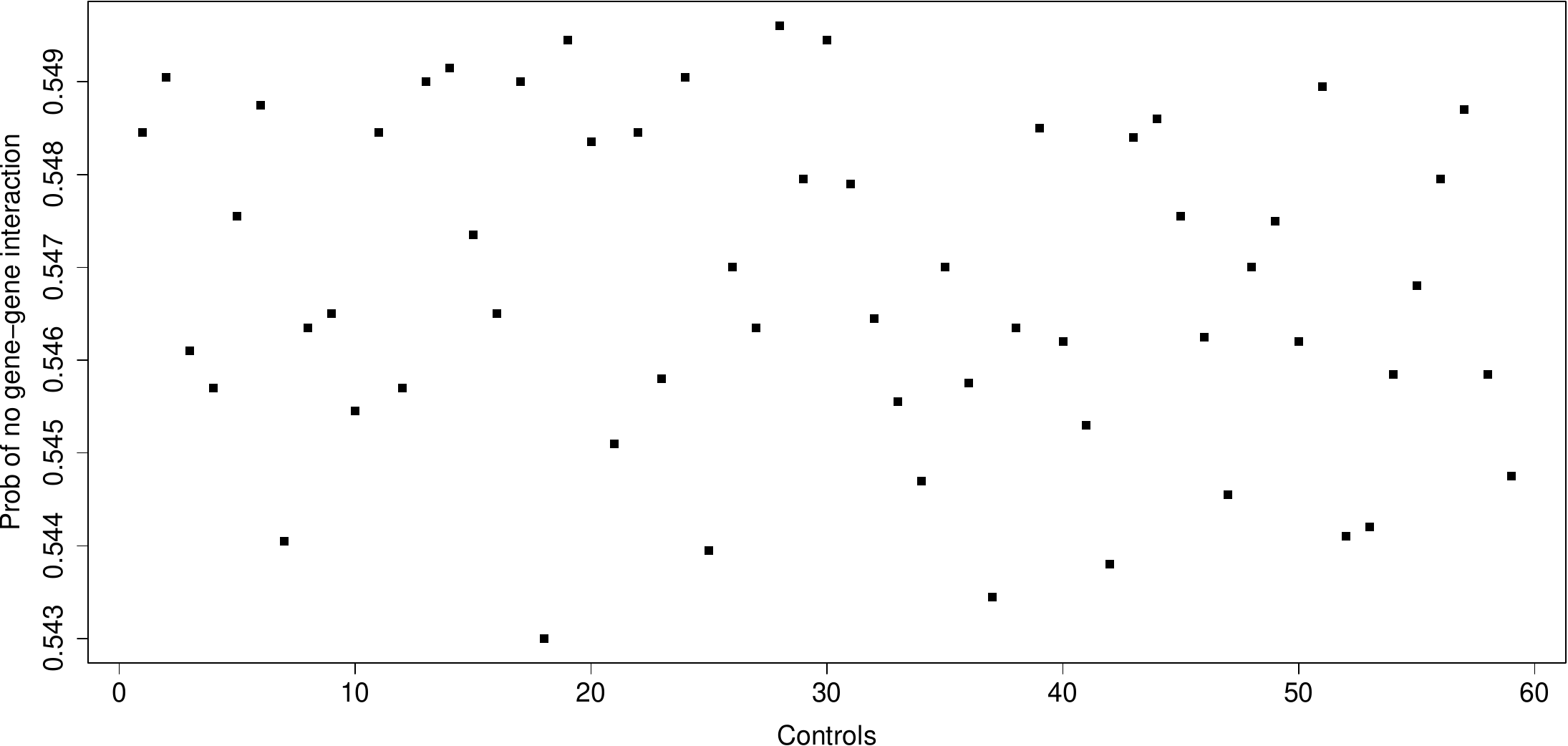}}\\
\vspace{4mm}
\subfigure[Posterior probability of no genetic effect with respect to cases.]{ \label{fig:case_ggi_2}
\includegraphics[width=15cm,height=8cm]{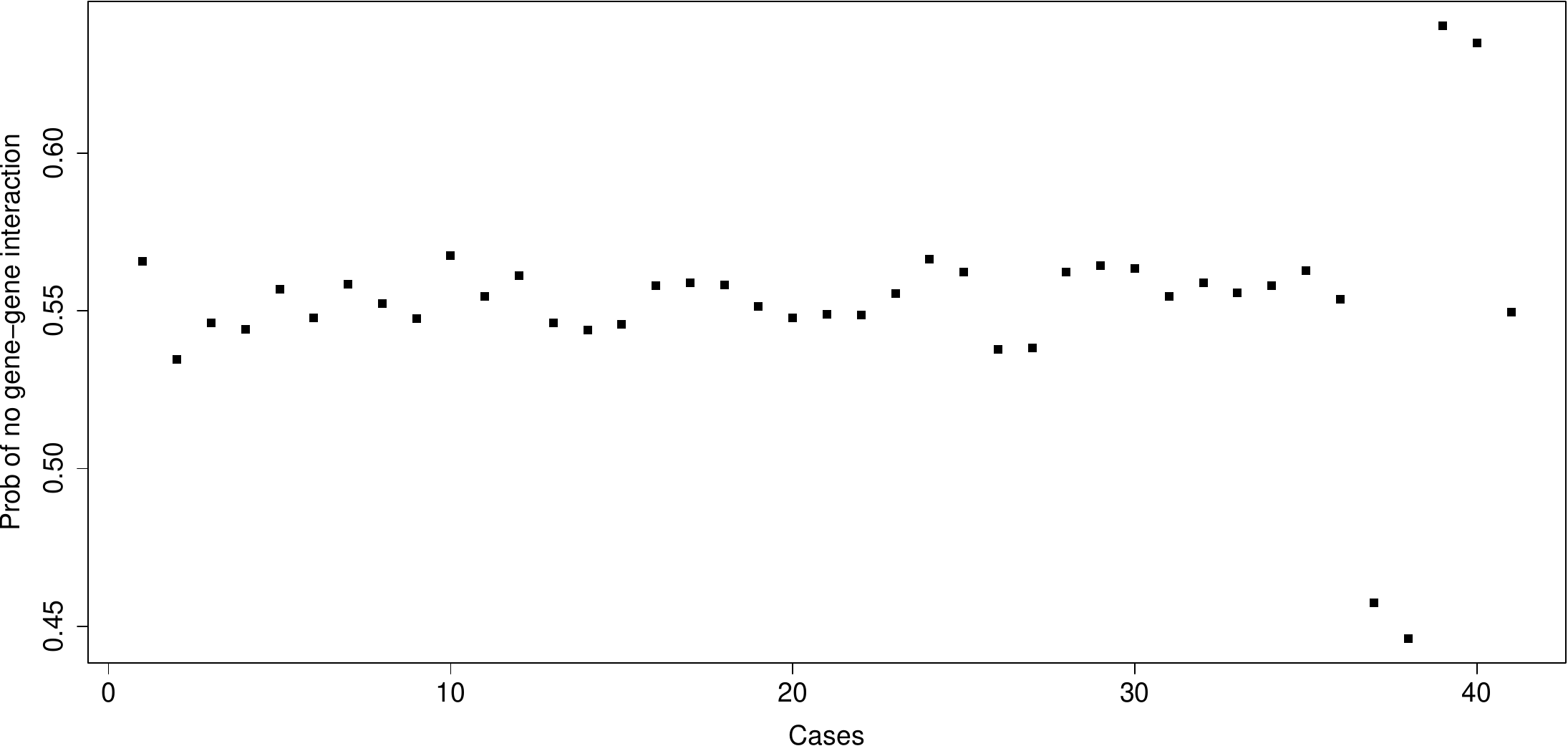}}
\caption{{\bf Presence of genetic and gene-gene interaction effects but absence of environmental effect:}
Index plots of the posterior probabilities of no gene-gene interactions in (a) controls
and (b) cases, with respect to the two genes.}
\label{fig:prob_no_ggi_2}
\end{figure}

Here we obtain
$P\left(|\beta_{G}|<\varepsilon_{\beta_{G}}|\mbox{Data}\right)\approx 0.549$, 
$P\left(|\beta_{G_0}|<\varepsilon_{\beta_{G_0}}|\mbox{Data}\right)\approx 0.542$
and $P\left(|\beta_{H}|<\varepsilon_{\beta_{H}}|\mbox{Data}\right)\approx 0.552$, correctly suggesting insignificance 
of the environmental variable with respect to its effect on the genetic structure.
Using logistic regression, \ctn{Bhattacharya16} conclude that the environmental variable 
has no role to play in the case-control status. 
Furthermore, we obtain 
$P\left(d^*<\varepsilon_{d^*}|\mbox{Data}\right)\approx 0.390$, 
$P\left(\hat d_1<\varepsilon_{\hat d_1}|\mbox{Data}\right)\approx 0.336$
$P\left(\hat d_2<\varepsilon_{\hat d_2}|\mbox{Data}\right)\approx 0.324$.
so that importance of genes is correctly indicated by our tests. 
Figure \ref{fig:prob_no_ggi_2} shows the posterior probabilities of no gene-gene interactions for controls and cases.
Interestingly, there seems to be no gene-gene interaction in the control group and only two (marginal) instances of gene-gene
interaction among the cases. 

\begin{figure}%[htp]
\centering
\subfigure[Index plot for the first gene]{ \label{fig:index_plot_gene1_gg}
\includegraphics[width=8cm,height=8cm]{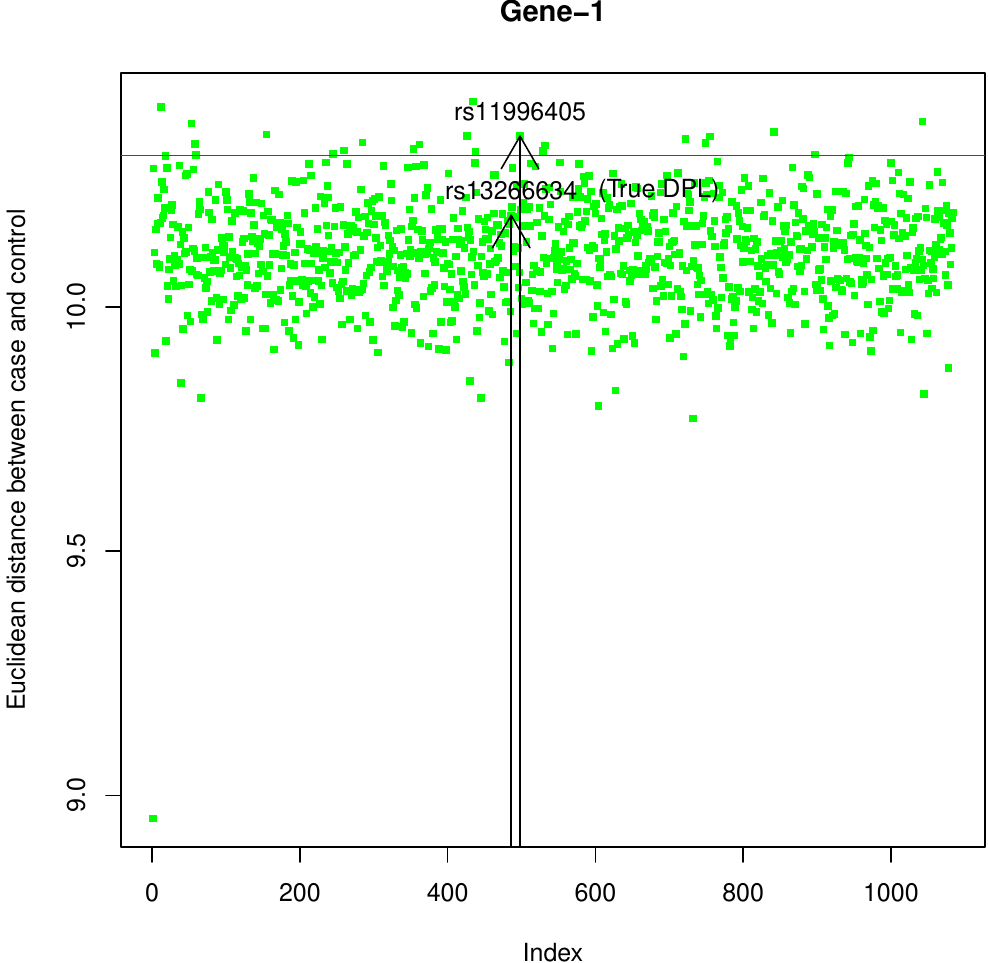}}
%\vspace{2mm}
\subfigure[Index plot for the second gene.]{ \label{fig:index_plot_gene2_gg}
\includegraphics[width=8cm,height=8cm]{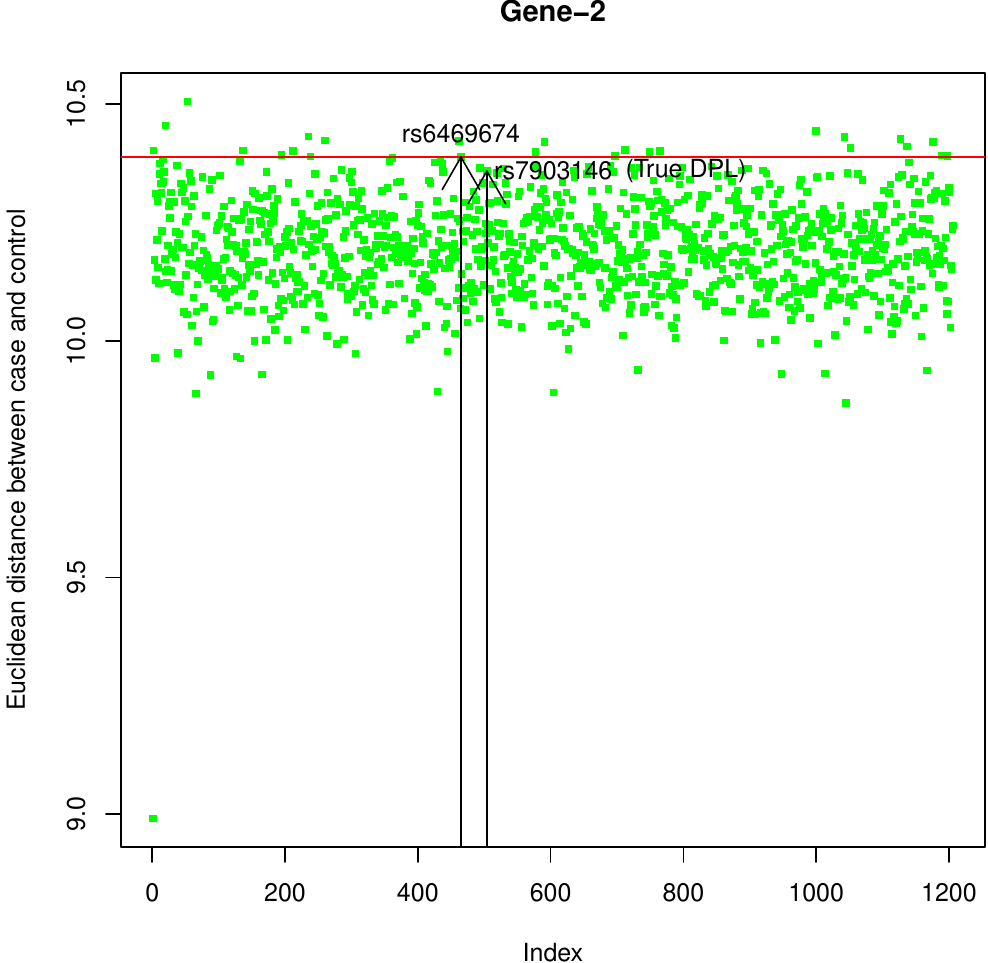}}
\caption{{\bf Presence of genetic and gene-gene interaction effects
but absence of environmental effect:} Plots of the Euclidean distances 
$\left\{d^r_j\left(\mbox{logit}\left(\bp^r_{i_0jk=0}\right),\mbox{logit}\left(\bp^r_{i_1jk=1}\right)\right);
~r=1,\ldots,L_j\right\}$
against the indices of the loci, for $j=1$ (panel (a)) and $j=2$ (panel (b)).}
\label{fig:index_plots_gg}
\end{figure}

Figure \ref{fig:index_plots_gg} shows the plots of Euclidean distances between cases and controls for the loci
of the two genes. In this case, Gene-1 has been located quite precisely, and for Gene-2 the Euclidean
distance for even the true DPL is very close to the red, horizontal line, indicating encouraging performance.

%For Gene-1, the Euclidean distance for the actual disease predisposing locus, very encouragingly, 
%falls much above the cut-off line. For Gene-2, the Euclidean distance falls slightly below the red, horizontal line.
%Overall, the results have been much more encouraging compared to the first simulation study concerning both gene-gene
%and gene-environment interaction. This improved performance may perhaps be attributed to the simpler dependence structure
%of the true model.

\subsection{{\bf Fifth simulation study: independent and additive genetic and environmental effects}}
\label{subsec:fifth_simulation_study}

As in \ctn{Bhattacharya16}, we consider the situation where the genetic and environmental effects
are independent of each other and additive; the data consists of 57 cases and 43 controls. 

\begin{figure}%[htp]
\centering
\subfigure[Posterior probability of no gene-gene interactions in control subjects.]{ \label{fig:control_ggi_3}
\includegraphics[width=15cm,height=8cm]{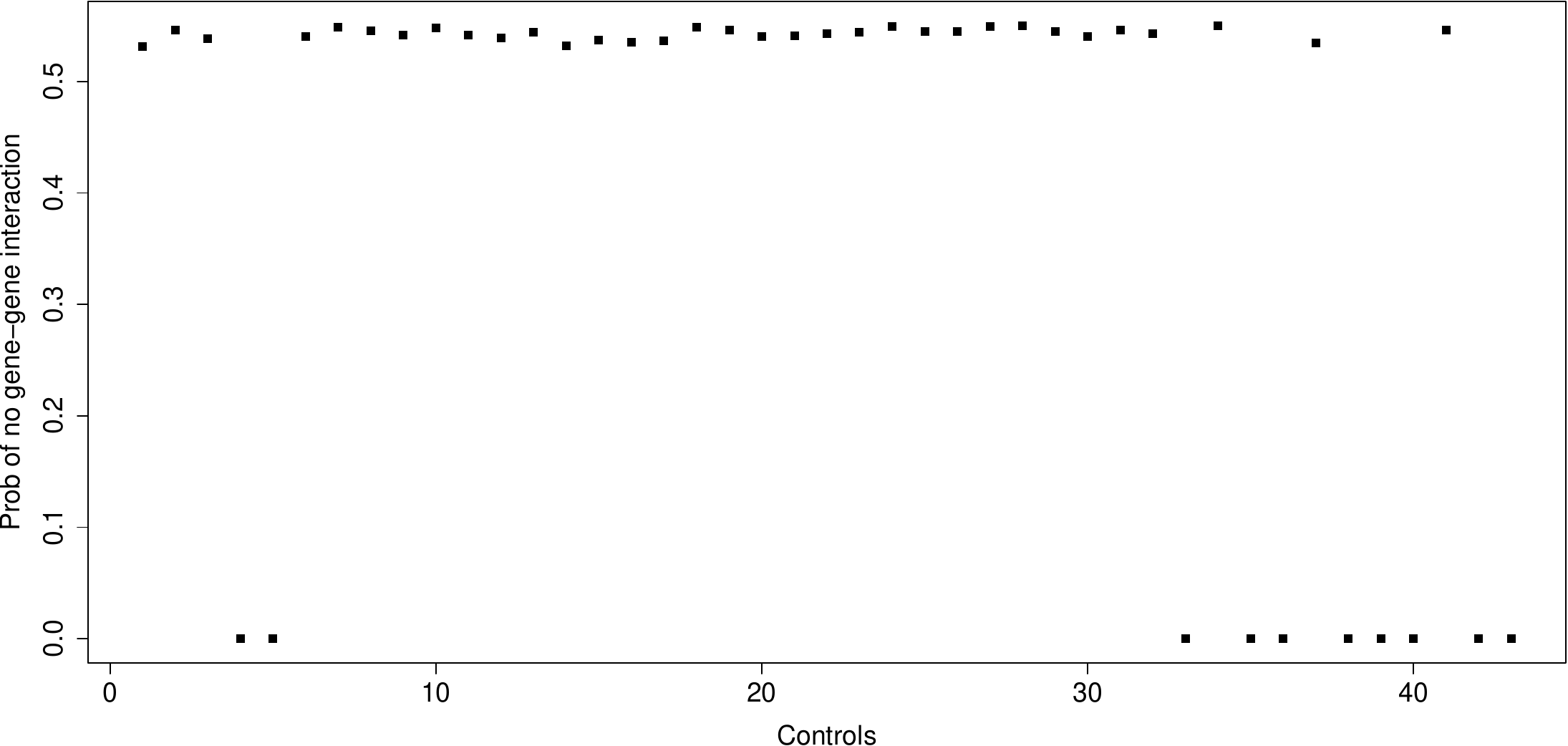}}\\
\vspace{4mm}
\subfigure[Posterior probability of no genetic effect with respect to cases.]{ \label{fig:case_ggi_3}
\includegraphics[width=15cm,height=8cm]{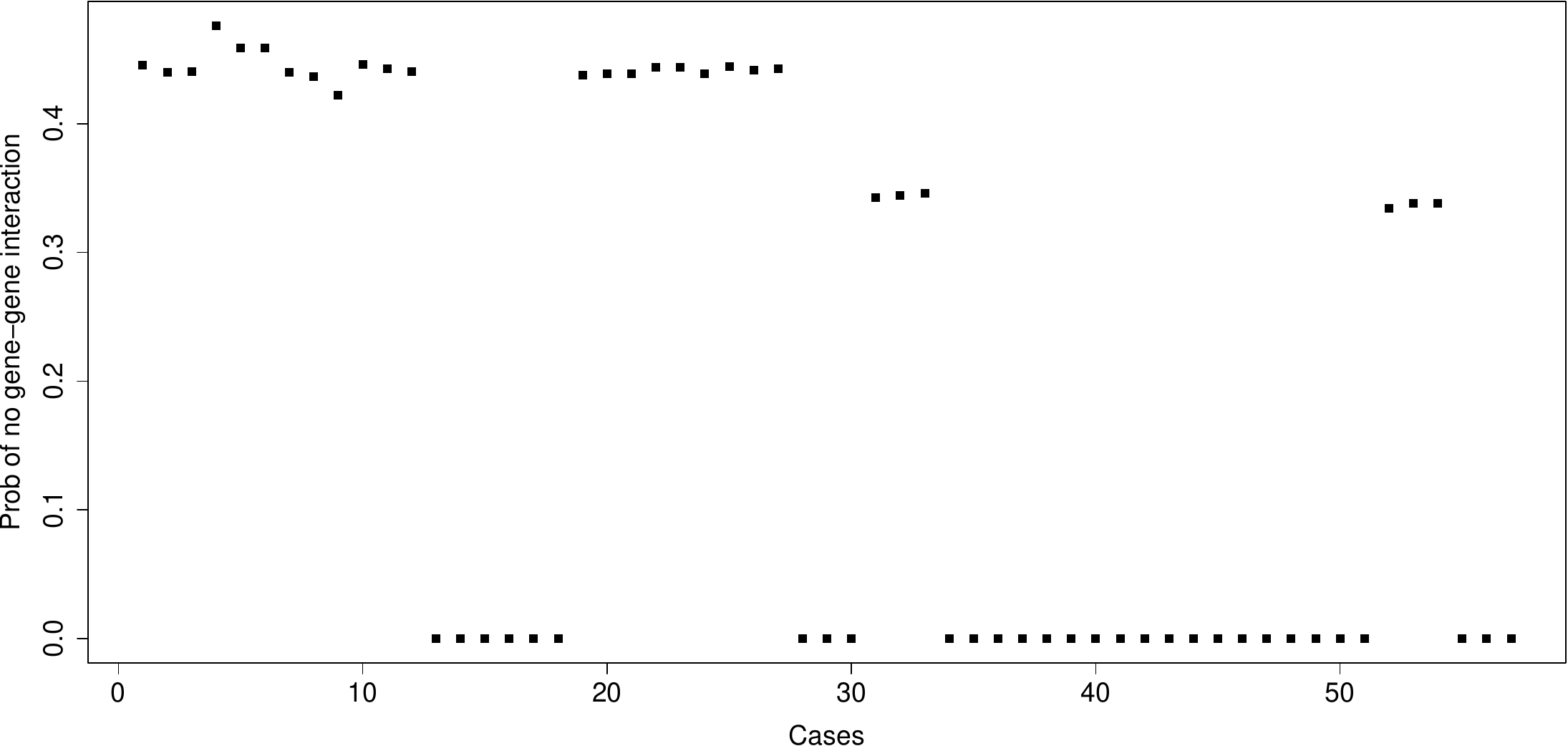}}
\caption{{\bf Presence of genetic and gene-gene interaction effects but absence of environmental effect:}
Index plots of the posterior probabilities of no gene-gene interactions in (a) controls
and (b) cases, with respect to the two genes.}
\label{fig:prob_no_ggi_3}
\end{figure}

Note that, as in \ctn{Bhattacharya16}, in our current HDP-based Bayesian model also 
there is no provision for additivity of genetic and environmental effects. As such, it is not expected
to capture the true data-generating mechanism accurately.
Indeed, here we obtain
$P\left(d^*<\varepsilon_{d^*}|\mbox{Data}\right)\approx 0.389$, 
$P\left(\hat d_{1}<\varepsilon_{\hat d_{1}}|\mbox{Data}\right)\approx 0.337$
and $P\left(\hat d_{2}<\varepsilon_{\hat d_{2}}|\mbox{Data}\right)\approx 0.331$,
indicating significance of the genes. However, the test with $d^*_E$ does not yield overwhelming evidence
against the null. Our tests of gene-gene interaction, as depicted in Figure \ref{fig:prob_no_ggi_3}, indicate significant
interactions for controls and particularly for cases.
Also, $P\left(|\beta_{G}|<\varepsilon_{\beta_{G}}|\mbox{Data}\right)$, 
$P\left(|\beta_{G_0}|<\varepsilon_{\beta_{G_0}}|\mbox{Data}\right)$
and $P\left(|\beta_{H}|<\varepsilon_{\beta_{H}}|\mbox{Data}\right)$ are given, approximately, by $0.547$, $0.550$ and 
$0.367$, the last value showing that the environmental variable does affect gene-gene interaction.
The lack of the additivity provision in our model seems to have forced the gene-environment interaction
in this case.

\begin{figure}%[htp]
\centering
\subfigure[Index plot for the first gene]{ \label{fig:index_plot_gene1_add}
\includegraphics[width=8cm,height=8cm]{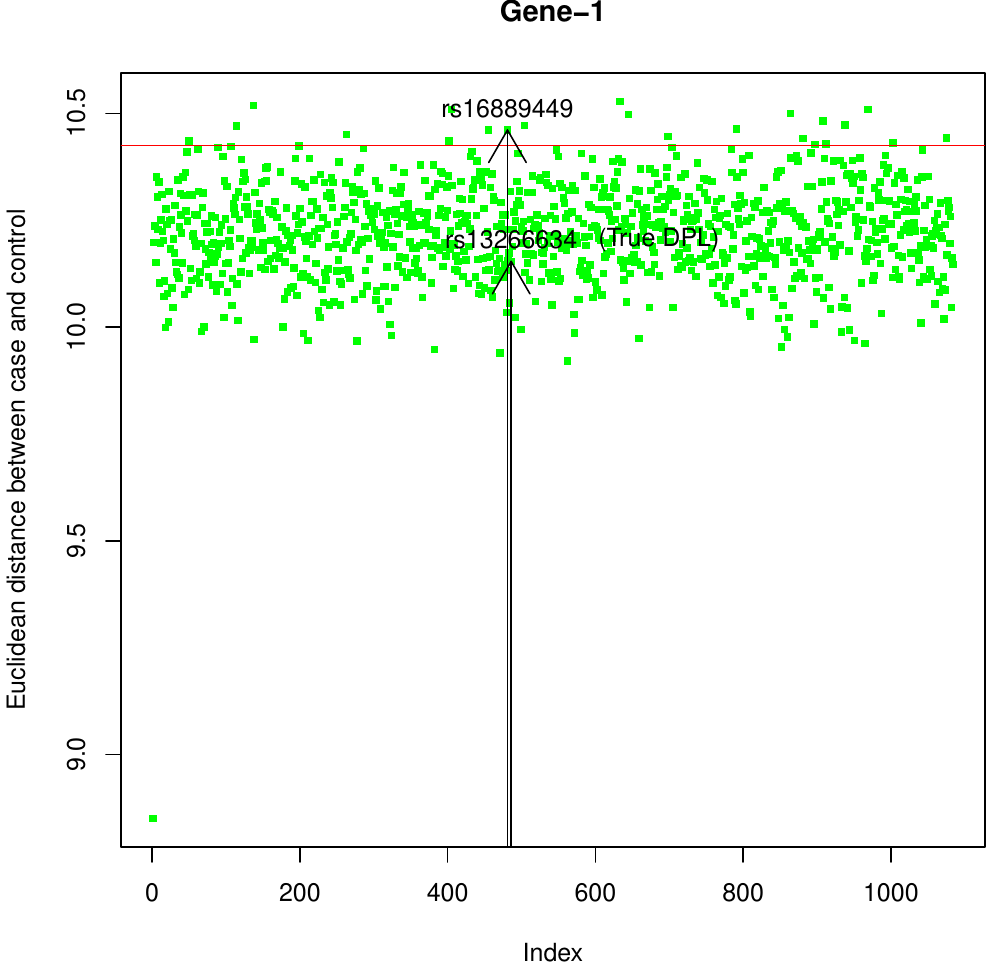}}
%\vspace{2mm}
\subfigure[Index plot for the second gene.]{ \label{fig:index_plot_gene2_add}
\includegraphics[width=8cm,height=8cm]{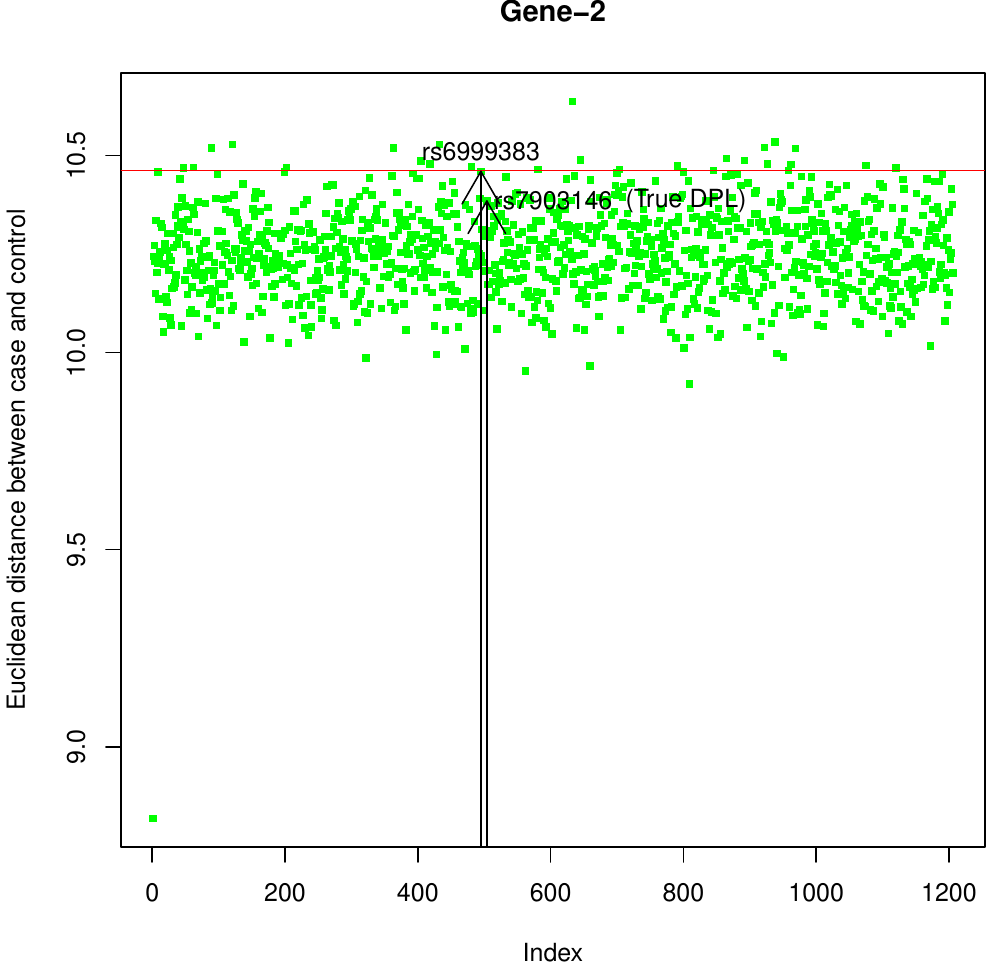}}
\caption{{\bf Independent and additive genetic and environmental effects:} Plots of the Euclidean distances 
$\left\{d^r_j\left(\mbox{logit}\left(\bp^r_{i_0jk=0}\right),\mbox{logit}\left(\bp^r_{i_1jk=1}\right)\right);
~r=1,\ldots,L_j\right\}$
against the indices of the loci, for $j=1$ (panel (a)) and $j=2$ (panel (b)).}
\label{fig:index_plots_add}
\end{figure}

In spite of the lack of additivity of our model %which forces the artificial gene-environment interaction also seems to 
the Euclidean distances between cases and controls for the gene-wise SNPs are not adversely affected, and the actual DPLs
are detected quite accurately; see Figure \ref{fig:index_plots_add}. This brings forth the generality and usefulness of our
nonparametric dependence structure.
As before, 5 sub-populations receive significant posterior probabilities.